\documentclass[10pt]{ClemsonThesis}
\pdfoutput=1
\title{The Pursuit for Gamma-ray Emitting Pulsar Wind Nebulae with the Fermi--Large Area Telescope}
\department{College of Science}
\documentType{Dissertation}
\major{Physics}
\degrees{Doctor of Philosophy}
\graduationMonth{August}
\graduationYear{2022}
\author{Jordan L. Eagle}
\committeeChair{Dr. Marco Ajello}
\committeeMemberOne{Dr. Daniel Castro, co-Chair}
\committeeMemberTwo{Dr. Dieter Hartmann}
\committeeMemberThree{Dr. Joan Marler}

\hypersetup{
    colorlinks,
    linkcolor={black},
    citecolor={black},
    filecolor={black},
    urlcolor={black},
    pdftitle={\theTitle},
    pdfauthor={\theAuthor},
    pdfsubject={\theDocumentType},
    pdfkeywords={Clemson University, \theDepartment, \theDocumentType, \theMajor, \theDegree},
    pdfstartpage={1},
}

\newcommand{\cha}{\textit{Chandra }}
\def\xmm{{XMM-{\it Newton }}}
\def\fermi{{{\it Fermi}--LAT }}

\usepackage{hyperref}
\usepackage[numbers]{natbib}
\usepackage{bibentry}
\usepackage{rotating}
\usepackage{pdflscape}
\usepackage{url}
\usepackage{amsmath,amssymb}
\usepackage{longtable}
\usepackage{gensymb}
\usepackage{float}
\usepackage{color}
\usepackage{graphicx}
\usepackage[utf8]{inputenc}
\usepackage{enumitem}
\usepackage{multirow,helvet}
\usepackage{todonotes}
\usepackage{lipsum,wrapfig} 
\usepackage{epstopdf}

\begin{document}
    \frontmatter 

    \addtotoc{Title Page}{\maketitle}          
    \doublespacing                             
    \setcounter{page}{2}                       
    \addtotoc{Abstract}{\pdfoutput=1
\chapter*{Abstract}
Pulsar wind nebulae are highly magnetized particle winds, descending from core collapse supernovae (CC SNe), and each powered by an energetic, rapidly rotating neutron star. There are at least 125 Galactic pulsar wind nebulae (PWNe) that have been discovered from radio wavelengths to TeV $\gamma$-rays, the majority of which were first identified in radio or X-ray surveys. An increasing number of PWNe are being identified in the TeV band by ground-based air Cherenkov Telescopes such as HESS, MAGIC, and VERITAS such that they constitute the dominant source class of Galactic TeV emitters. High-energy sources like PWNe may be responsible for producing the bulk of Galactic cosmic rays (CRs) with energy up to $E \sim 10^{15}$\,eV. Combining available MeV--GeV data with observations in the TeV band is critical for precise characterization of high-energy emission from the relativistic particle population in PWNe, thus revealing the capability to produce a significant fraction of the detected Galactic CR flux. However, MeV--GeV PWN counterparts are still largely lacking even after 12 years of continuous observation of the entire sky. Less than a dozen PWNe are currently identified by the {\it Fermi}--LAT in the MeV--GeV band. Most PWNe are located along the Galactic plane embedded within the prominent, diffuse Galactic $\gamma$-ray emission, which makes these sources difficult to disentangle from the bright diffuse background. Moreover, nearly 300 rotation-powered pulsars that are capable of generating PWNe can also emit brightly in the \fermi $\gamma$-ray band, potentially outshining and obscuring their fainter PWNe. The capability to identify more $\gamma$-ray PWNe is greatly improved by the recent upgrades in the event processing of the \fermi data, which provides better spatial resolution and sensitivity of the instrument. Taking advantage of the recent upgrade, we present a systematic search for $\gamma$-ray counterparts to known PWNe in the 300\,MeV -- 2\,TeV energy band using 11.5 years of \fermi data. For the first part of this search, we target the locations of PWNe previously identified across the electromagnetic spectrum that are not powered by pulsars previously detected by the \fermi as pulsating $\gamma$-ray signals, which includes 6 {\it Fermi} PWNe and 7 {\it Fermi} PWN associations. We report the analysis of 58 total regions of interest and provide all firm and tentative detections along with their morphological and spectral characteristics. There are 11 unidentified $\gamma$-ray sources that we classify as firm PWN counterparts, which doubles the PWN population detected by the {\it Fermi}--LAT, and 22 $\gamma$-ray sources that are PWN candidates. This will represent a catalog of \fermi PWNe, named the \fermi PWN catalog, or 1PWN. 

Understanding the PWN population and the interactions that take place are essential for identifying how relativistic particles are injected into the ISM, how they contribute to replenishing the Galactic CR population, and whether they are responsible for local enhancements in the e$^-$e$^+$ flux. For two newly-detected PWNe for which multiwavelength data exist, we apply developed emission models in order to expand our understanding of PWN evolution and hence the underlying particles. The $\gamma$-ray data are combined with available multiwavelength data and compared to the intrinsic properties of the associated systems, such as the supernova explosion energy and pulsar characteristics, in order to establish basic energetic and evolutionary trends for the PWN population. 

}  

    %
    %
    \addtotoc{Dedication}{\pdfoutput=1
\chapter*{Dedication}
For all the pulsar wind nebulae that have not yet been cataloged anywhere.... }

    %
    %
    \addtotoc{Acknowledgments}{\pdfoutput=1
\chapter*{Acknowledgments}
First, I wish to thank my partner, Noah Channell, for his support and love throughout all of this. You have given me a sense of familiarity, comfort, and safety that has enabled me to grow as a person and scientist in my field. Thank you for continuing to be there even as I moved up to Boston to complete the predoctoral fellowship. I am truly grateful for you and the life we have created! I must also specifically acknowledge our three pets Ruca, Mars, and Bella for the small cuddly moments and for Ruca's invaluable running partnership that kept me sane. The three of you should be sure to thank Noah as well for his support because it certainly extended to helping take proper care of you. 

Second, I need to thank my mentors and advisors. Daniel Castro, thank you for your patience, understanding, expertise, and leadership. I have learned so much from you! From pulsar wind nebulae to collaborating to presenting to even the smallest things like designing budget proposals and ever tinier things like writing a particular e-mail or finding a colleague's current contact information.... I don't know if I can articulate properly how grateful I am for everything. I am eager to continue our shared research interests as collaborators moving forward! I equally look forward to continued collaboration with colleagues Dan introduced me to and have also learned a great deal from including Patrick Slane, Joseph Gelfand, Samayra Straal, Tea Temim, and Matthew Kerr. My ``home'' advisor, Marco Ajello, I wish to also thank. If it weren't for Marco messaging me over two years ago with the link to the Chandra X-ray Center predoctoral fellowship saying ``I am trying to get you to leave Clemson forever'', I would have never known to even apply to such an opportunity, which I feel ended up catapulting my career into an exciting path of discovery and knowledge! Marco has been a continued resource for me even as I began focusing on Daniel's pulsar wind nebula project. Marco regularly encouraged me and other graduate students in his research group to apply for awards, scholarships, grants, and more. He has truly turned us all into proposal writing and reviewing machines who now look out for their own opportunities to apply for. Marco and Dan share this trait and have undoubtedly moulded well-rounded researchers because of it. I feel really, stupidly grateful for the mentorship experiences I have had from Marco, Dan, and countless others. It has been a constructive and fun experience. Honestly, I feel that without it I would not be sitting here writing the acknowledgement section to a Ph.D. thesis right now.

Lastly, I would like to thank my family for the unconditional support! Mom, Dad, my older sister Jamie, my nephew Brantley, Grandma, Tom, and Oma and Opa! I would also like to thank my Grandpa, even though he has passed, since he was and remains a pivotal influence to me. }

    \singlespacing                             
    \tableofcontents \clearpage                

    %
    %
    \addtotoc{List of Tables}{\listoftables}   
    \addtotoc{List of Figures}{\listoffigures} 

    %
    %

    \mainmatter 
    \doublespacing 

    %
    %
    
    \pdfoutput=1
\chapter{Introduction}
\section{Pulsar Wind Nebulae}
\begin{wrapfigure}{r}{0.57\textwidth}
\centering
\includegraphics[width=1.0\linewidth]{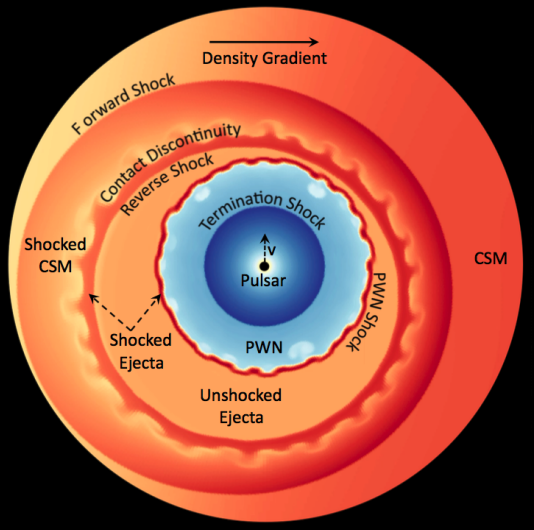}
\vspace{-.5cm}
\caption{Density image from a hydrodynamical simulation of a PWN expanding into an SNR
that is evolving into a circumstellar medium (CSM) density gradient increasing to the right. The reverse shock is propagating inward, approaching the PWN preferentially from the upper right due to the combined effects of the pulsar motion and the CSM density gradient. Adapted from \citep{slane2017}.}\label{fig:slane}
\end{wrapfigure}

A core collapse supernova (CC SN) is triggered by the collapse of a massive star and ejects several solar masses worth of stellar ejecta at thousands of kilometers per second into the surrounding interstellar medium (ISM, \citep{slane2017}). The ejected material forms the outer layer, the supernova remnant (SNR) shell also known as the forward shock, which compresses and heats up ambient gas that is swept up as it expands. The forward shock decelerates as it collects pressurized ISM gas, generating a reverse shock that accelerates back into the cold SN ejecta, in the SNR interior \citep{slane2006}. Often times, a rapidly spinning, highly magnetic neutron star is left behind at the explosion site. The neutron star loses angular momentum as rotational energy is converted to a relativistic particle wind made up of mostly electrons and positrons. Hence, pulsar wind nebulae (PWNe) are descendants of core collapse supernovae (CC SNe), each powered by an energetic, rapidly rotating neutron star. A diagram illustrating the dynamic layers of a CC SNR is displayed in Figure~\ref{fig:slane}.

SNRs and their PWNe are commonly categorized in distinct groups based on their observed radio morphology: shell-type remnants, Crab-like remnants, and composite remnants \citep{dubner2015,weiler1988}. Shell-type remnants are characterized by a limb-brightened shell which represents the forward shock of the SNR \citep{weiler1988}. 
Crab-like remnants, named after the famous Crab nebula, are also known as pulsar wind nebulae or plerions. Plerionic SNRs such as the Crab are characterized by a centrally-filled core (i.e., the PWN) and are often linearly polarized in radio \citep{dubner2015}. 
Finally, composite remnants are observed to have both an SNR shell and a centrally-peaked core. 

\singlespacing
\doublespacing
SNRs and PWNe may be able to explain the bulk of Galactic cosmic rays (CRs) having energies up to $10^{15}\,$eV \citep[e.g.,][]{slane2006,hillas2006,reynolds2008}. Both the forward shock of the SNR and the termination shock of the PWN are potential sites for efficient particle acceleration. The termination shock of a PWN is generated
where the ram pressure of the cold pulsar wind is balanced by the pressure of the highly relativistic plasma. Altogether, the nebula is confined by the surrounding SN ejecta (see e.g., Figure~\ref{fig:slane}, \citep{slane2017}). The termination shock is where particles injected by the central pulsar are accelerated as they enter the non-thermal pool of relativistic particles in the nebula. Synchrotron emission from the relativistic electrons accelerating in the magnetic fields within the nebula is observed from the majority of PWNe, from radio wavelengths to hard X-rays, while CR electrons are thought to scatter off of local photon fields, resulting in Inverse Compton (IC) emission at $\gamma$-ray energies.

Historically, PWNe like the Crab were discovered in droves in the radio and X-ray wavelengths and now $\gamma$-rays are becoming key to finding and characterizing PWNe. Cherenkov telescopes are particularly effective with the majority of identified Galactic TeV emitters classified as PWNe\footnote{\url{tevcat.uchicago.edu}} \citep{tevcat2008}. The observed TeV $\gamma$-ray spectra from many of these sources imply that the peak occurs in the MeV--GeV band, making the {\it Fermi}--Large Area Telescope (LAT) the ideal PWN hunting tool. MeV--GeV observations are also effective in identifying active pulsars that peak in the GeV band, but which may have faint PWNe in other wavelengths \citep[e.g.,][]{ackermann2011}. Lastly, combining \fermi and TeV data are critical for precise characterization of the high-energy emission spectra from PWNe which reveals the properties of the underlying particle population. While the radiation processes observed across the electromagnetic spectrum from PWNe are better known,
the particle acceleration processes that must be occurring within the PWN to generate the highest energy particles and subsequent radiation remain difficult to determine precisely. Understanding the PWN population are essential for identifying how the relativistic particles are injected into the ISM, how they contribute to the Galactic CR population \citep{karg2013,renaud2009}, and whether they are responsible for local enhancements in the e$^-$e$^+$ flux \citep{malyshev_2009}. 

The termination shock of a PWN is considered a likely site for efficient particle acceleration, notably the diffusive shock acceleration (DSA) process and in turn produces Galactic CRs \citep{sironi2017}. Particle-in-cell (PIC) simulations imply the DSA process may not be the only acceleration mechanism occurring at the shock boundary and also indicate that efficient particle acceleration is not confined to only the termination shock of a PWN \citep[e.g.,][]{sironi2017,sironi2008}. In general, DSA theory cannot easily explain observations of PWNe alone. As an example, the broadband spectrum of the Crab nebula and the implied electron distribution deviates from the DSA framework. A single power-law particle injection spectrum with an index greater than 2 is expected if particles are undergoing DSA, however, the observed spectrum from the Crab implies a particle index of $\sim 1.5$ \citep{lyubarsky2003,sironi2011}. 
Recent work \citep{lyutikov2019} has shown that two electron populations can explain the broadband spectrum, one which could be undergoing efficient particle acceleration via DSA in the equatorial regions (i.e. at the termination shock) and a second component where efficient particle acceleration occurs from magnetic reconnection layers within the nebula. 

\begin{wrapfigure}{l}{0.57\textwidth}
\centering
\vspace{-1.15cm}
\includegraphics[width=1.0\linewidth]{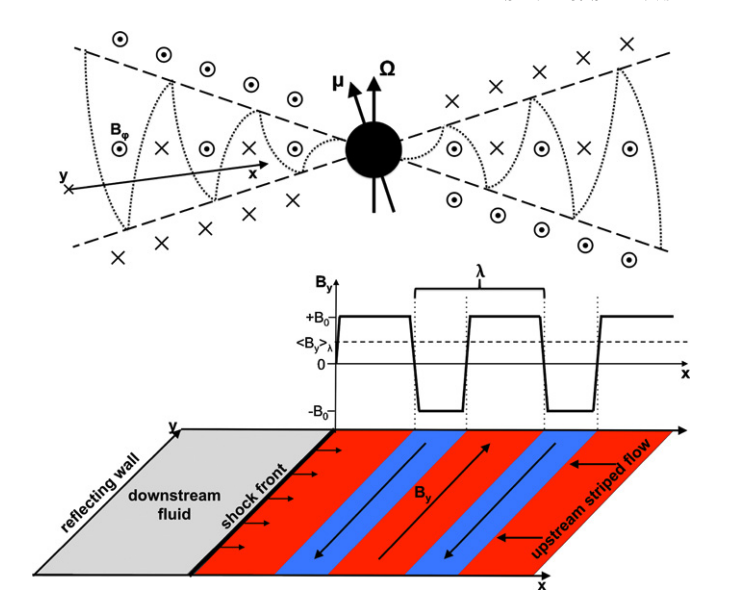}
\vspace{-.5cm}
\caption{The top panel illustrates the poloidal structure of a striped pulsar wind such as the Crab Nebula. Vectors indicate the pulsar rotational ($\Omega$) and magnetic axis ($\mu$) respectively. The stripes consist of toroidal magnetic field lines that alternate in polarity and are separated by current sheets (dotted lines). Alternate polarity is only expected within the equatorial region, marked by the dashed lines. The bottom panel is a 2D simulated geometry of the pulsar wind. The colder pulsar wind flows along $-\hat{x}$ and the termination shock is receding from the reflecting wall toward $+\hat{x}$. The magnetic field lines are perpendicular to both in the $y$-direction. The alternate polarity is indicated as red and blue colors. There is a non-zero net average magnetic field when choosing larger values of the magnetic field strength for only one polar direction. Adapted from \citep{sironi2011}.}\label{fig:sironi}
\end{wrapfigure}

Magnetic reconnection has been explored in sources like PWNe by using a ``striped wind'' model. The striped wind refers to the upstream flow structure as alternating stripes of magnetic field polarity, separated by current sheets of hot plasma, and is injected into the termination shock. See Figure~\ref{fig:sironi} for a diagram of a striped wind \citep{sironi2011}. If the magnetic field axis and the pulsar's rotational axis are not aligned as is suspected with the Crab Nebula, this could generate a striped pulsar wind. It has been proposed that DSA can occur in the equatorial regions so long as magnetic reconnection can transfer the bulk of the magnetic field energy to the particles at the termination shock (e.g., the ``sigma ($\sigma$) problem'', \citep{sironi2011}). In the case for the Crab nebula, shock-driven magnetic reconnection enables diffusive shock acceleration and can readily explain the X-ray spectrum \citep{lyutikov2019}. The broadband spectrum for the Crab can be appreciably characterized when combining this with a second acceleration component. The second acceleration mechanism is attributed to particles injected at higher latitudes than the equatorial wedge (Figure~\ref{fig:sironi}) that are accelerated by turbulent plasma interactions in reconnecting current sheets within the nebula. 

\singlespacing
\doublespacing
The findings suggest the synchrotron radiation from particles undergoing diffusive shock acceleration dominate in the 0.1--10\,keV X-ray range. Thereafter the magnetic reconnection layers contribute to particle acceleration, producing the MeV bump in the Crab nebula spectrum first observed by COMPTEL. In this scenario, the particles undergoing DSA dominate at lower energies (optical, UV, and X-ray), while the $\sim 100\,$keV--1\,MeV shape indicates both DSA and magnetic reconnection-driven acceleration processes, and the GeV emission is attributed to the CR electrons produced in the magnetic reconnection layers that then interact with the ambient photon fields. Magnetic reconnection-driven acceleration can also reasonably reproduce the unusually steep radio spectrum and $\gamma$-ray flares observed from the Crab \citep{crabflare2011}. Even though PWNe such as the Crab are some of the most studied and well-known objects in the high-energy sky, precise evolution paths of these nebulae have not been established. In particular, we lack clear characterization of the particle injection mechanism by the central pulsar and the subsequent particle evolution. Therefore, a systematic search and spectral characterization for MeV--GeV counterparts to PWNe detected at other wavelengths are imperative to revealing the underlying relativistic particle spectra. In the following section, we explore 
the current understanding of PWN evolution and the limitations.

\subsection{Pulsar Wind Nebula Evolution}
\begin{figure}
\begin{minipage}[b]{1.0\linewidth}
\hspace{-.75cm}
\includegraphics[width=1.10\linewidth]{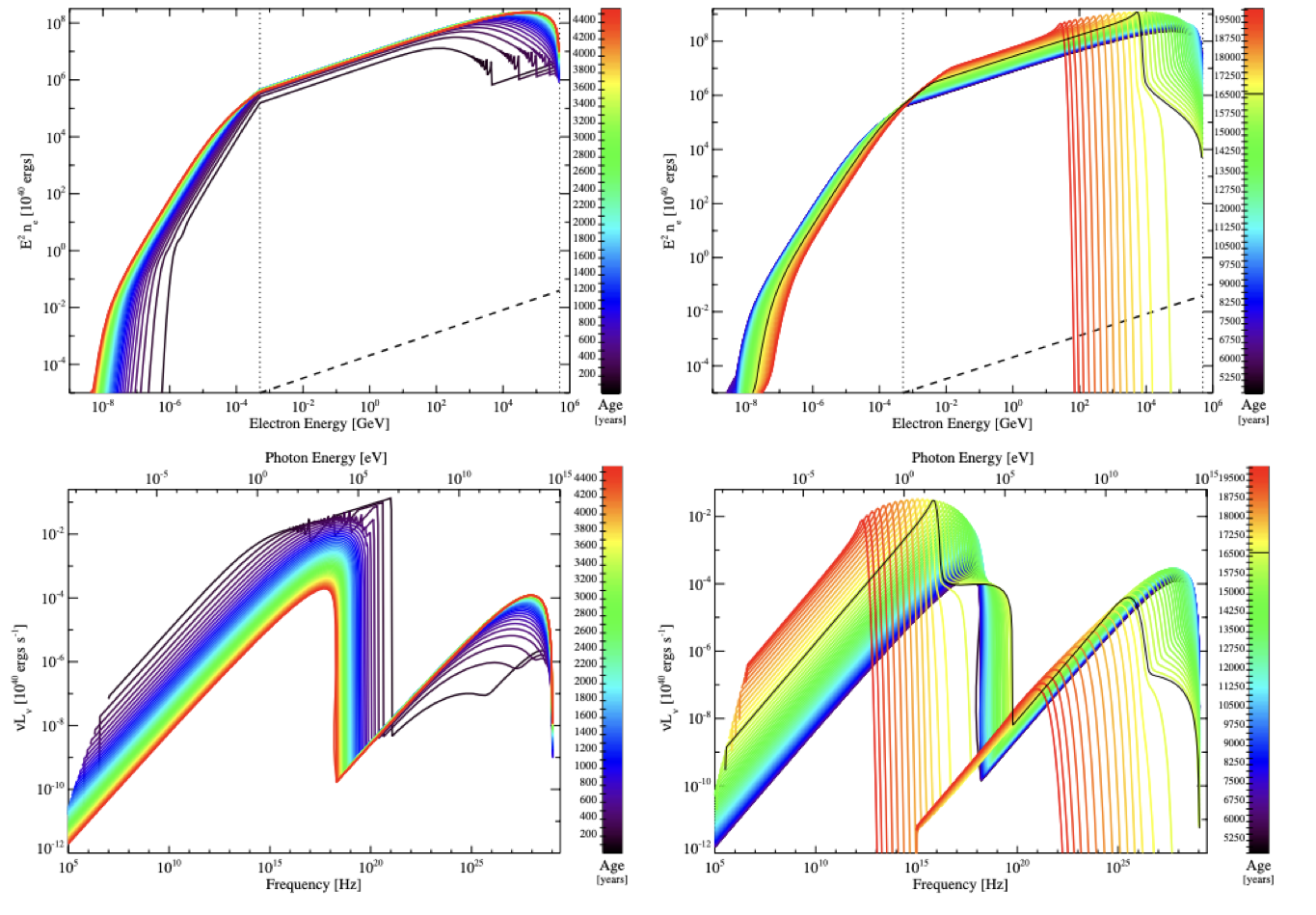}
\end{minipage}
\caption{{\it Top Panels:} The energy spectrum of electrons and positrons inside the PWN during the initial expansion (left) and compression (right) evolutionary phases predicted by the model developed in \citep{gelfand2009}. The vertical dotted lines indicate the minimum (left) and maximum (right) particle energy and the dashed line corresponds to the initial injection spectrum, assumed to be a simple power-law.  {\it Bottom Panels:} The energy spectrum of photons radiated by the particles during the initial expansion (left) and compression (right) evolutionary phases predicted by the model developed in \citep{gelfand2009}. In the right panels of both the electron and photon spectra, the black line represents the age (see color bar) when the pulsar leaves the PWN for the first time. In all panels, the different color lines correspond to the spectra at different ages. See \citep{gelfand2009} for details.}\label{fig:gelfand}
\end{figure}

A PWN evolves along with its host SNR and is influenced by the properties of the central pulsar, SNR, and the structure of the surrounding ISM \citep{slane2006}. 
How are the particles injected from the central pulsar into the nebula? How do these particles evolve once they enter the nebula? The shape of the injected particle spectrum is still poorly understood. Some models suggest that it is characterized as a simple power-law \citep{slane2008} while others predict a Maxwellian population with a power-law tail \citep{sironi2008} or a broken power law \citep{sironi2011}. Moreover, how the particle injection spectrum evolves as the PWN evolves is currently difficult to determine in part due to i) a lack of high-energy observations for the majority of PWNe, particularly in the \fermi energy range, and ii) limitations of modern simulation tools. Semi-analytic models have been developed to explore the PWN evolution inside a nonradiative SNR \citep{blondin2001, gelfand2009, swaluw2004}, each predicting efficient particle escape during the late-phase PWN evolution. Further, the evolutionary studies of PWNe indicate that the $\gamma$-ray luminosity will increase with time (e.g., Figure~\ref{fig:gelfand}), suggesting that many evolved PWNe may be emitting in the \fermi band. So far we have indeed seen that the majority of identified \fermi PWNe are relatively old, $\tau \gtrsim 5\,$kyr \citep[see e.g.,][]{devin2018,hess2012,principe2020}. We describe the semi-analytic model developed in \citep{gelfand2009} below, which is applied to the new \fermi PWN B0453--685 in Chapter~\ref{sec:b0453}.  

The dynamical and radiative properties of a PWN predicted by the evolutionary model represent a combination of neutron star, pulsar wind, supernova explosion, and ISM properties that can reasonably reproduce the observed properties of a PWN. The evolution of the nebula's power depends on the rate at which the neutron star injects energy into the PWN, the content of the pulsar wind, and the evolution of the particle and magnetic components of the PWN. For the characteristic timescale $t_{ch}$ of a pulsar \citep[see][]{slane2006,pacini1973}, the age $t_{age}$ is defined as
\begin{equation}
    t_{age} = \frac{2t_{ch}}{p-1} - \tau_{sd}
\end{equation}
and the initial spin-down luminosity $\dot{E_0}$ is defined as
\begin{equation}
    \dot{E}(t) = \dot{E_0}\big(1 + \frac{t}{\tau_{sd}}\big)^{-\frac{p+1}{p-1}}
\end{equation}
and are chosen for a braking index $p$ and spin-down timescale $\tau_{sd}$ to reproduce the pulsar's characteristic age and spin-down luminosity $\dot{E}$. A fraction $\eta_\gamma$ of this luminosity is responsible for $\gamma$-ray emission from the neutron star's magnetosphere, the rest $(1-\eta_\gamma)$ is injected into the PWN in the form of a magnetized, highly relativistic outflow, i.e., the pulsar wind. The pulsar wind is injected into the PWN at the termination shock, where the ram pressure of the unshocked pulsar wind equals the pressure of the nebula $P_{\text{PWN}}$. The termination shock (TS) location $r_{ts}$ is defined as \citep[e.g.,][]{slane2006}:  
\begin{eqnarray}
 r_{ts} = \sqrt{\frac{\dot{E}}{4\pi\xi c P_{PWN}}}
\end{eqnarray}
where $c$ is the speed of light and $\xi$ is the filling factor. $\xi = 1$ for an isotropic wind. The evolutionary model does not account for the pulsar wind before the termination shock $r < r_{ts}$ since the conditions of the wind near the neutron star are likely very different from the pulsar wind immediately downstream the termination shock. The rate of magnetic energy $\dot{E}_B$ and particle energy $\dot{E}_P$ injected into the PWN is assumed to be:
\begin{eqnarray}
\label{eqn:edotb1}
\dot{E}_B(t) & \equiv & (1-\eta_\gamma)\eta_{\rm B}\dot{E}(t) \\
\dot{E}_P(t) & \equiv & (1-\eta_\gamma)\eta_{\rm P}\dot{E}(t)
\end{eqnarray}
where $\eta_B$ is the magnetization of the wind and defined to be the fraction of the pulsar's spin-down luminosity injected into the PWN as magnetic fields and $\eta_P$ is the fraction of spin-down luminosity injected into the PWN as particles. The particle injection spectrum at the termination shock can take on a number of possible shapes. Here we assume it to be well-described by a simple power law distribution:
\begin{equation}
    \frac{d\dot{N}_{e^\pm}(E)}{dE} = \dot{N}_0 \big(\frac{E}{E_0}\big)^{-p}
\end{equation}
where $\dot{N}_{e^\pm}$ is the rate that electrons and positrons are injected into the PWN, and $\dot{N}_0$ is calculated using
\begin{equation}
    (1-\eta_B)\dot{E} = \int_{E_{min}}^{E_{max}} E \frac{d\dot{N}_{e^\pm}(E)}{dE}
\end{equation}
where $E_{min}$ and $E_{max}$ are the minimum and maximum particle energies for particles injected into the nebula.

The evolutionary model assumes the dominant radiative processes are synchrotron emission and Inverse Compton (IC) scattering of ambient photons. For a particle with energy $E$, mass $m$, and charge $q$, the synchrotron loss rate $P_{synch}$ is
\begin{equation}
    P_{synch} = \frac{2q^4}{3m^4c^7} B_{pwn}^2 \sin^2\theta E^2
\end{equation}
$\theta$ is the angle between the particle's velocity and the magnetic field $B_{pwn}$. For stochastic synchrotron radiation, $\sin^2\theta = 2/3$. Synchrotron self-absorption is likely to be most important for the very early PWN lifespan which is not relevant in this case, so we do not account for it (see \citep{gelfand2009,reynolds1984} for details).

The IC emission is characterized by
\begin{equation}
    P_{IC} = \frac{32\pi cq^4}{9(mc^2)^4} u_{rad}^2 E^2 f(E)
\end{equation}
where $u_{rad}$ is the energy density of the ambient photon field and $f(E)$ is the IC scattering cross section relative to the Thomson cross section. This parameter is most sensitive to the particle energy and spectrum of the ambient photon field. For most PWNe, the cosmic microwave background (CMB) is believed to be the dominant ambient photon field. In special cases where other photon fields are present such as nearby starlight, additional photon fields are included in the emission model. For additional photon fields, one can define the temperature $T_{IC}$ and normalization $K_{IC}$, such that the energy density of the photon field $u_{IC}$ is
\begin{equation}
    u_{IC} = K_{IC}a_{BB}T^4_{IC}
\end{equation}
where $a_{BB} = 7.5657 \times 10^{-15}$\,erg cm$^{-3}$ K$^{-4}$. The evolutionary model can also investigate the potential for a pulsar contribution to \fermi $\gamma$-ray data by adding a second emission component from the pulsar. 

\singlespacing
\doublespacing

The semi-analytic model evolves the relevant properties of a PWN from time $t$ to time $t + \Delta t$. 
The radius of the PWN $R_{pwn}$, magnetic field $B_{pwn}$, the electron spectrum and subsequent average particle energy, and the properties of the host SNR at time $t + \Delta t$ are calculated from the final model and can then be compared to the observationally constrained properties of the system. The evolutionary model, when inputting parameters measured for the Crab nebula, predicts four evolutionary stages for the PWN:
\begin{enumerate}
    \item Initial Expansion: occurs early on within the first few thousand years from the SN explosion ($\tau \sim 5\,$kyr), but depends on the ambient ISM density and consequently, the return of the reverse shock. The initial expansion stage ends when the reverse shock first collides with the PWN. Until then, the PWN expansion is driven by the PWN pressure being much larger than the pressure of the surrounding SN ejecta. For the majority of the first expansion stage, the pulsar radiates more energy than the PWN loses from radiative losses and adiabatic losses from expansion. The magnetic field will decline as the PWN expands, resulting in a rapid and steep decline in synchrotron luminosity with time. This enables the adiabatic losses to dominate the total energy loss until the collision with the reverse shock, ending the initial expansion phase. The particle spectrum is expected to be dominated by previously injected particles for all energies for the majority of PWN expansion. Similarly, the corresponding photon spectrum will exhibit a two-peaked structure dominated by previously injected particles: a lower-energy peak corresponding to synchrotron radiation and a higher-energy peak attributed to IC emission, see Figure~\ref{fig:gelfand}, left panels.
    \item Compression: The reverse shock accelerates inward, re-heating cold SN ejecta until it collides with the PWN. The SN ejecta now has a pressure exceeding that of the PWN such that the nebula becomes compressed. The compression will cause the synchrotron luminosity of the PWN to increase from an increasing magnetic field strength. As a consequence, the synchrotron lifetime of the highest energy particles will become significantly less than the age of the PWN, which will introduce an energy break in the particle spectrum above which only the recently injected particles (i.e., those with the highest energy) dominate. The first compression stage for the electron and corresponding photon spectra are displayed in the right panels of Figure~\ref{fig:gelfand}.
    \item Re-expansion: For the spherically symmetric case, a series of re-expansions and compressions are predicted as the PWN achieves pressure equilibrium with the SNR interior, which occurs when the SNR becomes radiative. During the second expansion, the maximum particle energy will decrease as well as the magnetic field strength. Correspondingly the photon spectrum will shift to lower energies from the initial expansion photon spectrum, such that the majority of the PWN emission is observed in the radio and soft and hard X-rays. Regardless of where the pulsar is, however, the pulsar continues to inject high-energy particles into its surroundings. If the pulsar has not yet exited the PWN or becomes overtaken during re-expansion then the resulting electron and photon spectra will appear with two distinct components: a lower energy population composed of particles injected at earlier times and a higher-energy population composed of recently injected particles. The oldest (i.e., previously injected) particles suffer the most from adiabatic losses while the youngest particles (i.e., recently injected) suffer the most from synchrotron losses.
    \item Second compression: By this stage, the synchrotron emission could extend to $\sim 10\,$keV due to adiabatic losses. The pulsar may contribute to the soft and hard X-ray luminosities, and the IC emission from recently injected particles may extend to TeV energies, increasing the $\gamma$-ray luminosity. This stage is expected to continue into the SNR's radiative phase of evolution, to which the model no longer applies.
\end{enumerate}
The electron spectra and corresponding photon spectra for the first two evolutionary stages are shown in Figure~\ref{fig:gelfand}. While the semi-analytic model can reasonably reproduce the spectral properties of several PWNe including the Crab, there are some oversimplifications that should be noted. The evolutionary model is limited to the one-dimensional spherically symmetric expansion for both the PWN and SNR into a uniform density and does not consider hydrodynamical instabilities. The spherical symmetry assumption additionally implies that more than one expansion and compression stage for the PWN occur, but may not be the case for SNRs that expand asymmetrically (e.g., SNR~G327.1--1.1, see also Chapter~\ref{sec:g327}). Despite these limitations, the spectral and dynamical properties for several PWNe appear to be reasonably predicted using this method (e.g., G327.1--1.1 \citep{temim2015}, G21.5--0.9 \citep{hattori2020}, Kes~75 \citep{gotthelf2021,straal2022}, HESS~J1640--465 \citep{mares2021}, and G54.1+0.3 \citep{gelfand2015}). In Chapter~\ref{sec:g327}, we compare the results of the best evolutionary model for G327.1--1.1 presented in \citep{temim2015} to those of a simpler radiative computation broadband model using the Python package \texttt{NAIMA} \citep{naima}. We apply the semi-analytic modeling method described here to a new \fermi PWN in Chapter~\ref{sec:b0453} and also compare the physical implications to a set of \texttt{NAIMA} broadband models. In the final section of this chapter, we describe the sample construction of known PWNe identified in radio, X-ray, or TeV surveys that are then systematically targeted in search for their \fermi counterparts. 

\section{Source Selection}\label{sec:sample}
Several PWNe with no associated $\gamma$-ray bright pulsar have been identified by the \fermi through spatial coincidence with observations in other wavelengths. An analysis was conducted in 58 regions around TeV PWNe and unidentified TeV sources within 5\,$^\circ$ of the Galactic plane using 45 months of \fermi data \citep{acero2013} that resulted in the detection of 30 \fermi sources, 3 of which were clearly identified as PWNe and 11 as PWN candidates. Since then, recent upgrades in event processing of the \fermi\ data have significantly improved the spatial resolution and sensitivity of the instrument \citep[dubbed Pass~8,][]{atwood2013}. We expand the \fermi PWN search effort by taking advantage of the new improvements and analyzing 138 months of \fermi data in the direction of all PWNe identified across the electromagnetic spectrum. The systematic search reported here targets the locations of 58 PWNe and PWN candidates that lack an associated detected $\gamma$-ray pulsar using 11.5\, years of \fermi data. 

\begin{figure}[htbp]
\includegraphics[width=1.0\linewidth]{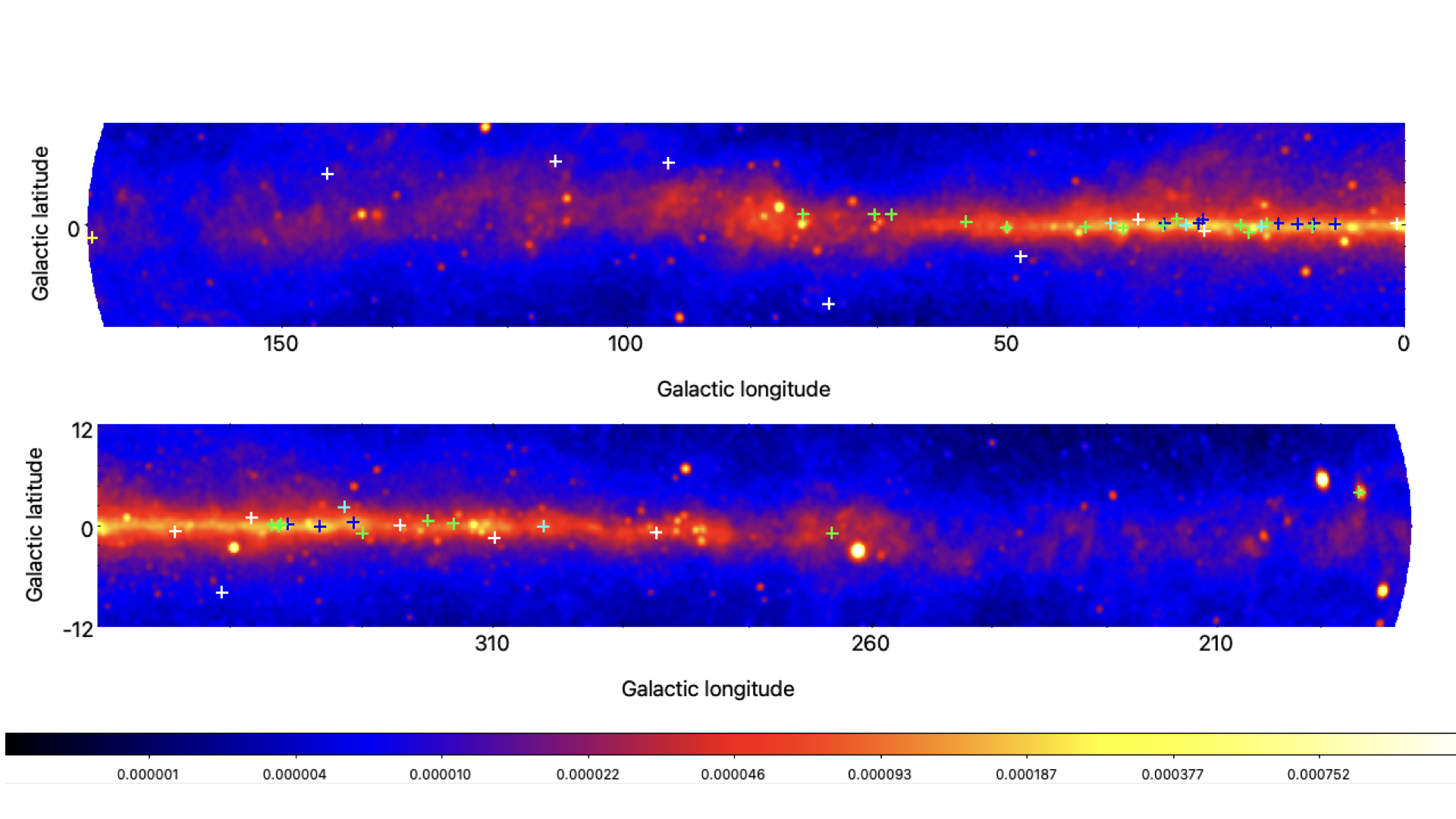}
\caption{The \fermi intensity map of the Galactic plane for $|b| < 12\,\degree$ from 12\,years of observation for energies $E > 1$\,GeV. Based on \texttt{P8R3\_SOURCE\_V2} class and \texttt{PSF3} event type. 56 ROIs are indicated as crosses and their color indicates whether the sources are detected (point-like sources in green and extended in blue) and nondetections in white. Two ROIs are located in the LMC and are not shown (see instead Figure~\ref{fig:lmc}). The units of the color scale are ph cm$^{-2}$ s$^{-1}$ sr$^{-1}$.
}\label{fig:12yr_allsky_map}
\end{figure}

There are at least 125 Galactic PWNe that have been discovered from radio to TeV $\gamma$-rays, the majority of which were first identified in radio or X-ray surveys\footnote{\url{http://snrcat.physics.umanitoba.ca/index.php?}} \citep{manitoba2012} with an increasing number of discoveries in the TeV band \citep{hess2018}. Indeed, the majority of the TeV Galactic source population is found to originate from PWNe as observed by Imaging Air Cherenkov Telescopes \citep[IACTs, e.g.][]{acero2013}. However, \fermi PWN counterparts are still lacking even after 12 years of observing the entire sky every 3 hours, with only 11 PWNe currently noted as associated with sources in the comprehensive \fermi source catalog, 4FGL \citep[data release 2 (DR2),][]{4fgl-dr2}. Most of these objects are located along the Galactic plane embedded within the prominent Galactic diffuse $\gamma$-ray emission (e.g., Figure~\ref{fig:12yr_allsky_map}), which makes these sources difficult to find. Additionally, nearly 300 rotation-powered pulsars that are capable of generating a PWN also emit brightly in the \fermi energy range, potentially outshining and obscuring their fainter PWN\footnote{A publicly available list that is continuously updated can be found here: \url{https://confluence.slac.stanford.edu/display/GLAMCOG/Public+List+of+LAT-Detected+Gamma-Ray+Pulsars}} \citep[e.g.,][]{2pc}. 

Of the 125 PWNe and PWN candidates currently known, 58 of these have no detected $\gamma$-ray pulsar. To avoid pulsar contamination and the need to consider pulsar timing solutions, we do not consider the 61+ regions of interest (ROIs) that have \fermi $\gamma$-ray pulsars. The 58 ROIs analyzed in this catalog are listed with relevant parameters in Table~\ref{tab:all_rois}. Each ROI is indicated on the 12-year \fermi all-sky map for $E > 1$\,GeV in Figure~\ref{fig:12yr_allsky_map}. A second approach in the systematic search for $\gamma$-ray emitting PWNe will involve studying the off-pulse phases of \fermi detected pulsars for the presence of an obscured PWN (future work). In Chapter \ref{sec:instrument} we describe the Fermi--LAT instrument and specifications. In Chapter \ref{sec:fermi_analysis} we describe the \fermi data selection, reduction, and analysis. 
In Chapter \ref{sec:results} we report the results of the 58 analyzed ROIs with detailed discussions. We provide an overview and conclusions in Section \ref{sec:conclude} of Chapter \ref{sec:results}. Two newly classified \fermi PWNe are studied in further detail incorporating available multiwavelength data in Chapters~\ref{sec:g327} and \ref{sec:b0453}. A final synopsis is provided in Chapter~\ref{sec:final}.

\pagebreak

\singlespacing
\setlength\LTcapwidth{1.0\linewidth}
\setlength{\tabcolsep}{0.04cm}
\small{
\begin{longtable}{ccccccc}
\centering
ROI & Galactic PWN Name & 4FGL Name (CLASS1) & R.A. & Dec. & Extent & $\lambda$  \\
\hline
\hline
			1 & G0.9+0.10  & -- & 266.836 & --28.154 & 0.02 & Radio \\
			2 & G8.4+0.15 & J1804.7--2144e (spp) & 271.180 & --21.740 & 0.25 & TeV \\
			3 & G11.0--0.05 & J1810.3--1925e (spp) & 272.518 & --19.413 & 0.05 & Radio \\
			4 & G11.1+0.08 & J1810.3--1925e (spp) & 272.457 & --19.207 & 0.05 & Radio \\
			5 & G11.2--0.35 & J1811.5--1925 (psr) & 272.876 & --19.435 & 0.038 & Radio \\
			6 & G12.8--0.02 & J1813.1--1737e (spp) & 273.400 & --17.830 & 0.05 & TeV \\
			7 & G15.4+0.10 & J1818.6--1533 (spp) & 274.500 & --15.470 & $<$ 0.04 & TeV \\
			8 & G16.7+0.08 & J1821.1--1422 (spp) & 275.239 & --14.329 & 0.015 & Radio \\
			9 & G18.0-0.69 & J1824.5--1351e (PWN) & 276.260 & --13.970 & 0.46 & TeV \\
			10 & G18.9--1.10 & J1829.4--1256 (spp) & 277.363 & --12.970 & 0.15 & Radio \\
			11 & G20.2--0.20 & J1828.0--1133 (spp) & 277.034 & --11.587 & 0.05 & Radio \\
			12 & G23.5+0.10 & -- & 278.408 & --8.454 & 0.017 & X-ray \\
			13 & G24.7+0.60 & J1834.1--0706e (SNR) & 278.550 & --7.040 & 0.5 & Radio \\
			14 & G25.1+0.02 & J1838.9--0704e (pwn) & 279.513 & --6.926 & 0.02 & X-ray \\
			15 & G25.2--0.19 & J1836.5--0651e (pwn) & 279.339 & --6.874 & 0.02 & X-ray \\
			16 & G26.6--0.10 & J1840.9--0532e (PWN) & 280.142 & --5.720 & 0.4 & TeV \\
			17 & G27.8+0.60 & J1840.0--0411 (spp) & 279.980 & --4.290 & 0.11 & Radio \\
			18 & G29.4+0.10 & J1844.4--0306 (unk) & 281.137 & --3.120 & 0.06 & Radio \\
			19 & G29.7--0.30 & J1846.9--0247c (unk/blank) & 281.600 & --2.970 & $<$ 0.015 & TeV \\
			20 & G32.6+0.53 & -- & 282.240 & --0.040 & 0.09 & TeV \\
			21 & G34.6--0.50 & -- & 284.044 & 1.222 & 0.05 & X-ray \\
			22 & G36.0+0.10 & J1857.7+0246e (PWN) & 284.340 & 2.760 & 0.613 & TeV \\
			23 & G39.2--0.32 & J1903.8+0531 (spp) & 286.020 & 5.453 & 0.01 & X-ray \\
			24 & G47.4--3.90  & -- & 293.031 & 10.921 & 0.08 & X-ray \\
			25 & G49.2--0.30 & J1922.7+1428c (unk/blank) & 290.702 & 14.274 & 0.02 & X-ray \\
			26 & G49.2--0.70 & J1922.7+1428c (unk/blank) & 290.830 & 14.060 & 0.02 & X-ray \\
			27 & G54.1+0.30 & J1930.5+1853 (DR3) (pwn) & 292.610 & 18.840 & $<$ 0.031 & TeV \\
			28 & G63.7+1.10 & J1947.7+2744 (pwn) & 296.992 & 27.741 & 0.13 & Radio \\
			29 & G65.7+1.18 & J1952.8+2924 (spp) & 298.210 & 29.480 & 0.1 & TeV \\
			30 & G74.0--8.50  & -- & 312.334 & 29.018 & 0.05 & X-ray \\
			31 & G74.9+1.11 & J2016.2+3712 (snr) & 304.038 & 37.187 & 0.03 & X-ray \\
			32 & G93.3+6.90  & -- & 313.058 & 55.289 & 0.01 & X-ray \\
			33 & G108.6+6.80  & -- & 336.419 & 65.602 & 0.05 & X-ray \\
			34 & G141.2+5.00  & -- & 54.296 & 61.887 & 0.10 & Radio \\
			35 & G179.7--1.70  & -- & 84.604 & 28.285 & 0.05 & X-ray \\
			36 & G189.1+3.00 & -- & 94.282 & 22.368 & 0.02 & X-ray \\
			37 & G266.9--1.10 & -- & 133.901 & --46.738 & 0.04 & Radio \\
			38 & G279.6--31.70 & J0537.8--6909 (pwn) & 84.447 & --69.172 & 0.008 & X-ray \\
			39 & G279.8--35.80 & -- & 73.410 & --68.489 & 0.006 & X-ray \\
			40 & G290.0--0.93 & -- & 210.191 & --63.429 & 0.0017 & X-ray \\
			41 & G304.1--0.24 & J1303.0--6312e (PWN) & 195.760 & --63.190 & 0.19 & TeV \\
			42 & G310.6--1.60  & -- & 230.798 & --57.096 & 0.043 & Radio \\
			43 & G315.8--0.23 & J1435.8--6018 (spp) & 219.345 & --60.019 & 0.17 & Radio \\
			44 & G318.9+0.40 & J1459.0--5819 (unk) & 224.710 & --58.412 & 0.05 & Radio \\
			45 & G322.5--0.10  & -- & 251.750 & --43.775 & 0.034 & X-ray \\
			46 & G326.2--1.70 & J1552.4--5612e (PWN) & 238.110 & --56.210 & 0.08 & Radio \\
			47 & G327.1--1.10 & J1554.4--5506 (DR3) (pwn) & 238.671 & --55.082 & 0.05 & Radio \\
			48 & G328.4+0.20 & J1553.8--5325e (blank) & 238.886 & --53.285 & 0.02 & X-ray \\
			49 & G332.5--0.3 & J1616.2--5054e (PWN) & 244.400 & --50.940 & 0.20 & Radio \\
			50 & G332.5--0.28 & J1616.2--5054e (PWN) & 244.410 & --51.039 & 0.02 & X-ray \\
			51 & G336.4+0.10 & J1631.6-4756e (pwn) & 247.980 & --47.770 & 0.182 & TeV \\
			52 & G337.2+0.1 & -- & 248.979 & --47.317 & 0.025 & X-ray \\
			53 & G337.5--0.1 & J1638.4--4715c (DR3) (blank) & 249.510 & --47.231 & 0.006 & X-ray \\
			54 & G338.2--0.00 & J1640.6--4632 (DR1), J1640.7--4631e (DR3) (spp) & 250.189 & --46.525 & 0.13 & Radio \\
			55 & G341.2+0.90  & -- & 260.906 & --37.578 & 0.05 & X-ray \\
			56 & G350.2--0.80  & -- & 284.146 & --37.908 & 0.002 & X-ray \\
			57 & G358.3+0.24  & -- & 265.320 & --30.380 & 0.004 & TeV \\
			58 & G358.6--17.2 & -- & 165.445 & --61.023 & 0.01 & X-ray \\
\hline
\hline
\caption{All PWN and PWNe candidate ROIs analyzed in this paper. Coincident 4FGL sources and their classifications are listed in the third column, considering also the 4FGL--DR3 catalog from \url{https://fermi.gsfc.nasa.gov/ssc/data/access/lat/12yr_catalog/}. The observed right ascension (R.A.) and declination in J2000 equatorial degrees are listed in the fourth and fifth columns as well as the observed extent (radius) in degrees in the sixth column. The last column specifies the wavelength of the observed positions and extents. TeV data is taken from \url{http://tevcat.uchicago.edu/} \citep{tevcat2008} and radio or X-ray data from \url{http://snrcat.physics.umanitoba.ca/SNRtable.php} \citep{manitoba2012}.} \label{tab:all_rois}
\end{longtable}}
\doublespacing \clearpage

    \pdfoutput=1
\chapter{The {\it Fermi}--LAT}\label{sec:instrument}
\normalsize
The Large Area Telescope (LAT) is the primary instrument on board the \textit{Fermi} Gamma-ray Space Telescope and surveys the entire sky from energies as low as 20\,MeV and as high as 2\,TeV with the latest Pass 8 event level analysis reconstruction upgrade \citep{atwood2009, atwood2013}. The \fermi was constructed by an international group of space agencies and academic institutes and universities in France, Italy, Sweden, United States, and Japan.  Launched by NASA in June 2008, the LAT's original mission was to determine the nature of high-energy sources such as blazars, pulsars, gamma-ray bursts (GRBs) and more that comprise the high-energy universe. Since then, the \fermi has made several {\it thousand} new discoveries of high-energy sources that have been reported in various catalogs since observations began\footnote{See: \url{https://fermi.gsfc.nasa.gov/ssc/data/access/} for a complete list of the \fermi catalogs.}.

\begin{figure}[htbp]
\centering
\includegraphics[width=0.9\linewidth]{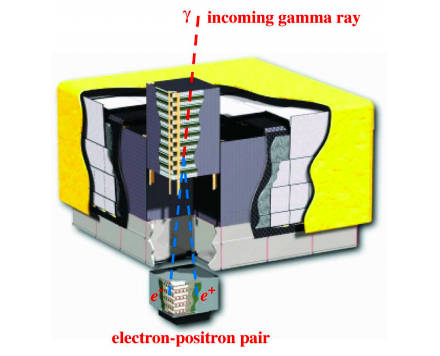}
\caption{The basic design of the LAT. The interior is composed of 16 identical towers placed into a $4 \times 4$ array. Each tower includes a precision tracker and a calorimeter. The yellow outer layer represents the anti-coincidence shield, which protects the tracker array from contamination of bombarding cosmic rays. A simple schematic illustrates the pair conversion process of a $\gamma$-ray photon detected by the LAT. Adapted from \citep{atwood2009}.}\label{fig:fermi}
\end{figure}

The design of the LAT has seven major goals: 1) permitting rapid notification of GRBs and transients and facilitating the monitoring of these variable sources, 2) yielding a large catalog (up to several thousand high-energy sources discovered during an all-sky survey), 3) measuring spectra from 20\,MeV to more than 50\,GeV for several hundred sources, 4) optimizing uncertainties for point sources to just 0.3--2\,arcminutes, 5) mapping and obtaining the spectra of extended sources (e.g. supernova remnants, molecular clouds, nearby galaxies, etc.), 6) measuring the diffuse isotropic $\gamma$-ray background up to 2\,TeV, and finally, 7) exploring the discovery space for dark matter \citep{atwood2009}. 

\section{Instrumental Design}
The LAT measures the directions, energies, and arrival times of incident $\gamma$-rays over a large field of view (FoV) while simultaneously rejecting background from cosmic rays. The LAT is a pair-conversion telescope with a precision converter-tracker and a calorimeter, a segmented anti-coincidence detector, and a programmable trigger and data acquisition system. High-energy $\gamma$-rays are not easily reflected or refracted, instead they only interact with matter by the conversion of the $\gamma$-ray photon into an electron and positron pair. Pair conversion is therefore an efficient method for $\gamma$-ray detection. See Figure~\ref{fig:fermi} for a schematic of the LAT design alongside a simple diagram illustrating the pair conversion process. 

The \fermi is composed of 16 identical towers disposed in a $4 \times 4$ array \citep[see][for details]{atwood2009}. Each tower has a precision tracker and a calorimeter. The towers are aligned on a low-mass Aluminum grid design and the segmented anti-coincidence detector covers the 16 towers. Each module in the precision tracker is accompanied by a series of 18 vertical tracking planes including two layers of single-side silicon strip detectors (SSDs) and a high-Z converter material (tungsten) per tray. The SSDs act as small position-sensitive detectors to record the passage of charged particles and measuring the paths of particles that result from pair conversion. This is done by the $e^{-}e^{+}$ pair hitting the SSDs causing ionization and thus creating a small electrical pulse that is then detected by the SSD. In this way, the tracker can reconstruct the direction and path of the incident $\gamma$-ray photon.

The pair-converted particles travel through the layers of the tracker before they enter the calorimeter. The calorimeter has the same design of a $4 \times 4$ array of 16 modules, each made up of 96 Cesium iodide scintillator crystals instead of SSDs. With this design, the longitudinal and transverse information about the deposited energy are both known and this is what enables the high-energy reach of the LAT. The calorimeter's depth and segmentation also significantly contribute to the background rejection capabilities. A larger field of view $\sim$ 2.4\,str ($\sim$20\% of sky) is possible due to the aspect ratio of the tracker which also ensures that the majority of all pair-conversion showers initiated in the tracker will pass into the calorimeter for energy measurement. 

The segmented anti-coincidence detector (ACD) covers the tracker array in order to provide protection from contamination of cosmic rays. For this reason it is designed to have a high detection efficiency of charged particles. It is reported in \citep{atwood2009} that the efficiency of the ACD reaches 0.9997 for a single charged particle detection. The programmable trigger and data acquisition system utilizes prompt signals available from the precision tracker, calorimeter, and anti-coincidence detector subsystems to form a trigger. Once an event is triggered, the data acquisition system (DAQ) initiates the read out of the 3 subsystems and uses on-board event processing to reduce the rate of events transmitted to the ground to a rate compatible with the 1\,Mbps bandwidth that is available to the LAT. The maximized number of events triggered by $\gamma$-rays are transmitted to the ground while the on-board processing rejects events triggered by cosmic background particles. Heat pipes are used throughout the grid to keep the unit from over-heating. 

In summary, the \fermi is designed to provide good angular resolution for source localization and broadband investigations as well as a high sensitivity over a large field of view to monitor variability and detecting transients. A sensitive calorimeter is installed over an extended energy band (50\,GeV--2\,TeV) to study spectral breaks and energy cut-offs. Reliable calibration and stability offers long term flux measurement. The normal to the front of the instrument is on alternate orbits (i.e. ``scanning'' mode), rocking northward in one orbit and southward the next, in order to measure the entire sky in almost uniform sensitivity after about 2 orbits (which takes the \fermi about 3\,hours at 565\,km and a $25.5\degree$ inclination).

\section{Pass 8}

Event level analysis is the process of reconstructing a photon event detected by the LAT using the measurements from the three sub-systems. 
Before the Pass 8 event-level analysis ability, the LAT framework was largely developed through Monte-Carlo simulations. The initial event-level analysis upon launch was Pass 6 to which Pass 7 followed with the first improvements to the data reduction.  The main effort for the Pass 8 event-level analysis was to target the loss of effective area of the LAT due to residual signals coming from out-of-time cosmic ray events \citep{atwood2013}. In the Pass 8 update, the main areas for improvement include the Monte Carlo simulation of the detector, event reconstruction, and background rejection.

The reconstruction improvements were made to the tracker, calorimeter, and ACD. The tracker reconstruction code was modified to fix four areas of operation. The problematic areas begin with the track-following algorithm. First, an initial direction is required to start the track-following for the path of the converted $e^{-}e^{+}$ pair to trace back where the photon actually underwent conversion. The second problem arises in the track model that includes multiple Coulomb scattering, but requires an estimate of the track energy which is measured from the calorimeter. These two issues make the tracker dependent on the accuracy of the calorimeter reconstruction which means the residual signals coming from out-of-time cosmic rays (i.e. ghost signals) that confuse the calorimeter also become a problem for the tracker. The third area for improvement is more accurate pair conversion measurement, since a single pair conversion can often develop electromagnetic showers and cause multiple detections from the same event. The last area that the tracker reconstruction of Pass 8 aims to rectify is that when off-axis photons deposit large energies into the calorimeter, this can cause particles to move upwards in the calorimeter and start randomly hitting strips which can wash out some of the real signal. These effects can cause a loss of events that fail to reconstruct at all, or mix events from the center of the point spread function (PSF) to the outer edges due to the poorly reconstructed tracks, and can ultimately confuse good $\gamma$-ray events as background. 

Pass 8 addresses all of the outlined issues by using a global approach called tree-based tracking. When a shower of particles begins, the tree-based tracking looks for the conversion in the tracker to try to model the process of the shower as the electron and positron interact with the tracker and radiate energy. Once the trajectories of the particles are known, they are fit using the Kalman Filter technique and this will account for multiple scatterings. This part of the Pass 8 reconstruction reduces the fraction of mis-tracked events and boosts the high-energy acceptance by 15--20\% while also providing improvement for the off-axis effective area of the instrument \citep[see][for details]{atwood2013}. The Pass 8 reconstruction for the calorimeter introduces a clustering stage which helps the calorimeter to identify ghost signals and recover any loss in the effective area. This technique is borrowed from graph theory and has proved to be optimal with Pass 8 up to 1\,TeV (above 1\,TeV, saturation becomes an issue). 

The first major improvement to the ACD is the incorporation of the calorimeter's information when configuring incident particle direction from the deposited energy. Now, directional information measured from the calorimeter clusters is propagated to the ACD, accompanied by the paths measured by the tracker. This aids in identifying background information at high energies or large incident angles as this is more susceptible to tracking errors. Secondly, when associating tracks and clusters with energy depositions in the ACD, the ACD utilizes event-by-event directional uncertainties and this new approach will significantly enhance information about background rejection. At low energies, the use of trigger information in background rejection removes the ghost signals from the ACD and substantially increases effective area. This is done by taking advantage of the fast ACD signals from the LAT trigger.

The Pass 8 event level reconstruction can increase acceptance (relative to Pass 7)\footnote{see \url{https://www.slac.stanford.edu/exp/glast/groups/canda/lat_Performance.htm} for a review of the Pass~8 improvements compared to Pass~7.} by $\sim25\%$ at high energies and, at energies below 300\,MeV, as high as a factor of 3. The improvements of the 3 subsystems on the LAT have enabled the detection of multi-photon events whereas previously, the lack of calorimeter clustering along with background rejection almost completely washed away any meaningful signal the LAT might have had with such events. Hence, Pass~8 has extended the energy reach of the LAT both below 100\,MeV and above 1\,TeV. All information described in this chapter can be found in more detail in \citep[instrument design:][]{atwood2009} and \citep[Pass~8:][]{atwood2013}. \clearpage

    \pdfoutput=1
\chapter{{\it Fermi}--LAT Data Selection and Analysis}\label{sec:fermi_analysis}
The {\it Fermi} Gamma-ray Space Telescope houses both a Gamma-ray Burst Monitor \citep[GBM,][]{meegan2009} and the Large Area Telescope \citep[LAT,][]{atwood2009}. The LAT instrument is sensitive to $\gamma$-rays from 50\,MeV to 2\,TeV \citep{4fgl2020} and has been continuously surveying the entire sky every 3\,hours since beginning operation in August 2008. The latest upgrade to the event reconstruction process and instrumental response functions \citep[Pass 8,][]{atwood2013} has improved the effective area, point spread function, and background rejection capabilities (see Chapter~\ref{sec:instrument}). These upgrades have enabled an improved effective energy and angular resolution. The point spread function (PSF) is energy-dependent, becoming large at lower energies but declining at increasing energies. The PSF reaches optimal resolution at $\gtrsim10$\,GeV with a PSF $\sim0.1$\,$\degree$. A major advantage to the Pass 8 event reconstruction upgrade is the classification of detected photon events based on the quality of the angular reconstruction. The events are divided into four categories: \texttt{PSF0}, \texttt{PSF1}, \texttt{PSF2}, or \texttt{PSF3} types. \texttt{PSF0} has the worst reconstruction quality and each type increases in reconstruction quality such that \texttt{PSF3} contains the highest-quality events. 

For the analysis of coincident residual emission that is point-like and faint, we must consider all events. For the analysis of coincident residual emission that is significantly detected and/or exhibits extension, we consider only events of \texttt{PSF3} type. Of the 58 ROIs analyzed here, only four have faint, point-like detections that require all events: B0453--685, G327.1--1.1, G54.1+0.3, and G266.9--1.10 (see Section~\ref{sec:results}).

\section{\fermi Data Selection}
For all 58 ROIs, we perform two binned likelihood analyses using 11.5\,years (from 2008 August to 2020 January) of \texttt{P8R3\_SOURCE\_V3} photons with energy between 300\,MeV--2\,TeV considering 1) all events and 2) \texttt{PSF3} events only. We utilize the latest Fermitools package\footnote{\url{https://fermi.gsfc.nasa.gov/ssc/data/analysis/software/}.} (v.2.0.8) and FermiPy Python 3 package \citep[v.1.0.1,][]{py3,fermipy2017} to perform data reduction and analysis. 
We organize the events by PSF type, as described above, using \texttt{evtype=4,8,16,32} to represent \texttt{PSF0, PSF1, PSF2}, and \texttt{PSF3} components. For the all-event analyses, a binned likelihood analysis is performed on each event type and then combined into a global likelihood function for the ROI to represent all events\footnote{See FermiPy documentation for details: \url{https://fermipy.readthedocs.io/en/0.6.8/config.html}}. For \texttt{PSF3}-event analyses, the global likelihood function is a binned likelihood analysis performed on only the \texttt{evtype=32 (PSF3)} events. Photons detected at zenith angles larger than 100\,$\degree$ were excluded to limit the contamination from $\gamma$-rays generated by CR interactions in the upper layers of Earth's atmosphere. The data were additionally filtered to remove time periods where the instrument was not online (e.g., when flying over the South Atlantic Anomaly). The $\gamma$-ray data is modeled with the latest comprehensive \fermi source catalog, 4FGL--DR2 \citep{4fgl-dr2}, the LAT extended source template archive for the 4FGL catalog\footnote{\url{https://fermi.gsfc.nasa.gov/ssc/data/access/lat/8yr_catalog/}.}, and the latest Galactic diffuse template to model the Galactic interstellar emission (\texttt{gll\_iem\_v07.fits}). For the four ROIs that require consideration of all events, we employ the isotropic diffuse template \texttt{iso\_P8R3\_SOURCE\_V3\_v1.txt}. For the remainder of ROIs that only consider \texttt{PSF3} events, the isotropic diffuse template employed is \texttt{iso\_P8R3\_SOURCE\_V3\_PSF3\_v1.txt}\footnote{LAT background models and appropriate instrument response functions: \url{https://fermi.gsfc.nasa.gov/ssc/data/access/lat/BackgroundModels.html}.}.

\section{\fermi Data Analysis}
\subsection{Building the Source Models}
For most sources, a $10 \degree \times 10 \degree$ ROI is used to construct the source model considering 4FGL sources and backgrounds within $15 \degree$ of the ROI center which is always chosen to be the PWN position. For all sources, we fit the ROI with 4FGL point sources and extended sources along with the diffuse Galactic and isotropic emission templates. The only sources that require larger ROIs ($15-20 \degree$ while considering sources within $20-25 \degree$ of ROI center) are SNR~S~147 (G179.7--1.70) and SNR~G74.0--8.5. For these two ROIs, we instead consider events within $20 \degree$ for S~147 and within $15 \degree$ for G74.0--8.5 due to their uniquely large SNR radii and coincident extended 4FGL sources. The shell of SNR~S147 is nearly 5 degrees in diameter and is an identified $\gamma$-ray emitter, 4FGL~J0540.3+2756e, characterized best using the $r \sim 2.5\,\degree$ H-$\alpha$ SNR shell structure to model the $\gamma$-ray emission. SNR~G74.0--8.5 is nearly 8 degrees in diameter and has a large extended source, 4FGL~J2051.0+3049e \citep[outer annulus radius $\sim 1.6 \degree$,][]{4fgl-dr2}, overlapping in location that may be associated to the SNR shell. 

\begin{figure}[htbp]
\hspace{-.25cm}
\singlespacing
\includegraphics[width=1.05\linewidth]{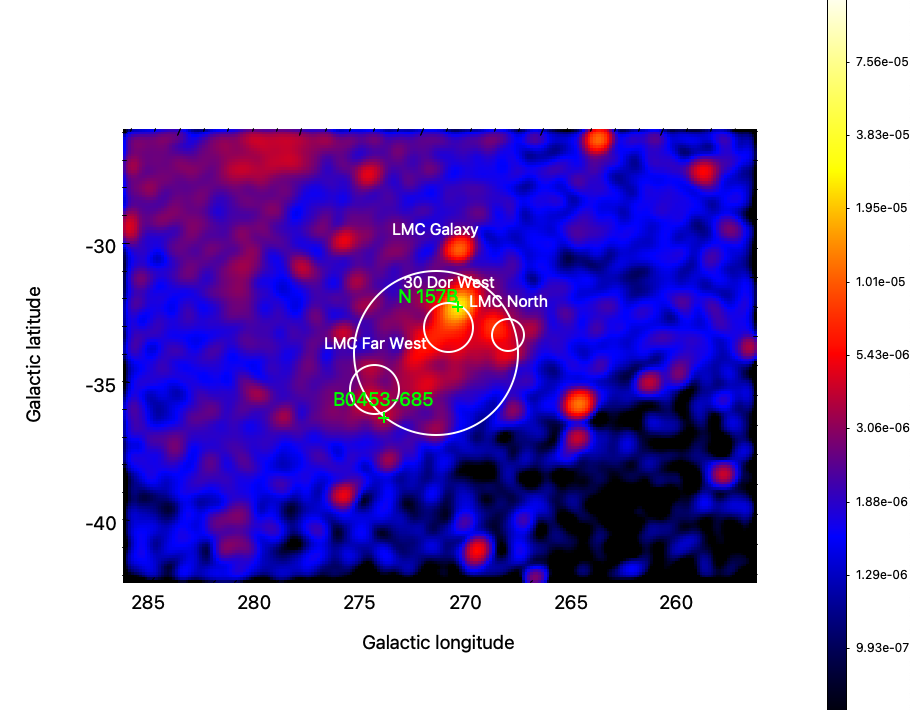}
\caption{The \fermi intensity map for $E > 1\,$GeV of the Large Magellanic Cloud centered on ($l$,$b$) = (278.76$\,\degree$, --33.46$\,\degree$). The two ROIs analyzed in the LMC are indicated as green crosses. The four extended source components representing the diffuse LMC emission (following the {\it{emissivity model}} developed in \citep{lmc2016}) are indicated in white. Based on \texttt{P8R3\_SOURCE\_V2} class and \texttt{PSF3} event type. The units of the color scale are ph cm$^{-2}$ s$^{-1}$ sr$^{-1}$.
}\label{fig:lmc}
\end{figure}

Two of the 58 ROIs are located within the Large Magellanic Cloud (LMC, see Figure~\ref{fig:lmc}): N~157B (4FGL~J0537.8--6909) and newly detected B0453--685 and as such require accounting for a third diffuse emission component from the LMC. The source models additionally employ four extended source components to reconstruct the {\it{emissivity model}} developed in \citep{lmc2016} to represent the diffuse LMC emission. The four additional sources are 4FGL~J0500.9--6945e (LMC Far West), 4FGL~J0519.9--6845e (LMC Galaxy), 4FGL~J0530.0-6900e (30 Dor West), and 4FGL~J0531.8--6639e (LMC North).

\doublespacing

\subsection{Detection Method}
With the source models described above, we allow the background components and sources with TS $\geq 25$ and a distance from the ROI center (chosen to be the PWN position) $\leq3.0$\,$\degree$ to vary in spectral index and normalization. We compute a series of diagnostic test statistic (TS) and count maps in order to search for and understand any residual $\gamma$-ray emission. The TS is defined to be the natural logarithm of the as shown in between the likelihood of one hypothesis (e.g. presence of one additional source) and the likelihood for the null hypothesis (e.g. absence of source):
\begin{equation}
    \text{TS} = 2 \times \log\big({\frac{{\mathcal{L}_{1}}}{{\mathcal{L}_{0}}}}\big)
\end{equation}
The TS quantifies the significance for a source detection with a given set of location and spectral parameters and the significance of such a detection can be estimated by taking the square root of the TS. TS values $>25$ correspond to a detection significance $> 4 \sigma$ (for 4 degrees of freedom or DOF) \citep{mattox1996}. We generated the count and TS maps by the following energy ranges: 300\,MeV--2\,TeV, 1--10\,GeV, 10--100\,GeV, and 100\,GeV--2\,TeV. The motivation for increasing energy cuts stems from the improving PSF of the \fermi instrument with increasing energies\footnote{See \url{https://www.slac.stanford.edu/exp/glast/groups/canda/lat_Performance.htm} for a review on the dependence of PSF with energy for Pass 8 data.}. We inspect the TS maps for additional sources that coincide with the PWN position. In most cases, unassociated, unknown, or plausibly PWN/SNR-related (i.e., ``spp'' class) 4FGL sources were already found to be coincident and likely counterparts to these systems (Table~\ref{tab:all_rois}).

\subsection{Localization and Extension}\label{sec:ext}
To model residual $\gamma$-ray emission coincident in location that meets the \fermi significance threshold ($\text{TS}=25$ for 4 DOF \citep{mattox1996}), we first add a fixed point source characterized by a power-law spectrum and index $\Gamma = 2$ to the global source model after removing any unidentified 4FGL sources coincident in location. We localize the point source executing \texttt{GTAnalysis.localize} to find the best-fit position. If a new best-fit position is preferentially found and the 95\% uncertainty radius remains to be positionally coincident with the PWN position, we keep the new location. In some analyses, no better position is found and therefore the position remains fixed to the PWN location. 

For all detected ROIs, we run extension tests in FermiPy utilizing \texttt{GTAnalysis.extension} and the two spatial templates supported in the FermiPy framework, radial disk and radial Gaussian. Both of these extended templates assume a symmetric 2D shape with width parameters radius $r$ and sigma $\sigma$, respectively. If the position of the point source is reasonable when free, we allow the position to remain free when finding the best-fit spatial extension for both templates, otherwise the position remains fixed to the PWN location.

As shown in Tables~\ref{tab:extent} and \ref{tab:points}, the majority of all extension tests are consistent with prior results. Of the 15 extended sources, 13 were already extended in the 4FGL catalogs. The extensions found for these 13 sources are comparable to the sizes reported in the 4FGL, where the only differences are in the best-fit template used. For all extended sources in this paper except for MSH~15--56, we find a radial Gaussian can better fit or fit just as well the $\gamma$-ray emission as the radial disk template. MSH 15--56 is the only extended source that is best-fit with a custom spatial template (i.e., the PWN shape in radio, \citep{devin2018}). For the sources that are characterized best as a radial Gaussian in this catalog such as 4FGL~J1836.5--0651e (best-fit as $\sigma = 0.53 \degree$), the spatial extensions are comparable to the extension when fit as a radial disk in the 4FGL catalogs. In the 4FGL--DR2 release, 4FGL~J1836.5--0651e is fit as a radial disk, $r = 0.5\,\degree$ \citep{4fgl-dr2}. We also report that the new best-fit for 4FGL~J1836.5--0651e can model well the extended emission coincident to the TeV PWN~HESS~J1837--069 \citep[$r = 0.36\,\degree$,][]{hess2018} such that the TS of the nearby extended source 4FGL~J1838.9--0704e, a second possible PWN association to HESS~J1837--069, subsequently declines to TS\,$ = 0$. We therefore suggest that there is only one extended source associated to HESS~J1837--069, and it is most likely to be 4FGL~J1836.5--0651e, though two X-ray PWNe are coincident in location with this extended source, see Section~\ref{sec:new}. Similarly, we report not one but two X-ray PWNe that coincide with the Fermi PWN 4FGL~J1616.2--5054e, as a second PWN is only $\sim 0.02 \degree$ from RCW~103 that question the true X-ray counterpart for 4FGL~J1616.2--5054e and the associated TeV PWN HESS~J1616--508 (see Section~\ref{sec:new}). For other sources such as 4FGL~J1631.6--4756e and 4FGL~J1813.1--1737e, we find that the radial Gaussian is better fit with an extent that is $\sim 1.5$ times smaller than when fit as a radial disk. This is an expected result \citep{fges2017, lande2012}.

There are two new extended sources in Table~\ref{tab:extent} that are not considered extended in the 4FGL catalogs: 4FGL~J1818.6–-1533 (PWN candidate G15.4+0.10) and 4FGL~J1844.4--0306 (PWN candidate G29.4+0.10). Some crowded regions have inconclusive extension test results including W44, IC~443, PSR~J0855--4644/G266.9--1.10, G337.2+0.1, and G337.5--0.1. For this reason, we only consider the point-source detection as a tentative detection for these ROIs. The smallest extended source is MSH 15--56 with a radius $r \sim 0.08\, \degree$ and the largest extended source is 4FGL~J1824.5--1351e (HESS~J1825--137) with  $\sigma = 0.83\,\degree \pm 0.074$. The average extension modeled as a radial Gaussian for the sample of 15 extended sources is $\sigma = 0.37\,\degree$. Of the 39 total detections, 15 are found to be extended, which make up 38\% of the entire sample reported in this catalog. We explore the systematics on the extension in Section~\ref{sec:sys}.  The summary of the best-fit parameters for the extended sources including the measured systematic error on the extension are listed in Table \ref{tab:extent}.

\subsection{Fluxes and Upper Limits}
For detected sources, we measure the energy flux from 300\,MeV--2\,TeV in seven energy bins using \texttt{GTAnalysis.sed}, where only the background components and the source of interest are left free to vary in spectral index and normalization. For most sources, we assume the spectrum can be characterized as a power-law\footnote{For a review of Fermi source spectral models see \url{https://fermi.gsfc.nasa.gov/ssc/data/analysis/scitools/source_models.html}.}:
\begin{equation}
    \frac{dN}{dE} = N_{0} \big(\frac{E}{E_0}\big)^{-\Gamma}
\end{equation}\label{eq:1} 
setting $\Gamma = 2.0$ and allow the index and normalization $N_0$ to vary in subsequent fits. Even for 4FGL sources that were fit as a Log Parabola spectrum in the 4FGL catalog,
\begin{equation}
    \frac{dN}{dE} = N_{0} \big(\frac{E}{E_b}\big)^{-(\alpha+\beta\log{E/E_b})}
\end{equation}\label{eq:2}
it is found that a power-law spectrum can adequately model the $\gamma$-ray emission with comparable or better results (i.e., spectral index and detection significance). Only two sources in this catalog keep their log-parabola spectral model: MSH 15--56 and 4FGL~J1459.0--5819 (G318.9+0.4 PWN candidate). The remainder of source detections are characterized by a power-law spectrum.  

We additionally measure the flux for each best-fit extension template for all sources (point source, radial disk, and radial Gaussian templates) and find that the spectral flux for all detected sources are nearly independent of their morphology. We measure the systematic error on the flux for all 39 source detections and describe the method and results in Section~\ref{sec:sys}. The systematic error for the spectral flux per energy bin are listed in the Appendix, Table~\ref{tab:spectral_fluxes}.

For non-detections, we place a point source at the location of the PWN systems, which corresponds to the size of the systems as observed by the \fermi (see Table~\ref{tab:all_rois}) and perform a global fit with only the spectral parameters of the backgrounds and the point source free to vary. If the point source has TS\,$< 25$ for 4 DOF, we measure the 95\% confidence level (C.L.) upper limit fluxes for the 300\,MeV--2\,TeV energy range and are listed in the final column of Table~\ref{tab:nondetect}. We also calculate the 95\% C.L. upper limit flux values for 300\,MeV--2\,TeV across nine energy bins. All upper limit flux values for the 16 undetected sources are listed in the Appendix, Table~\ref{tab:spectral_ul_fluxes}.

\subsection{Sources of Systematic Errors}\label{sec:sys}
We account for systematic uncertainties introduced by the choice of the interstellar emission model (IEM) and the instrument response functions (IRFs), which mainly affect the spectrum of the measured $\gamma$-ray emission. We have followed the prescription developed in \citep{acero2016, depalma2013}, which generated eight alternative IEMs using a different approach than the standard IEM \citep[see][for details]{acero2016}. For this analysis, we employ the eight alternative IEMs (aIEMs) that were generated for use on Pass 8 data in the \fermi Galactic Extended Source Catalog \citep[FGES,][]{fges2017}. 
We re-generate the best-fit model and perform independent fits for each ROI using the 8 aIEMs for a total of 9 fits per ROI including the standard model fit. We then compare the spectral flux values to the standard model following equation (5) in \citep{acero2016}. 

We estimate the systematic uncertainties introduced by the uncertainty in the effective area\footnote{\url{https://fermi.gsfc.nasa.gov/ssc/data/analysis/LAT_caveats.html}} while enabling energy dispersion as follows: $\pm 3\%$ for $E < 100$\,GeV, $\pm 4.5 \%$ for $E = 175\,$GeV, and $\pm 8\%$ for $E = 556\,$GeV. Since the IEM and IRF systematic errors are taken to be independent, we can evaluate both and perform the quadratic sum for the total systematic error. This method has been rigorously applied in two prior catalogs and both reported systematic errors on the same order as the $1 \sigma$ statistical errors in flux \citep[see Fig.~29 in][]{acero2016}, which is consistent with our results (see Table~\ref{tab:spectral_fluxes} in the Appendix). In general, it is found that the systematics introduced by the IEM and IRF are most important for sources that lie along the Galactic plane $|b| < 1.0\,\degree$ and for energies $E < 5\,$GeV.

For the two sources located in the LMC (Figure~\ref{fig:lmc}): N~157B (4FGL~J0537.8--6909) and newly detected B0453--685, we find that the systematic errors are negligible for all energy bins, which is not surprising given the location of the LMC with respect to the bright, diffuse $\gamma$-ray emission along the Galactic plane. However, we must also account for the systematic error that is introduced by having an additional diffuse background component for the two LMC ROIs. This third component is attributed to the cosmic ray (CR) population of the Large Magellanic Cloud interacting with the LMC ISM and there are limitations to the accuracy of the background templates used to model this emission, similar to the Galactic diffuse background. We can probe these limitations by employing a straightforward method described in \citep{lmc2016} to measure systematics from the diffuse LMC. This requires replacing the four extended sources that represent the diffuse LMC in this analysis \citep[the {\it emissivity model},][]{lmc2016} with four different extended sources to represent an alternative template for the diffuse LMC \citep[the {\it analytic model},][]{lmc2016}. The $\gamma$-ray point sources coincident with N~157B and SNR~B0453--685 are both refit with the alternative diffuse LMC template to obtain a new spectral flux that we then compare with the results of the {\it emissivity model} following equation (5) in \citep{acero2016}. The uncertainty in the choice of the diffuse LMC template is similar in impact to the standard IEM, where the systematic errors are on the same order as the $1 \sigma$ statistical errors in flux. Similarly, we find the errors are largest in the energy bins below 5\,GeV, but are negligible for higher-energy bins.

To explore the uncertainties on the extension influenced by the IEM, we follow the same method outlined for the spectral flux above: We re-generate the best-fit point-source model for all extended sources and perform independent extension tests. Each ROI is tested for extension considering the 8 aIEMs for a total of 9 extension tests per ROI including the standard model fit. We then compare the best-fit extension and symmetric error values to the standard model following equation (5) in \citep{acero2016}. This method has been similarly applied in prior catalogs \citep{acero2016,fges2017} and both reported similar trends on the systematic error on extension as was found on the flux. The systematic errors for extension are generally also on the same order as the $1 \sigma$ statistical errors \citep[e.g., Section~2.4.2 in][]{acero2016}, which is generally consistent with our results that are displayed in Table~\ref{tab:extent}. For crowded regions, a larger error on both the flux and the extension is expected \citep[see Tables~\ref{tab:extent}, A.\ref{tab:spectral_fluxes} and][]{fges2017}. 

We note that the systematic uncertainties on extension for 4FGL~J1631.6--4756e and 4FGL~J1824.5--1351e are incompatible to prior systematic studies. For 4FGL~J1631.6--4756e, we find this source is best-fit as a radial Gaussian with $\sigma = 0.19 \pm 0.027 \pm 0.83\,\degree$, where the first quoted error represents the $1 \sigma$ statistical error and the latter the systematic error. For comparison, the FGES catalog found an extension using a radial disk template $r = 0.26 \pm 0.02 \pm 0.08\,\degree$ for 4FGL~J1631.6--4756e. However, this source lies in a crowded region among several other extended sources, including a larger, unknown extended 4FGL source (J1633.0--4746e) coincident in location, and therefore the large systematic errors are not unexpected. As for 4FGL~J1824.5--1351e, which we find is better fit as a radial Gaussian with $\sigma = 0.83 \pm 0.074 \pm 0.10\,\degree$, has a smaller systematic error than what is reported in the FGES catalog by almost a factor of 3, but this is likely due to the much larger radius found for the extension of 4FGL~J1824.5--1351e which was a radial disk with $r = 1.05\,\degree$ in the FGES catalog. For the remainder of the extended sources, we find compatible systematic uncertainties on the extension as those reported in the FGES catalog, see Table~\ref{tab:extent}.  

\subsection{Caveats}\label{sec:caveats}
The \fermi PWN catalog is limited by the complexity of the $\gamma$-ray data including the systematic uncertainties from the diffuse Galactic background and the uncertainties from the effective area. We emphasize here that while we attempt to account for these effects, the uncertainties explored here may not fully represent the range of systematics involved \citep[see e.g., Section 2.4.2 in][]{acero2016}.  

Important caveats are demonstrated in the crowded ROI for Vela Junior. The PWN G266.9--1.10 is displayed in Figure~\ref{fig:psrj0855-4644}, both panels, and is also discussed in Section~\ref{sec:pwnc}. The Vela Jr. SNR is a GeV emitter, 4FGL~J0851.9--4620e and is nearly $1\,\degree$ in radius. There is some evidence for a PWN contribution in TS maps that include the SNR in the background (see Figure~\ref{fig:psrj0855-4644}), but is exclusively detected $E < 1\,$GeV. This would not be unusual for \fermi pulsars, but may also be explained as structure associated with the Galactic diffuse emission that is not being accounted for properly by the background. While we consider it a tentative detection, it is unclear whether this emission is truly from a Galactic source such as the PWN G266.9--1.10 or PSR~J0855--4644. The $\gamma$-ray bright Vela pulsar is additionally $\sim 4\,\degree$ from Vela Jr. such that the pulsar's emission is intensely bright even up to $E \sim 40\,$GeV, potentially becoming a source of contamination. The large systematic errors on the flux (see Table~\ref{tab:spectral_fluxes} in the Appendix) and the inconclusive extension tests encourage skepticism for the authenticity of this detection. 

\fermi data analysis is even more complicated near the ($\sim 1.0\,\degree$) Galactic center, where at least four PWN candidates are located: SNR~G00.0+0.0, PWN~G0.13--0.11, PWN~G359.9--0.04, and PWN~G358.5--0.96. The last source may have a possible extended GeV counterpart 4FGL~J1745.8--3028e. The ROIs are heavily crowded due to the proximity to the Galactic center, introducing convergence problems in the global fits as well as making source identification extremely difficult if not impossible with the current capabilities of the \fermi instrument. Hence we do not analyze or discuss any of the PWN candidates located within $\sim 1.0\,\degree$ from the Galactic center. 

In general, uncertainties from the interstellar emission model remain to be the most important, while consideration of the location to nearby bright sources especially those of \fermi pulsars can also be critical. Though an incomplete systematic study as mentioned above, this work still represents the most detailed investigation on the systematic uncertainties for fluxes and extensions of \fermi detected PWNe. \clearpage

    \pdfoutput=1
\chapter{Results: The \fermi PWN Catalog}\label{sec:results}

\singlespacing
\begingroup
\begin{table*}
\hspace{-.55cm}
\scalebox{0.95}{
\begin{tabular}{ccccccccc}
\hline
\hline
Galactic PWN Name & 4FGL Name & R.A. & Dec. & TS & TS$_{\text{ext}}$ & $\sigma$ ($\degree$) & 95\% U.L. $\sigma$ ($\degree$) \\
\hline
G18.0--0.69 (HESS~J1825--137) & J1824.5--1351e & 275.931 & 	--13.878 & 211.46 & 88.29 & $0.83 \pm 0.074 \pm 0.10$ & 1.01   \\
G26.6--0.10 & J1840.9--0532e & 280.185 & --5.631 & 411.41 & 74.32 & $0.36 \pm 0.033 \pm 0.038$ & 0.41   \\
G36.0+0.10 & J1857.7+0246e & 284.212 & +2.763 & 310.84	& 93.71 & $0.45 \pm 0.052 \pm 0.16$ & 0.55   \\
G304.1--0.24 (HESS~J1303--631) & J1303.0--6312e & 195.812 & --63.158 & 161.85 & 79.45 & $0.35 \pm 0.033 \pm 0.12$ & 0.40   \\
G326.2--1.70 (MSH 15--56) & J1552.4--5612e & 238.110 & --56.210 & 90.77 & -- & -- & --   \\
\hline
G8.4+0.15 & J1804.7--2144e & 271.114 & --21.742 & 274.96 & 59.65 & $0.29 \pm 0.025 \pm 0.11$ & 0.34   \\
G24.7+0.60 & J1834.1--0706e & 278.529 & --7.119 & 353.70 & 72.57 & $0.19 \pm 0.016 \pm 0.025$ & 0.22   \\
G25.1+0.02 (HESS~J1837--069) & \multirow{2}{*}{J1836.5--0651e} & \multirow{2}{*}{279.243} & \multirow{2}{*}{--6.9090} & \multirow{2}{*}{1736.74} & \multirow{2}{*}{616.23} & \multirow{2}{*}{$0.53 \pm 0.018 \pm 0.058$} & \multirow{2}{*}{0.56}   \\
G25.2--0.19  (HESS~J1837--069) & & & & & & &   \\
G332.5--0.28 & \multirow{2}{*}{J1616.2--5054e}$\dag$ & \multirow{2}{*}{244.205} & \multirow{2}{*}{--50.990} & \multirow{2}{*}{720.43} & \multirow{2}{*}{187.30} & \multirow{2}{*}{$0.31 \pm 0.025 \pm 0.026$} & \multirow{2}{*}{0.35}   \\
G332.5--0.30 (RCW~103) & & & & & & &   \\
G336.4+0.10 & J1631.6--4756e &  248.137 &  --47.912 & 183.85 & 24.56 & $0.19 \pm 0.027 \pm 0.83$ & 0.23   \\
\hline
G11.0--0.05 & \multirow{2}{*}{J1810.3--1925e} & \multirow{2}{*}{272.392} & \multirow{2}{*}{--19.423} & \multirow{2}{*}{113.27} & \multirow{2}{*}{25.88}& \multirow{2}{*}{$0.41 \pm 0.048 \pm 0.048$} &  \multirow{2}{*}{0.49} \\
G11.1+0.08 &  &  &  &  &  &  &    \\
G12.8--0.02 & J1813.1–1737e & 273.473 & --17.651 & 1032.79 & 195.63 & $0.41 \pm 0.021 \pm 0.027$ & 0.45   \\
G15.4+0.10 & J1818.6--1533 & 274.605 & --15.562 & 480.53 & 20.60 & $0.19 \pm 0.030 \pm 0.10$ & 0.24   \\
G29.4+0.10 & J1844.4--0306 & 281.147 & --3.125 & 338.88 & 18.77 & $0.27 \pm 0.040 \pm 0.17$ & 0.34   \\
G328.4+0.20 (MSH~15--57) & J1553.8--5325e & 238.610 & --53.385 & 1241.60 & 436.48 & $0.43 \pm 0.019 \pm 0.028$ & 0.46   \\ 
\hline
\hline
\end{tabular}}
\caption{{\bf (Top) Extended PWNe:} Results of the maximum likelihood fits for previously identified extended LAT PWNe along with the ROI name (PWN Name), right ascension (R.A.) and declination in J2000 equatorial degrees, the detection significance (TS) of the best-fit and the significance for extension TS$_{\text{ext}}$ using the radial Gaussian spatial template (see Section~\ref{sec:ext} for details). The last two columns quote the best-fit extension using the radial Gaussian template and the 95\% upper limit. The first quoted error on the extension ($\sigma$) corresponds to the symmetric $1 \sigma$ statistical error and the latter corresponds to the systematic error. {\bf (Middle) Newly Identified Extended PWNe:} Results of the maximum likelihood fits for new identifications of extended PWNe. {\bf (Bottom) Extended PWN candidates:} Results of the maximum likelihood fits for tentative detections that coincide with extended \fermi sources. \footnotesize{$\dag$ RCW~103 is a known {\it Fermi} PWN, but may have contamination from a nearby PWN, G33.2--0.28, overlapping in location, see text.}}
\label{tab:extent}
\end{table*}
\endgroup
\doublespacing

\noindent Six of the 11 \fermi sources classified as PWNe, each detected as extended, are included in this sample:
\singlespacing
\begin{itemize}
    \item G18.0--0.69 (4FGL~J1824.5--1351e)
    \item G26.6--0.10 (4FGL~J1840.9--0532e)
    \item G36.0+0.10 (4FGL~J1857.7+0246e)
    \item G304.1--0.24 (4FGL~J1303.0--6312e)
    \item G326.2--1.70 (4FGL~J1552.4--5612e)
    \item HESS~J1616--508 (4FGL~J1616.2--5054e)
\end{itemize}
and seven \fermi PWN associations are also analyzed here:
\begin{itemize}
    \item N~157B (4FGL~J0537--6909)
    \item G54.1+0.30 (4FGL~J1930.5+1853, DR3)
    \item G63.7+1.10 (4FGL~J1947.7+2744)
    \item G327.1--1.10 (4FGL~J1554.4–550, DR3)
    \item G336.4+0.10 (4FGL~J1631.6–-4756e)
    \item HESS~J1837--069 (4FGL~J1836.5–-0651e and 4FGL~J1838.9-0704e)
\end{itemize}
\doublespacing
The eighth known PWN association, G0.13--0.11 (4FGL~J1746.4–2852), is not considered here due to its proximity to the Galactic Center. The remainder of known \fermi PWNe: 4FGL~J0534.5+2201e (Crab), 4FGL~J0833.1--4511e (Vela), 4FGL~J1355.2--6420e, 4FGL~J1420.3--6046e, and 4FGL~J1514.2--5909e are all powered by \fermi detected pulsars and are not included in this catalog. 

Of the 58 regions analyzed, we detect 11 unidentified $\gamma$-ray sources that we classify as firm PWNe and are presented in the middle panel of Table~\ref{tab:extent} and in the top panel of Table~\ref{tab:points} based on whether the source detection is extended (4/11, Table~\ref{tab:extent}) or point-like (7/11, Table~\ref{tab:points}). These catalog results have doubled the \fermi PWN population from 11 to 22. The four newly classified extended $\gamma$-ray PWNe bring the total number of extended \fermi PWNe with no detectable $\gamma$-ray pulsar from six to 10, and for the entire extended \fermi PWN population from 11 to 15. There are currently no identified point-like \fermi PWNe in the 4FGL catalogs \citep{4fgl-dr2}, so the seven newly classified $\gamma$-ray point sources represent the first sample of point-like \fermi PWNe. We classify another 22 $\gamma$-ray sources as PWN candidates and are listed in the bottom panels of Tables~\ref{tab:extent} and \ref{tab:points} based on whether they are observed as extended (5/22) or not (17/22). There are 16 ROIs where no significant residual emission is detected and are listed in Table~\ref{tab:nondetect}. 

\section{Source Classification}
\noindent We classify the detected sources considering three important distinctions:
\singlespacing
\begin{itemize}
    \item Positional overlap with a PWN identified in radio, X-ray or TeV surveys
    \item Source extent as observed by the \fermi and the observed extent in radio, X-ray, or TeV surveys
    \item The energetics of the central pulsar, PWN, and host SNR
\end{itemize}
\doublespacing
The last distinction is the most challenging since it requires a combination of available multi-wavelength observations. Moreover, it is often the case that the central pulsar and(or) the SNR shell is not identified, which can prevent reliable source classification if based solely on the energetics of the local environment. Because source classification for high-energy PWNe is so difficult, we can only confidently identify a small number of $\gamma$-ray sources here as PWNe, while the remainder require in-depth multi-wavelength analyses to determine a likely origin and are therefore considered PWN candidates.

\singlespacing
\begingroup
\begin{table*}
\centering
\scalebox{1.0}{
\begin{tabular}{ccccccc}
\hline
\hline
Galactic PWN Name & 4FGL Name & R.A. & Dec. & R95 & TS \\
\hline
G29.7--0.30 (Kes~75) & J1846.9--0247c & 281.600 & --2.9700 & -- & 67.92 \\  
G54.1+0.30 	& J1930.5+1853 (DR3)  & 292.652 & +18.901 & 0.1030 & 33.90 \\
G65.7+1.18 & J1952.8+2924 & 298.131 & +29.456 &	0.0497 & 116.13 \\
G279.6--31.70 (N~157B) & J0537.8--6909 & 84.455 & --69.165  & -- & 34.49 \\
G279.8--35.80 (B0453--685) &  --  & 73.388 & --68.495 & 0.117 & 23.59 \\
G315.8--0.23 (Frying Pan) & J1435.8--6018 &	219.356 & --60.107 & 0.1400 & 38.83 \\
G327.1--1.10 & J1554.4–5506 (DR3) & 238.578 & --55.110 & 0.0986 & 33.76\\
\hline
G11.2--0.35 & J1811.5--1925 & 272.879 & --19.439 & 0.0458 & 58.40 \\
G16.7+0.08 & J1821.1--1422 & 275.239 & --14.329 & -- & 149.37 \\
G18.9--1.10 & J1829.4--1256 & 277.336 & --12.883 & 0.0602 & 85.92 \\
G20.2--0.20 & J1828.0--1133 & 277.020 & --11.571 & 0.0598 & 162.36 \\
G27.8+0.60 & J1840.0--0411 & 279.970 & --4.267 & 0.0399 & 130.98 \\
G34.6--0.50 (W44) & -- & 284.044 & +1.222 & -- & 55.94 \\
G39.2--0.32 & J1903.8+0531 & 286.020 & +5.453 & -- & 124.38 \\
G49.2--0.30 & J1922.7+1428c & 290.702  & +14.274 & -- & 653.81 \\
G49.2--0.70 (W~51C) & J1922.7+1428c & 290.830 & +14.060 & -- & 69.16 \\
G63.7+1.10 & J1947.7+2744 & 296.965 & +27.727 & 0.0668 & 83.70 \\
G74.9+1.11 & J2016.2+3712 & 304.038 & +37.187 & -- & 125.74 \\
G189.1+3.0 (IC~443) & -- & 94.282 & +22.368 & -- & 26.38 \\
G266.9--1.10 & -- & 133.901 & --46.738 & -- & 31.53 \\
G318.9+0.40 & J1459.0--5819 & 224.695 &	--58.379 &	0.1000 & 50.44 \\
G337.2+0.1 & -- & 248.979 & --47.317 & -- & 23.97\\
G337.5--0.1 & J1638.4--4715c (DR3) & 249.510 & --47.231 & -- & 55.57\\
G338.2--0.00 & J1640.6--4632 (DR1), J1640.7--4631e (DR3) & 250.187 & --46.570 & 0.1934 & 191.37 \\
\hline
\hline
\end{tabular}}
\caption{{\bf (Top) Newly Identified Point-like PWNe:} Results of the maximum likelihood fits for newly identified point-like LAT PWNe along with the ROI name (PWN Name), right ascension (R.A.) and declination in J2000 equatorial degrees, the 95\% uncertainty radius in degrees if applicable, and the detection significance (TS). {\bf (Bottom) Point-like PWN candidates:} Results of the maximum likelihood fits for point-like $\gamma$-ray source detections that coincide with known PWNe.}
\label{tab:points}
\end{table*}
\endgroup
\doublespacing

\section{Catalog Results}
\noindent We discuss the results of each new source detection and classification along with their morphological and spectral characteristics in this section.

\subsection{New PWNe}\label{sec:new}
\begin{figure*}[ht]
\begin{minipage}[b]{.5\linewidth}
\centering
\includegraphics[width=1.0\linewidth]{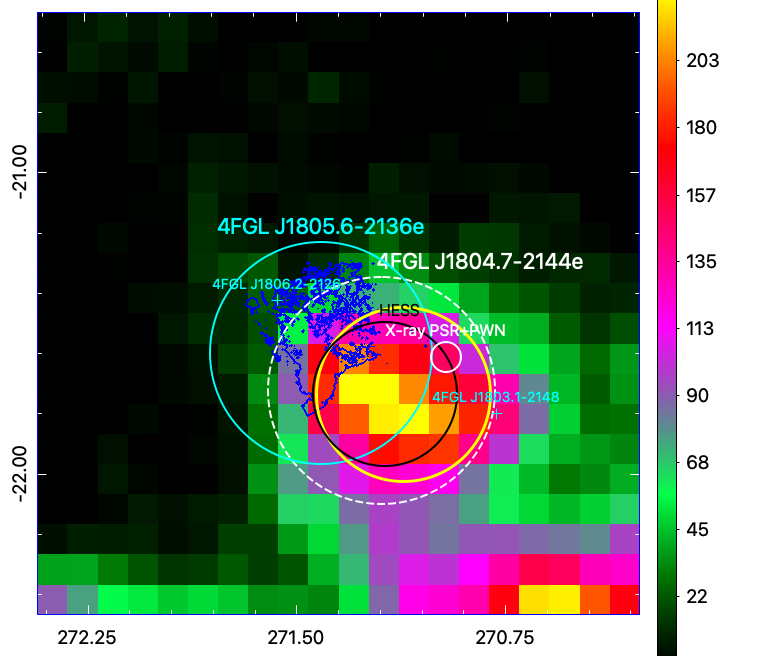}
\end{minipage}
\begin{minipage}[b]{.5\linewidth}
\centering
\includegraphics[width=1.0\linewidth]{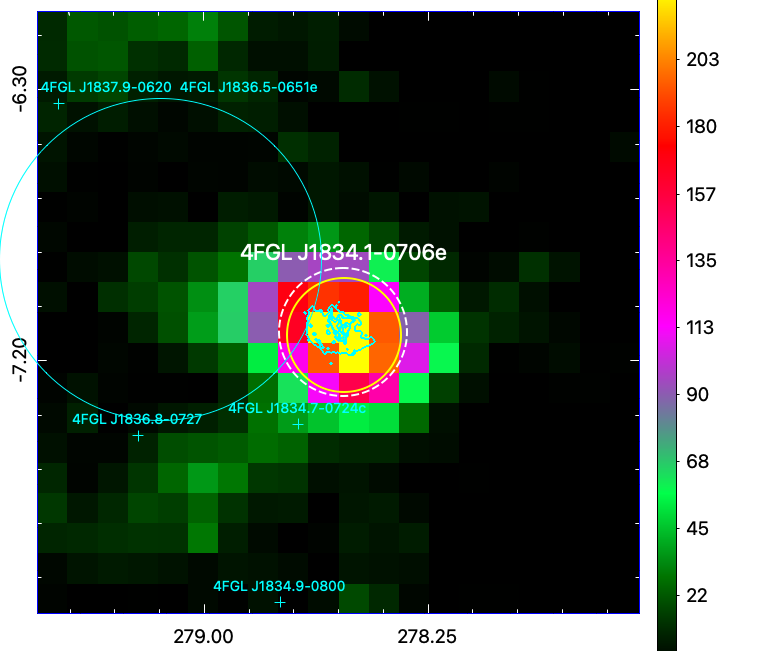}
\end{minipage}
\caption{{\it Left:} A $2\,\degree \times 2\,\degree$ 300\,MeV--2\,TeV TS map of \texttt{PSF3} events for PWN~G8.4+0.15. 4FGL sources in the field of view are labeled in cyan. The TeV PWN HESS~J1804--216 is displayed as a black circle, $r = 0.24\,\degree$. The position for PSR~J1803--2137 and the size of the extended nebula observed in X-ray are marked and labeled as the white circle. The blue contours represent the SNR~G8.7--1.4 in radio. The 4FGL~J1804.7--2144e position and extent is indicated by the white dashed circle but is not included in the source model. The best-fit position and extent for the radial Gaussian template is indicated by the yellow circle. The maximum TS is $\sim$ 238. {\it Right:} A $2\,\degree \times 2\,\degree$ 300\,MeV--2\,TeV TS map for PWN~G24.7+0.60. Unrelated 4FGL sources are indicated in cyan. The 4FGL~J1834.1--0706e position and extent is indicated by the white dashed circle but is not included in the source model. The best-fit position and extent for the radial Gaussian template is indicated by the yellow circle.  The cyan contours represent the Crab-like SNR in radio. The maximum TS at the PWN/SNR position is $\sim$ 266.}\label{fig:8.4_24.7}
\end{figure*}

{\bf G8.4+0.15:} First detected in the TeV band \citep{hess2005,hess2018}, HESS~J1804--216 is an extended unidentified source with semi-major and semi-minor axes 0.24\,$\degree$ and 0.16\,$\degree$, respectively. In \citep{liu2019}, the corresponding extended GeV emission is found to most likely originate from two unrelated sources: PWN~G8.4+0.15 and SNR~G8.7--1.4. 4FGL~J1805.6-2136e is plausibly the SNR~G8.7--1.4 and 4FGL~J1804.7-2144e is argued to be the PWN~G8.4+0.15 which can also explain the coincident TeV source HESS~J1804--216. \citep{liu2019} and the 4FGL catalogs characterize the GeV PWN counterpart as a radial disk with $r = 0.38\,\degree$. We find a slightly smaller extension using the radial Gaussian template $\sigma = 0.29 \,\degree$ characterizes the emission just as well, see Table~\ref{tab:extent}. The best-fit spectrum is a power-law and the spectral index is consistent with what is reported in the 4FGL catalogs, $\Gamma = 1.96 \pm 0.04$. The TeV emission is characterized as a power-law with $\Gamma = 2.69 \pm 0.04$ \citep{hess2018}, indicating that the peak of $\gamma$-ray emission is likely occuring in the \fermi band. Given the agreement between the HESS and \fermi extensions, spectral properties, and the positional coincidence with an identified PWN that is roughly 12$^{\prime\prime}$ in extent from pulsar PSR~J1803--2137 as observed by \cha \citep{karg2007}, we classify 4FGL~J1804.7--2144e as a PWN (see Figure~\ref{fig:8.4_24.7}, left panel).

\singlespacing
\doublespacing

\smallskip
{\bf G24.7+0.60:} is a Crab-like SNR, observed as a bright, centrally peaked core and is roughly $\sim 30^{\prime}$ in size in radio at 20\,cm \citep{becker1987}. The central pulsar is not known, but the radio morphology and spectral properties highly suggest a PWN origin such as a flat radio spectrum and linear polarization from the core \citep{becker1987}. A \fermi source 4FGL~J1834.1--0706e is positionally coincident with G24.7+0.60 and of similar size, but is classified as an SNR in the 4FGL catalogs \citep{4fgl-dr2}. Prior work has argued an SNR origin based on the possible association of a nearby extended TeV source, MAGIC~J1835--069 \citep{g24.72019}, motivated by similar spectral indices and the possibility of the SNR shell interacting with denser material. The best-fit spectral index characterizing the $\gamma$-ray emission is $\Gamma = 2.22 \pm 0.06$ for 4FGL~J1834.1--0706e and $\Gamma \sim 2.75$ for MAGIC~J1835--069 \citep{g24.72019}. It is argued that CRs accelerated by the SNR shell diffuse and illuminate a nearby cloud coincident to the TeV emission.  While it is possible the GeV and TeV sources are associated, we question the depicted scenario since the SNR shell is not firmly detected in any wavelength, whereas the polarized, centrally peaked radio core is observed and strongly suggests a PWN origin instead. The source extension for 4FGL~J1834.1--0706e is modeled as a radial disk with $r = 0.21\,\degree$ and we find the extended emission is fit marginally better as a radial Gaussian with $\sigma = 0.19\,\degree$. We classify this source as a PWN, instead of an SNR as in the 4FGL--DR2, based on the Crab-like nature of the SNR observed in radio.

\singlespacing
\begin{figure*}[ht]
\begin{minipage}[b]{.5\linewidth}
\centering
\includegraphics[width=1.0\linewidth]{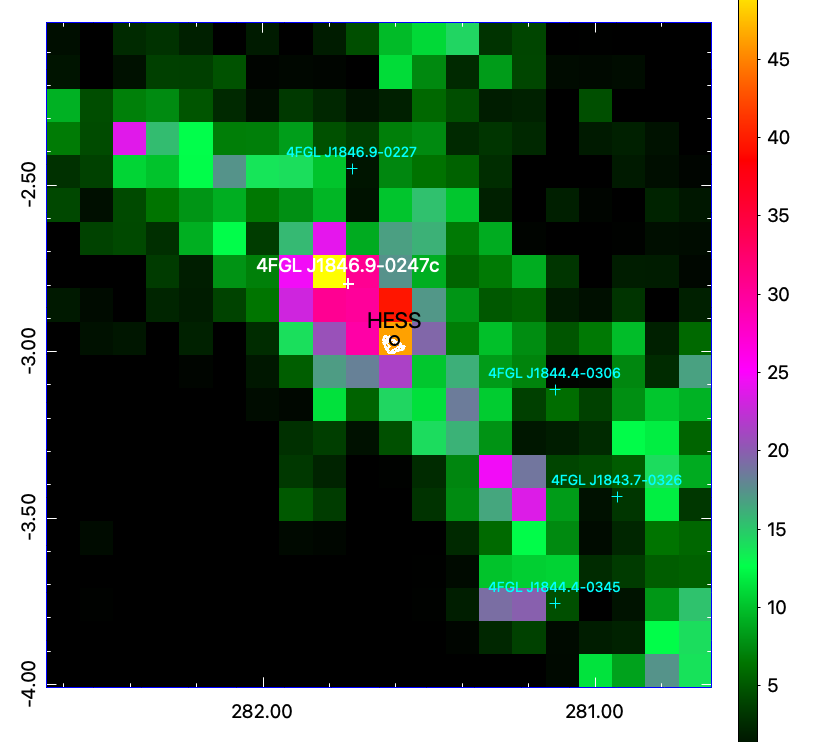}
\end{minipage}
\begin{minipage}[b]{.5\linewidth}
\centering
\includegraphics[width=1.0\linewidth]{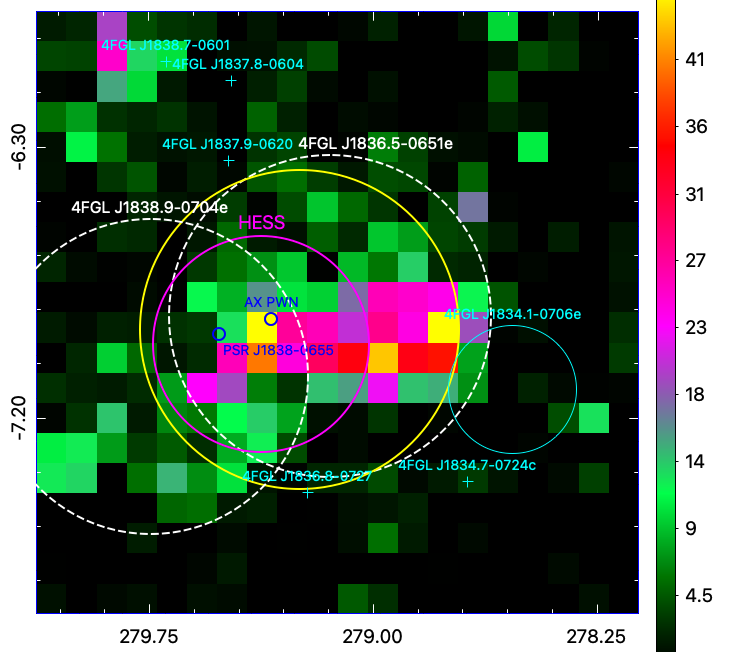}
\end{minipage}
\caption{{\it Left:} A $2\,\degree \times 2\,\degree$ 300\,MeV--2\,TeV TS map of \texttt{PSF3} events centered on Kes~75 (denoted by radio contours in white) and the 68\% confidence region for the TeV PWN HESS~J1846--029. 4FGL~J1846.9--0247c is replaced by a point source at the TeV PWN location. The maximum TS at the PWN position is $\sim$ 46. {\it Right:} A $2\,\degree \times 2\,\degree$ TS map for $E > 10\,$GeV of the TeV PWN HESS~J1837--069 (magenta circle). Unrelated 4FGL sources are indicated in cyan. The 4FGL~J1836.5--0651e and 4FGL~J1838.9--0704e positions and extents are indicated by the white dashed circles but are not included in the source model. The best-fit position and extent for the radial Gaussian template is indicated by the yellow circle.  The blue circles correspond to the location and size of the X-ray nebula AX~J1837.3--0652 (``AX PWN'') and of the nebula powered by PSR~J1838--0655. The maximum TS at the PWN/SNR position is $\sim$ 46. }\label{fig:29.7_hess_j1837}
\end{figure*} 
\doublespacing

\smallskip
{\bf G29.7--0.3 (Kes~75):} is the youngest known PWN in the Milky Way Galaxy with an estimated age $\tau \sim 500\,$yr \citep{kes752018}. It is powered by a very energetic pulsar that may be a magnetar based on its high surface magnetic field strength and having emitted several magnetar-like short X-ray bursts \citep{kes752018}. The PWN is visible in radio and X-ray as a bright, compact nebula encompassed by the incomplete SNR shell. The PWN is a known TeV emitter HESS~J1846--029. The radio contours of the PWN and SNR are shown in the Figure~\ref{fig:29.7_hess_j1837}, left panel, where the central contours correspond to the PWN and are coincident with the HESS 68\% confidence region. There is a nearby unknown source 4FGL~J1846.9--0247c that is plausibly associated to Kes~75. We replace this source with a point source at the PWN/SNR location and find a comparable statistical fit. It seems reasonable that the PWN of Kes~75 is the \fermi source 4FGL~J1846.9--0247c given the spatial agreement and the TeV PWN counterpart. Furthermore, a detailed broadband investigation was performed in \citep{straal2022} that similarly argues 4FGL~J1846.9--0247c is indeed the PWN from Kes~75. The best-fit photon index is $\Gamma = 2.61 \pm 0.11$. We classify Kes~75 as a \fermi PWN. 

\smallskip
{\bf HESS~J1837--069:} First detected as an unidentified TeV source \citep{hess2005}, it was investigated as a PWN candidate after the discovery of two compact ($\lesssim 1^\prime$) X-ray PWNe that coincide with the extension of the HESS source \citep[$\sim 0.36\,\degree$,][]{gotthelf2008,hess2018}. The subsequent detection of an extended \fermi counterpart \citep{acero2013} further supported HESS~J1837--069 as a strong TeV PWN candidate. The investigation into the local environment solidified the PWN origin as the most probable source class \citep{fujita2014}. 
Displayed in Figure~\ref{fig:29.7_hess_j1837}, right panel, is the HESS~J1837--069 TS map and all potential counterparts are indicated. 
Two X-ray PWNe are coincident in location with both the extended TeV and GeV sources. PWN AX~J1837.3--0652 (``AX PWN'' or G25.2--0.20) belongs to SNR~G25.5+0.0 and PSR~J1838--0655 (PWN~G25.1+0.02) powers a compact X-ray nebula closeby \citep{karg2012}. There are two extended sources associated with HESS~J1837--069 in 4FGL--DR2: 4FGL~J1838.9--0704e and 4FGL~J1836.5--0651e. Re-analyzing the extended emission associated to 4FGL~J1836.5--0651e suggests that 4FGL~J1838.9--0704e does not exist, see Section~\ref{sec:nondetect} for details. 4FGL~J1836.5--0651e is modeled with $r = 0.5\,\degree$ and 4FGL~J1838.9--0704e is similarly fit as $r = 0.52\,\degree$, both as a radial disk in 4FGL--DR2 \citep{4fgl-dr2}. We find that there is only one extended source required for the observed emission and corresponds to 4FGL~J1836.5--0651e, but is significantly better fit as a radial Gaussian extension with $\sigma = 0.53\,\degree$. The best-fit spectral index $\Gamma = 1.97 \pm 0.02$ is in good agreement with the TeV PWN spectral index $\Gamma = 2.54 \pm 0.04$. 

Given the PWN classification of the TeV source HESS~J1837--069 and the positional coincidence of the TeV and GeV emission with two compact X-ray PWNe, it seems likely the GeV counterpart is a PWN. The PSF of the \fermi instrument limits distinguishing how each X-ray nebula may contribute to the GeV emission, however. It is possible one or both X-ray nebulae contribute. Therefore we classify the extended emission modeled by 4FGL~J1836.5--0651e as a PWN, but multi-wavelength investigations are needed to determine the X-ray counterpart(s). 

\singlespacing
\begin{figure*}[ht]
\begin{minipage}[b]{.5\linewidth}
\centering
\includegraphics[width=1.0\linewidth]{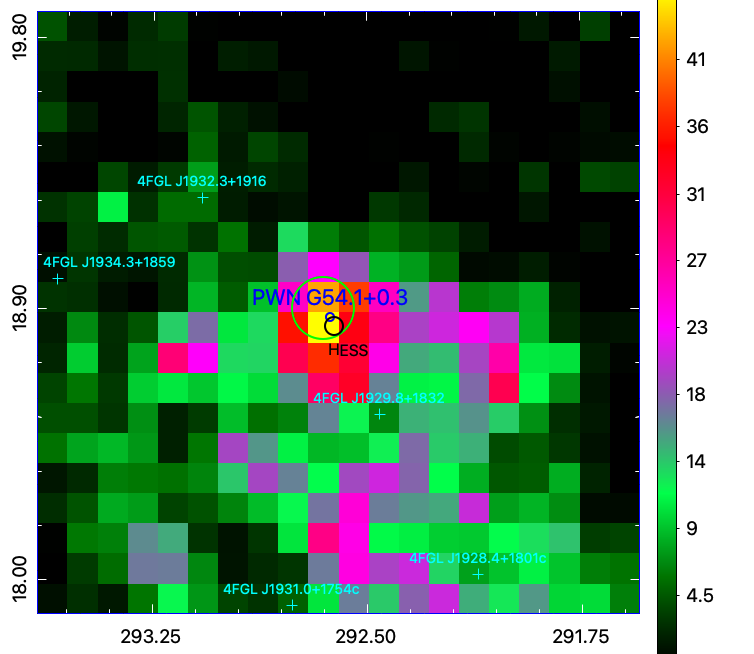}
\end{minipage}
\begin{minipage}[b]{.5\linewidth}
\centering
\includegraphics[width=1.0\linewidth]{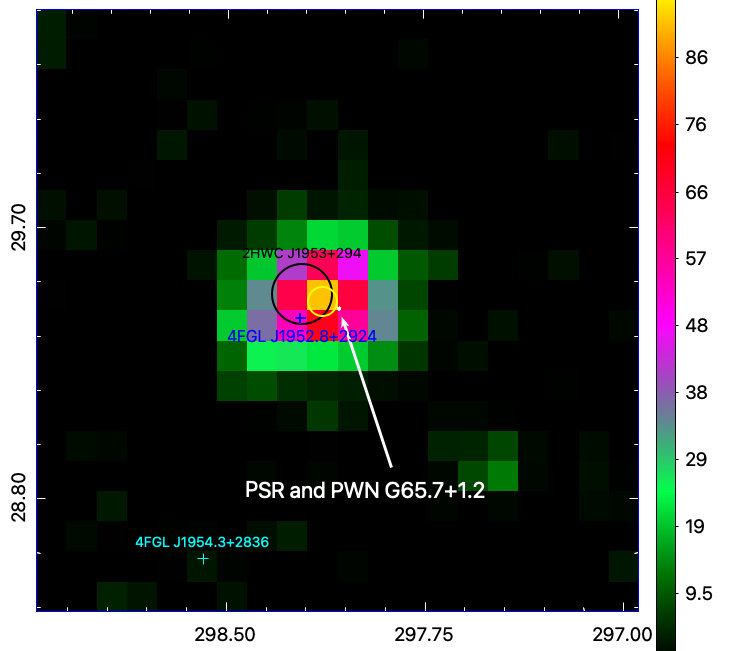}
\end{minipage}
\caption{{\it Left:} A $2\,\degree \times 2\,\degree$ 300\,MeV--2\,TeV TS map of \texttt{ALL} events for PWN~G54.1+0.30. Unrelated 4FGL sources are indicated in cyan. The 95\% uncertainty region of a point source at the PWN position is indicated in green. The blue circle represents the size and location of the X-ray PWN. The 68\% uncertainty region for the TeV PWN HESS~J1930+188 is indicated in black. The maximum TS at the PWN position is $\sim$ 45. {\it Right:} A $2\,\degree \times 2\,\degree$ 300\,MeV--2\,TeV TS map of \texttt{PSF3} events for PWN~G65.7+1.18. The pulsar and PWN in X-ray as observed by \cha are denoted with white contours and are highlighted using the white arrow and label. Unrelated 4FGL sources are indicated in cyan. The 95\% uncertainty region of a point source at the PWN position is indicated in yellow and corresponds to the best-fit position for 4FGL~J1952.8+2924. The maximum observed extension for the TeV PWN 2HWC~J1953+294 is indicated in black. The maximum TS at the PSR/PWN position is $\sim$ 91.
}\label{fig:54.1_65.7}
\end{figure*} 
\doublespacing

\smallskip
{\bf G54.1+0.30:} is a young, Crab-like SNR powered by pulsar J1930+1852 and is roughly $\sim 2^\prime$ in size as observed by \cha \citep{temim2010}. The central pulsar and PWN have been studied in detail in both radio and X-ray \citep[e.g.,][]{camilo2002,gelfand2015, temim2010}. Point-like TeV emission HESS~J1930+188 is identified as the TeV counterpart to the PWN \citep{hess2018}. Only in the DR3 of the 4FGL catalogs is there a new PWN association listed for G54.1+0.3, 4FGL~J1930.5+1853. In 4FGL--DR2 and prior catalogs, there is no cataloged source modeling residual emission associated to G54.1+0.30 \citep{4fgl-dr2}. However, using 11.5\, years of \fermi data, we still detect point-like $\gamma$-ray emission (TS $\sim 34$) not accounted for by 4FGL--DR2 global source models which corresponds to the 4FGL--DR3 source 4FGL~J1930.5+1853. 4FGL~J1930.5+1853 is probably the PWN~G54.1+0.30 considering the positional coincidence. The agreement for the position and spectral index between the GeV and TeV emission \citep[$\Gamma = 2.09 \pm 0.12$ in GeV and $\Gamma = 2.59 \pm 0.26 $ in TeV, respectively,][]{hess2018} confidently classifies the new DR3 source as a PWN. 

\smallskip
{\bf G65.7+1.18:} is another Crab-like PWN with an identified TeV counterpart. NuSTAR and \cha observations revealed a compact X-ray nebula within a radius $r \sim 20^{\prime\prime}$ overlapping in location with the previously unknown TeV source 2HWC~J1953+294 \citep{cover2019}. The TeV emission has no evidence for extension. An unidentified \fermi source 4FGL~J1952.8+2924 is plausibly the GeV counterpart due to positional coincidence (see Figure~\ref{fig:54.1_65.7}, right panel). The TeV counterpart's spectral index $\Gamma = 2.78 \pm 0.15$ \citep{hawc2017} is consistent with the GeV spectral index $\Gamma = 2.50 \pm 0.10$. We hence classify 4FGL~J1952.8+2924 as a PWN based on the similar position, extent, and energetics of the Crab-like SNR observed in each waveband.

\singlespacing
\begin{figure*}[htbp]
\begin{minipage}[b]{.5\linewidth}
\centering
\includegraphics[width=1.0\linewidth]{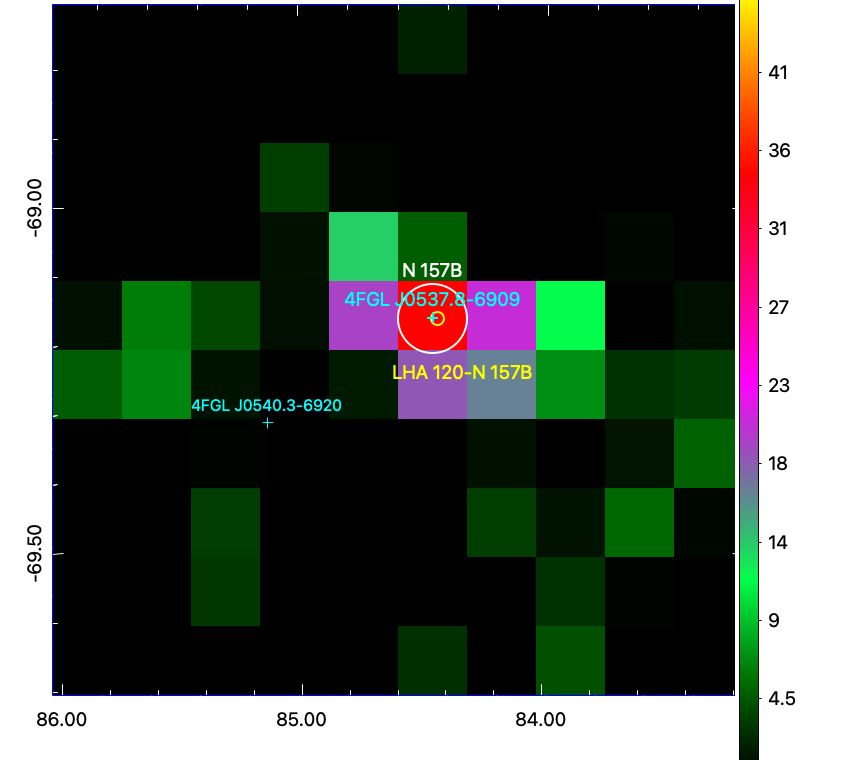}
\end{minipage}
\begin{minipage}[b]{.5\linewidth}
\centering
\includegraphics[width=1.0\linewidth]{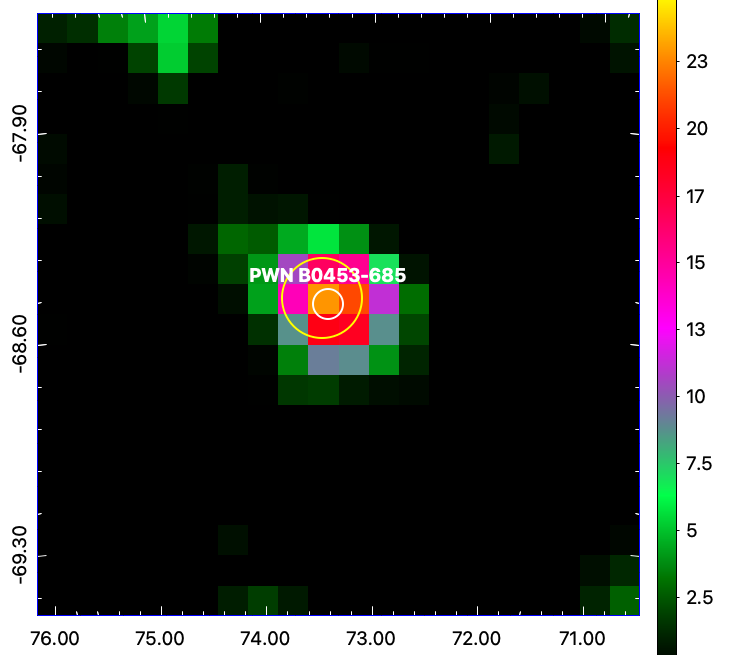}
\end{minipage}
\caption{{\it Left:} A $1\,\degree \times 1\,\degree$ 10\,GeV--2\,TeV TS map of \texttt{PSF3} events for PWN~N~157B. The 4FGL counterpart is 4FGL~J0537.8--6909, indicated in cyan, which is in spatial coincidence with the TeV PWN ``LHA 120--N 157B'' (yellow) and the X-ray nebula (white contours within the TeV PWN region). Unrelated nearby 4FGL sources are labeled. The maximum TS at the PWN/SNR position is $\sim$ 35 for $E > 10$\,GeV.  {\it Right:} A $2\,\degree \times 2\,\degree$ TS map of \texttt{ALL} events centered on B0453--685 (white circle) between 1--10\,GeV. The 95\% positional uncertainty for a point source at the PWN/SNR position is indicated as the yellow circle. The maximum TS at the PWN/SNR position is $\sim$23 in the 1--10\,GeV energy range.}\label{fig:n157b_b0453}
\end{figure*} 
\doublespacing

\smallskip
{\bf N~157B: } is a young PWN powered by PSR~J0537--6910 located within the LMC and is the first MeV--GeV  \citep{lmc2016,saito2017} and TeV  \citep{n157b2012} detection for $\gamma$-ray emission associated to a PWN outside of the Milky Way Galaxy. The pulsar has a spin-down luminosity comparable to that of the Crab pulsar ($L \sim 5 \times 10^{38}\,$erg s$^{-1}$), making it the most rapidly spinning and most powerful young pulsar known. The $\gamma$-ray detection of N~157B is possible given the central pulsar's enormous power output, in addition to potentially rich local photon fields. Point-like GeV emission coincident with the system could plausibly be from the energetic pulsar, the PWN, or the SNR \citep{lmc2016}. The morphology for the PWN/SNR is not clear: The lack of thermal X-rays and the missing limb-brightened outer SNR shell imply a Crab-like morphology, but the diffuse X-rays beyond the cometary nebula have a weak thermal component, implying possible re-heated SN ejecta from the passage of the reverse shock \citep{chen2006}.  Even though the observed \fermi emission cannot firmly rule out a SNR scenario when considered alone, it seems unlikely given the lack of observational evidence for an energetic SNR shell in other wavebands. Pulsations from the central pulsar were searched for in the observed $\gamma$-ray emission, but none were evident above a 1-$\sigma$ confidence level \citep{lmc2016}. The identified TeV counterpart for the PWN combined with no detectable pulses in the \fermi signal and a photon index similar to other \fermi PWNe ($\Gamma = 2.11 \pm 0.07$) guide us to classify this source as a PWN, see also Figure~\ref{fig:n157b_b0453}, left panel.

\smallskip
{\bf B0453--685:} SNR~B0453--685 is a middle-aged ($\tau \sim 14\,$kyr) composite SNR located in the LMC, on the opposite (Western) side from PWN N~157B \citep[e.g., Figure~\ref{fig:lmc},][]{gaensler2003}. \fermi $\gamma$-ray emission coincident with SNR~B0453--685 is detected at a significance level $>$4\,$\sigma$, see Figure \ref{fig:n157b_b0453}, right panel. The $\gamma$-ray emission displays no evidence for extension and the best-fit spectral index $\Gamma = 2.27 \pm 0.18$ characterizes a power-law spectrum. The central pulsar has not been detected though a detailed multi-wavelength investigation described in \citep{eagle2022a} (see also Chapter~\ref{sec:b0453}), finds the most likely origin to be the PWN with a possible pulsar contribution. The host SNR displays energetics that are inconsistent if the SNR is the $\gamma$-ray origin such as no detectable non-thermal emission in X-ray nor observational evidence suggesting an interaction with ambient media. The position, extent, and energetics instead favor a PWN origin. 

\begin{figure*}[htbp]
\begin{minipage}[b]{.5\linewidth}
\centering
\includegraphics[width=1.0\linewidth]{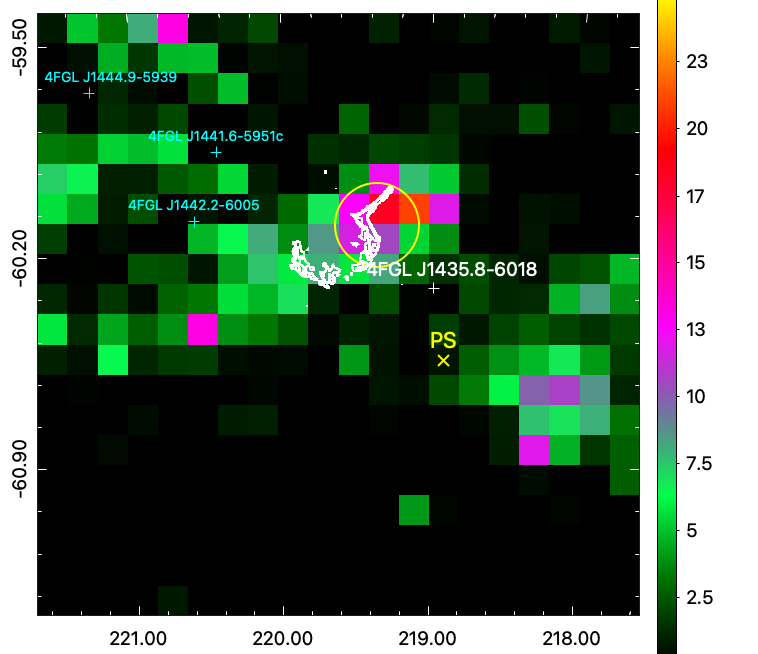}
\end{minipage}
\begin{minipage}[b]{.5\linewidth}
\centering
\includegraphics[width=1.0\linewidth]{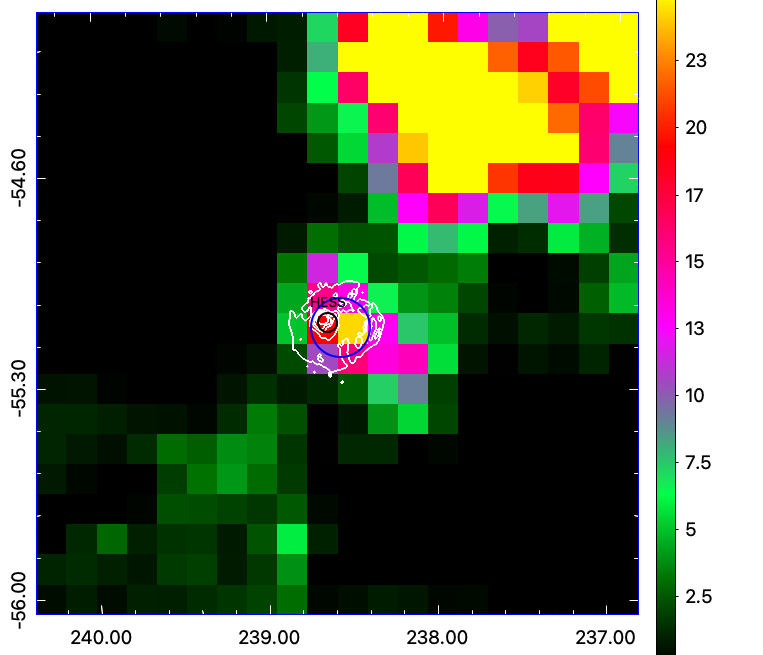}
\end{minipage}
\caption{{\it Left:} A $2\,\degree \times 2\,\degree$ 300\,MeV--2\,TeV TS map of \texttt{PSF3} events for PWN~G315.8--0.23. There is one possibly associated \fermi source, 4FGL~J1435.8--6018. The 95\% uncertainty region of the new best-fit position for 4FGL~J1435.8--6018 is the yellow circle, which coincides with the ``handle'' of the Frying Pan radio morphologhy. The ``handle'' consists of the supersonic pulsar and the trailing PWN. The yellow cross labeled ``PS'' represents the addition of one point soure to model residual emission unrelated to the SNR. The maximum TS at the PWN/SNR position is $\sim$ 21. {\it Right:} A $2\,\degree \times 2\,\degree$ 1--10\,GeV TS map of \texttt{ALL} events centered on SNR~G327.1--1.1 (denoted by 843\,MHz radio contours in white) accompanied by the 95\% confidence regions for the TeV PWN HESS~J1554--550 (black) and for a point source modeling the residual MeV--GeV emission (blue). The maximum TS at the PWN/SNR position is $\sim$24 in the 1--10\,GeV energy range.}\label{fig:g315_g327}
\end{figure*}

\smallskip
{\bf G315.8--0.23:} Also known as the Frying Pan, this ancient system features a radio SNR shell and an exiting supersonic pulsar that powers a trailing PWN in its wake \citep{ng2012}. An unidentified source 4FGL~J1435.8--6018 is in close proximity to the radio position of the bow-shock PWN and pulsar. We investigate the possible association with PWN~G315.8--0.23 by replacing 4FGL~J1435.8--6018 with a point source at the PWN position, in addition to two other point sources to the Southwest of the SNR shell in order to model unrelated yet persisting diffuse residual $\gamma$-ray emission. The two point sources are required for a significantly better global fit in addition to localizing 4FGL~J1435.8--6018 to the ``handle'' of the Frying Pan SNR morphology, which is where the supersonic pulsar and its bow-shock nebula are located, see Figure~\ref{fig:g315_g327}, left panel.  The TS for a point source at the PWN/PSR position is TS $= 39$ with no evidence for extension. The placement of the $\gamma$-ray source with the bow-shock nebula, its spectral index $\Gamma = 2.76 \pm 0.16 $ being similar to other \fermi PWNe, as well as considering the energetic nature of the bow-shock nebula \citep{ng2012}, all support a PWN classification.

\singlespacing
\doublespacing

\smallskip
{\bf G327.1--1.10:} The first MeV--GeV detection of this source by the \fermi was reported in \citep{xiang2021}. The $\gamma$-ray emission is detected at $>4$\,$\sigma$ significance with the \fermi between 300\,MeV--2\,TeV and is best characterized with photon index $\Gamma = 2.50 \pm 0.12$. A detailed multi-wavelength investigation for the associated emission was analyzed in \citep{eagle2022b} (see also Chapter~\ref{sec:g327}), where the point-like MeV--GeV $\gamma$-ray emission is consistent with a PWN origin from G327.1--1.1, supported by its TeV counterpart HESS~J1554--550.  Similar to B0453--685, the position, extent, and energetics of the system favor a PWN origin. A 1--10\,GeV TS map demonstrating the source detection is displayed in Figure~\ref{fig:g315_g327}, right panel.

\begin{figure*}[htbp]
\begin{minipage}[b]{.5\linewidth}
\centering
\includegraphics[width=1.0\linewidth]{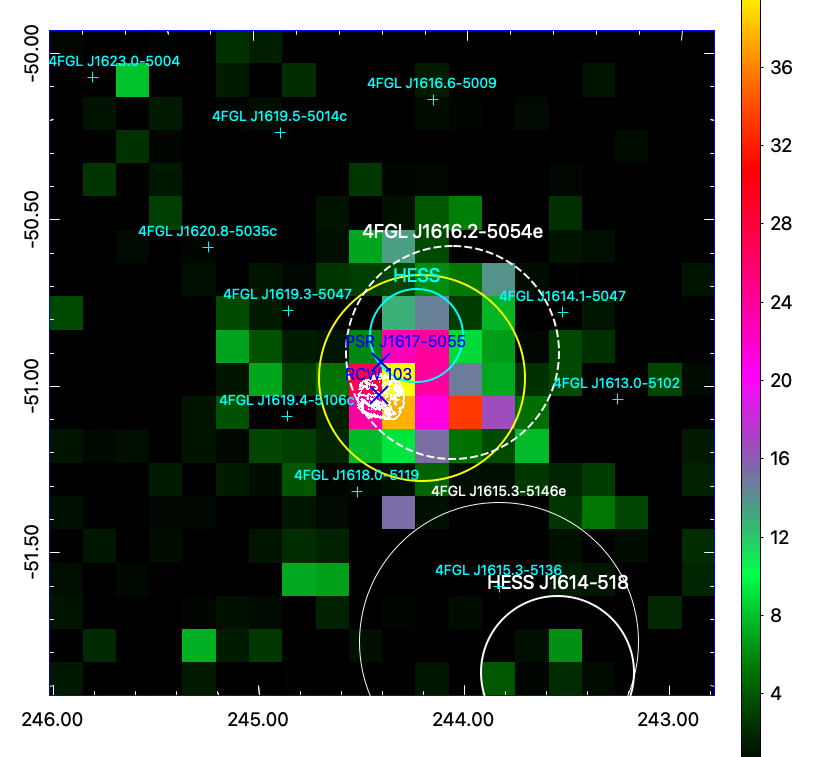}
\end{minipage}
\begin{minipage}[b]{.5\linewidth}
\centering
\includegraphics[width=1.0\linewidth]{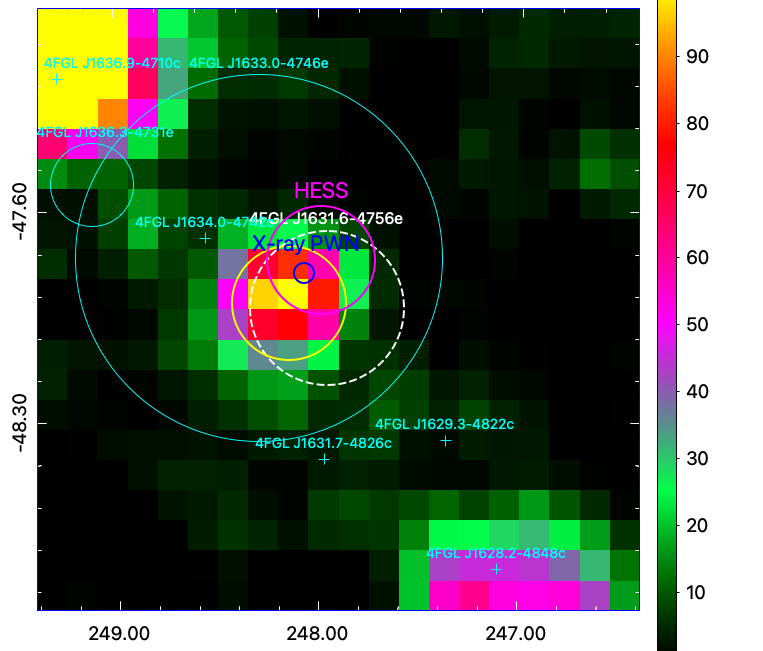}
\end{minipage}
\caption{{\it Left:} A $2\,\degree \times 2\,\degree$ 10\,GeV--2\,TeV TS map of \texttt{PSF3} events for both X-ray PWNe coincident with the \fermi PWN 4FGL~J1616.2--5054e: G332.5--0.3 (RCW~103, with \cha X-ray contours in white) and G332.5--0.28 (PSR~J1617--5055). 4FGL~J1616.2--5054e is indicated as the white dashed circle but is not included in the source model. The best-fit for the extended emission is marked as a yellow circle.  The TeV PWN counterpart corresponds to the cyan circle. Unrelated 4FGL sources are labeled in cyan. The maximum TS occurs between the two X-ray PWNe with value $\sim$ 50 for energies 1--10\,GeV. 4FGL~J1615.3--5146e (solid white circle) is listed in the 4FGL as the SNR shell for RCW~103, which seems unlikely given the large displacement.  {\it Right:} A $2\,\degree \times 2\,\degree$ 1--10\,GeV TS map of \texttt{PSF3} events for PWN~G336.4+0.10. Unrelated 4FGL sources are in cyan. 4FGL~J1631.6--4756e is indicated but not included in the source model. The best-fit radial Gaussian template for the extended emission is the yellow circle. The X-ray PWN location and extent is marked with a blue circle and the TeV PWN counterpart HESS~J1632--478 in magenta.  The maximum TS at the PWN position is $\sim$ 109 for energies 1--10\,GeV.}\label{fig:g332s_g336}
\end{figure*}

\smallskip
{\bf G332.5--0.3 (RCW~103) and G332.5--0.28:} The SNR~RCW~103 is listed as the extended \fermi source 4FGL~J1615.3--5146e which is coincident with the extended TeV shell HESS~J1614--518, but is not in positional agreement with the $\sim 5^\prime$ X-ray SNR shell as observed by \cha \citep{rcw1032015}. As displayed in Figure~\ref{fig:g332s_g336}, left panel, 4FGL~J1615.3--5146e is $\sim 1\,\degree$ from the X-ray PWN/SNR position, making the SNR shell from RCW~103 an unlikely origin for both the GeV and TeV extended emission there. However, RCW~103 and a second X-ray PWN powered by the pulsar J1617--5055 are positionally coincident with the \fermi PWN for RCW~103, 4FGL~J1616.2--5054e, which is characterized as a radial Gaussian template with $r = 0.32\,\degree$ in the 4FGL catalogs \citep{4fgl-dr2}. We report consistent extension results for this source, also using a radial Gaussian template with $\sigma = 0.31\,\degree$, but with an off-set position that is better localized to both PWNe. The PWN associated to PSR~J1617--5055 is $\sim 1^\prime$ in length observed in X-ray \citep{karg2009}, while RCW~103 is $\sim 10^\prime$ in size \citep{rcw1032015}. The two systems are only $0.02\,\degree$ from the other in location, making it impossible to distinguish the more likely $\gamma$-ray emitter. The TeV counterpart, HESS~J1616--508 displays similar extension to the GeV source, though notably smaller and more localized to PSR~J1617--5055 than to RCW~103. The best-fit spectral index for the GeV source is $\Gamma = 1.98 \pm 0.03$ which is comparable to the TeV source's index $\Gamma = 2.32 \pm 0.06$ \citep{hess2018}. It seems probable that the origin for the GeV and TeV emission is from one or both PWNe, in which case we still consider the extended source 4FGL~J1616.2--5054e a firm \fermi PWN, while noting there is potentially more than one X-ray counterpart. Future high-energy studies should explore this region further. 

\singlespacing
\doublespacing

\smallskip
{\bf G336.4+0.10:} HESS~J1632--478 was first discovered as an extended ($\sim 12^\prime$) unidentified TeV source \citep{aha2006}. An X-ray PWN was subsequently identified by \xmm observations of a point source accompanied by faint, diffuse non-thermal X-ray emission extending outward $\sim 32^{\prime\prime}$ in size \citep{balbo2010}. The X-ray nebula is  in positional coincidence with the HESS emission in addition to possible extended GeV emission.  The X-ray, GeV, and TeV extended emission favor a PWN scenario. In all \fermi catalogs, the extended GeV emission corresponding to 4FGL~J1631.6--4756e is fit as a radial disk with $r = 0.25\,\degree$. Re-analysis of the \fermi emission concludes that a radial Gaussian template of size $\sigma = 0.19\,\degree$ can significantly improve the fit. HESS~J1632--478 is $\sim0.21\,\degree$ in size \citep{hess2018}, in good agreement with the extension we find. The TeV spectral index $\Gamma = 2.52 \pm 0.06$ is also in good agreement with the spectral index measured by \fermi $\Gamma = 1.86 \pm 0.04$. Since it is highly likely the GeV source is the counterpart to the TeV PWN, we classify 4FGL~J1631.6--4756e as a firm \fermi PWN detection. 

\subsection{New PWN candidates}\label{sec:pwnc}
\begin{figure*}[htbp]
\begin{minipage}[b]{.5\linewidth}
\centering
\includegraphics[width=1.0\linewidth]{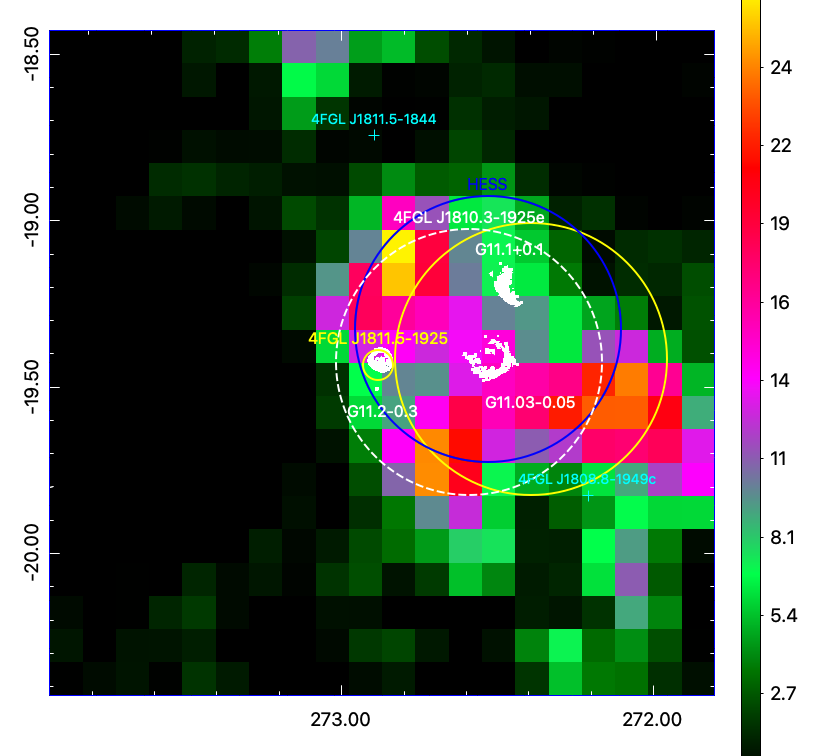}
\end{minipage}
\begin{minipage}[b]{.5\linewidth}
\centering
\includegraphics[width=1.0\linewidth]{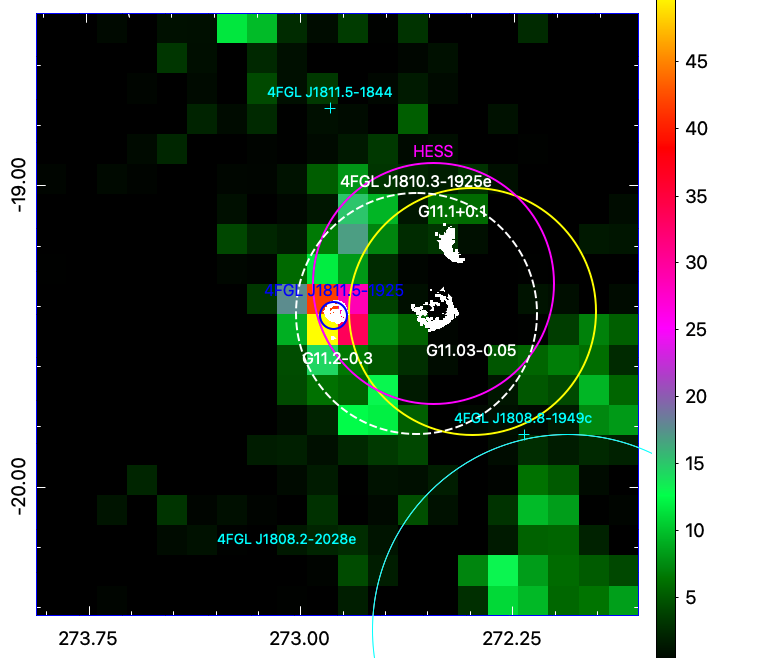}
\end{minipage}
\caption{{\it Left:} A $2\,\degree \times 2\,\degree$ 1--10\,GeV TS map of \texttt{PSF3} events for plerionic SNRs G11.0--0.05 and G11.1+0.10 (denoted by radio contours in white). G11.2--0.35 is also plotted with its radio contours and is coincident to 4FGL~J1811.5--1925. An unidentified, extended TeV source HESS~J1809--193 is coincident with all three PWNe, displayed in blue. 4FGL~J1810.3--1925e is indicated but not included in the source model. The best-fit radial Gaussian template for the extended emission is the yellow circle. The maximum TS is $\sim$27 in the 1--10\,GeV energy range. {\it Right:} A $2\,\degree \times 2\,\degree$ 300\,MeV--2\,TeV TS map of PWN~G11.2--0.35. The 95\% uncertainty region for 4FGL~J1811.5--1925 (blue circle) is coincident with the radio position of SNR~G11.2--0.3 (radio contours in white). Unidentified HESS~J1809--193 is displayed as the magenta circle with radius $r = 0.4\,\degree$. The maximum TS at the PWN/SNR position is TS $\sim$ 54. In both panels, unrelated 4FGL sources in the field of view are labeled in cyan.}\label{fig:g11.0_g11.1_g11.2}
\end{figure*}

{\bf G11.0--0.05 and G11.1+0.08:} These are two uncertain SNR/PWN candidates. Both were first identified in \citep{brogan2004} using Very Large Array (VLA) observations at 1465\,MHz along with Giant Metrewave Radio Telescope (GMRT) at 235\,MHz. The two SNR candidates in addition to their neighbor, the composite SNR~G11.2--0.3, are plotted as their VLA radio contours in white in both panels of Figure~\ref{fig:g11.0_g11.1_g11.2}. An extended source possibly associated to either or both SNRs is detected by the \fermi and is characterized using a radial disk template with $r = 0.5\,\degree$ \citep{4fgl-dr2}. The extended emission can also be reasonably modeled using a radial Gaussian template with a slightly smaller size, $\sigma = 0.41\,\degree$. An unidentified, extended TeV source HESS~J1809--193 is coincident with both PWN candidates and G11.2--0.3. HESS~J1809--193 has extension best-fit as a Gaussian, $\sigma = 0.40\,\degree$ \citep{hess2018}. 

In our analysis of the \fermi data, we tested for multiple point sources replacing 4FGL~J1810.3--1925e, but one extended \fermi source is clearly required here with only one additional point source, 4FGL~J1811.5--1925. It is possible that the coincident \fermi and TeV emission originate from one or both PWN candidates. HESS~J1809--193 may be unidentified, but it has spectral properties consistent with a PWN origin \citep[$\Gamma = 2.38 \pm 0.07$,][]{hess2018}. Combined with the positional coincidence with at least two plerionic SNRs, it seems likely the TeV emission originates from one or both PWN candidates, which also categorizes the coincident GeV emission with similar extent and spectral index ($\Gamma = 2.33 \pm 0.08$) as a PWN candidate. Both G11.0--0.05 and G11.1+0.08 are therefore considered possible PWN radio counterparts to the extended GeV emission reported here. 

\singlespacing
\doublespacing

\smallskip
{\bf G11.2--0.35:} belongs to composite SNR G11.2--0.3 (the ``Turtle'') and is powered by pulsar PSR~J1811--1925. This region is crowded both in radio \citep{brogan2004} and in \fermi $\gamma$-rays. Two other plerionic SNR candidates are nearby \citep{brogan2004}: G11.1+0.1 (PWN candidate G11.1+0.08) and G11.0--0.0 (PWN candidate G11.0--0.05). All three plerionic SNRs are coincident in location with an unidentified extended 4FGL source 4FGL~J1810.3--1925e. A second point source is additionally coincident to G11.2--0.35 (see Figure~\ref{fig:g11.0_g11.1_g11.2}, right panel). The extended source is required to model significant extended $\gamma$-ray emission, though with a location off-set from 4FGL~J1811.5--1925 and G11.2--0.35 and fit as a radial Gaussian template, $\sigma = 0.41\,\degree$. For comparison, the 4FGL catalogs model the extended emission as a radial disk with $r = 0.5\,\degree$, such that the extension coincides with all three SNRs. 

With the new template and location, 4FGL~J1810.3--1925e is found to overlap with the PWN candidates G11.1+0.08 and G11.0--0.05 and the unidentified TeV source HESS~J1809--193 that has a Gaussian extension $r = 0.40\,\degree$ \citep{hess2018}. 
The additional source 4FGL~J1811.5--1925 is best-fit as a power-law with $\Gamma = 2.03 \pm 0.01$ and is coincident with the position of the young and energetic SNR~G11.2--0.3. The radio-quiet central pulsar, PWN, and host SNR are detected in X-ray \citep{madsen2020}. The SNR shell exhibits non-thermal X-ray emission which suggests evidence for efficient particle acceleration of electrons to multi-TeV energies \citep{madsen2020}. However the \fermi data cannot readily distinguish a dominant emission component. The young age and the energetic nature of the pulsar, PWN, and SNR shell require a broadband investigation to determine the most probable scenario of the observed $\gamma$-ray emission.  

\begin{figure*}[ht]
\begin{minipage}[b]{.5\linewidth}
\centering
\includegraphics[width=1.0\linewidth]{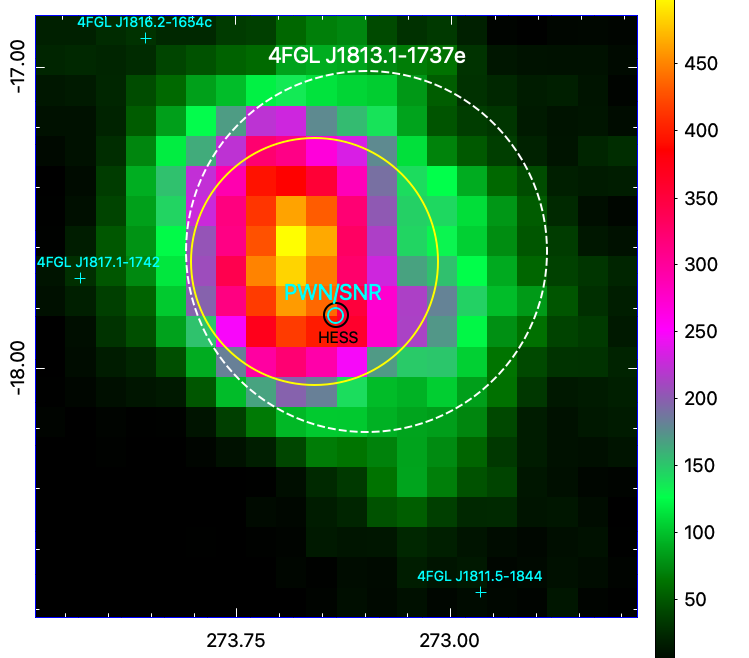}
\end{minipage}
\begin{minipage}[b]{.5\linewidth}
\centering
\includegraphics[width=1.0\linewidth]{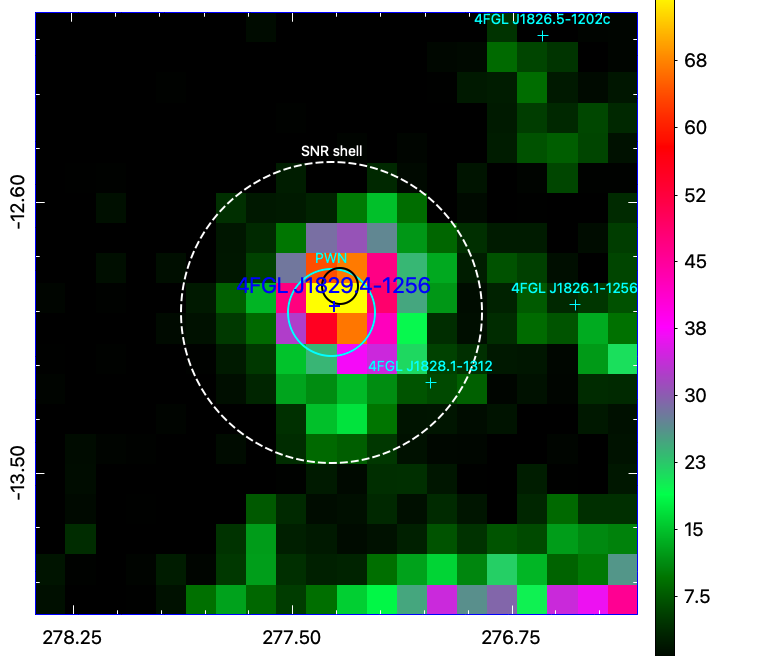}
\end{minipage}
\caption{{\it Left:} A $2\,\degree \times 2\,\degree$ 300\,MeV--2\,TeV TS map of \texttt{PSF3} events for PWN~12.8--0.02. The SNR shell is $\sim 3^\prime$ in diameter (cyan circle) which embodies PSR~J1813--1749 and the X-ray PWN. The TeV PWN HESS~J1813--178 location and size is marked in black. The 4FGL~J1813.1--1737e position and extent is indicated by the white dashed circle but is not included in the source model. The best-fit position and extent for the radial Gaussian template is indicated by the yellow circle. The maximum TS is $\sim$ 506. {\it Right:} A $2\,\degree \times 2\,\degree$ 300\,MeV--2\,TeV TS map for PWN~G18.9--1.10. The location and size of the radio PWN is represented as a cyan circle. The white dashed circle corresponds to the size of the SNR shell in radio. The 95\% positional uncertainty region for a point source modeling emission associated to 4FGL~J1829.4--1256 is marked as a black circle. Unrelated 4FGL sources are indicated in cyan.  The maximum TS at the PWN position is $\sim$ 80.}\label{fig:12_18.9}
\end{figure*}

\smallskip
{\bf G12.8--0.02:} is a composite SNR housing the central pulsar J1813--1749 that powers the nebula. The PWN was first detected as the TeV source HESS~J1813--178 \citep{hess2005} and later identified as a PWN after being identified in radio and X-ray \citep{g12_2005}. The TeV PWN is extended $r = 0.04\,\degree$ and is consistent with the full extent of the SNR shell $r \sim 0.025\,\degree$ in radio \citep{hess2018}. An extended \fermi source, 4FGL~J1813.1--1737e overlaps in location, but exhibits a much larger extension $r = 0.6\,\degree$ as a radial disk in the 4FGL catalogs \citep{araya2018,4fgl-dr2}. The extended emission can be better fit where the centroid is localized closer to the SNR position and modeled as a radial Gaussian and smaller extent, $\sigma = 0.41\,\degree$, see Figure~\ref{fig:12_18.9}, left panel. The larger extension in the GeV band compared to the X-ray and TeV morphologies is notably different, but the spectral index for the GeV source $\Gamma = 2.34 \pm 0.03$ is similar to the TeV PWN spectral index $\Gamma = 2.07 \pm 0.05$ \citep{hess2018}.

\singlespacing
\doublespacing

\smallskip
{\bf G18.9--1.10:} is a strong PWN candidate based on \cha X-ray observations of a possible point source accompanied by an extended nebula that is $\sim 8^\prime$ in size \citep{tullmann2010}. The central pulsar has not been detected though the larger SNR shell is and nearly $0.5\,\degree$ in size in radio \citep{tullmann2010}. An unidentified source 4FGL~J1829.4--1256 is coincident in position to the X-ray PWN, see Figure~\ref{fig:12_18.9}, right panel. Given the observed $\gamma$-ray emission is point-like, localized well within the SNR radio shell, and has a spectral index $\Gamma = 2.40 \pm 0.09$, we classify this source as a probable PWN, depending on confirming the X-ray counterpart as a PWN, though it is possible the central pulsar may have some contribution to the observed $\gamma$-ray emission.

\singlespacing
\begin{figure}[htbp]
\centering
\includegraphics[width=0.85\linewidth]{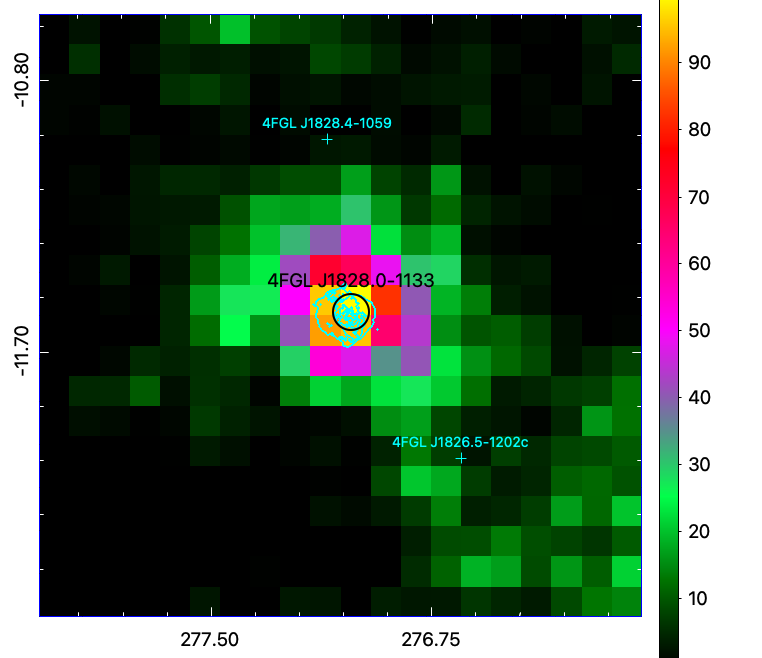}
\caption{ A $2\,\degree \times 2\,\degree$ 300\,MeV--2\,TeV TS map of \texttt{PSF3} events for PWN~G20.2--0.20. Unrelated 4FGL sources are indicated in cyan. The 95\% uncertainty region for the coincident source 4FGL~J1828.0--1133 is in black. The cyan contours represent the radio SNR and the central peak corresponds to the PWN. The maximum TS at the PWN/SNR position is $\sim$ 101.}\label{fig:g20.2}
\end{figure} 
\doublespacing

\smallskip
{\bf G20.2--0.20:} While the central pulsar has not been identified, the SNR shell and PWN are detected in X-ray, where the SNR morphology is probably the result of the shell interacting with ambient molecular clouds \citep[MCs,][]{petriella2013}. The thermal nature of the diffuse X-ray emission filling the SNR shell is not distinguished, but the overlapping nonthermal radio nebula favors a PWN scenario \citep{becker1985,petriella2013}. An unidentified 4FGL source 4FGL~J1828--1133 is coincident in both position and size for the entire SNR system, see Figure~\ref{fig:g20.2}. It is likely the source is associated to the PWN/SNR system in some way, but because the SNR may be interacting with its surroundings, a detailed analysis considering the broadband properties will be required to determine the most likely origin between the central pulsar, PWN, and SNR. The best-fit spectral index is $\Gamma = 2.59 \pm 0.07$. 

\begin{figure*}[htbp]
\begin{minipage}[b]{.5\linewidth}
\centering
\includegraphics[width=1.0\linewidth]{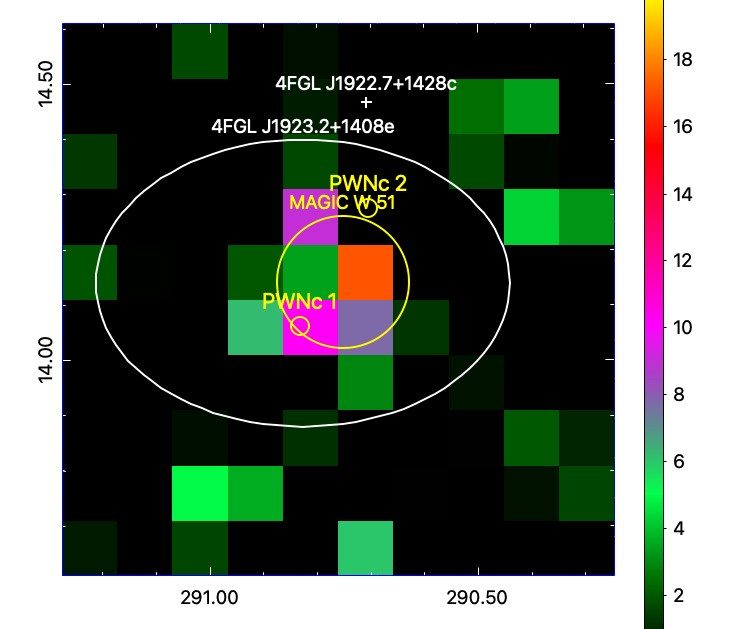}
\end{minipage}
\begin{minipage}[b]{.5\linewidth}
\centering
\includegraphics[width=1.0\linewidth]{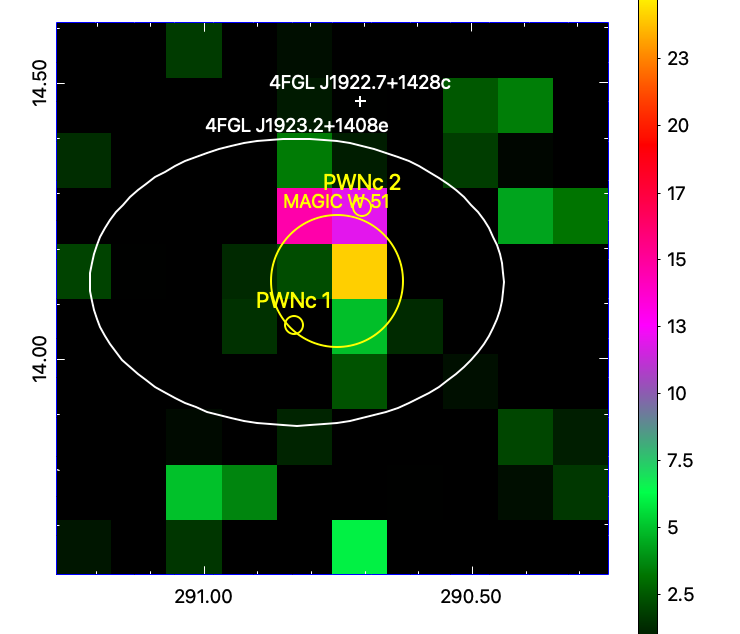}
\end{minipage}
\caption{{\it Left:} A $1\,\degree \times 1\,\degree$ 10\,GeV--2\,TeV TS map of \texttt{PSF3} events for PWN~G49.2--0.7, the possible PWN candidate to SNR~W~51C. The SNR is the GeV emitter 4FGL~J1923.2+1408e. There is one possibly associated \fermi source that is not included in the source model, 4FGL~J1922.7+1928c. The extended TeV source for the SNR is the yellow circle, which overlaps spatially with both PWN candidates. The maximum TS at the PWN position (``PWNc 1'')  is $\sim$17 for $E >10$\,GeV. {\it Right:} Same as the left panel, but instead displaying $\gamma$-ray emission accounted for by G49.2--0.3 (``PWNc 2''), the second PWN candidate to SNR W51~C. The maximum TS at the PWN position is $\sim$24 for $E >10$\,GeV.}\label{fig:w51c_g49.2}
\end{figure*}

\smallskip
{\bf G49.2--0.3 and G49.2--0.7 (W51~C):} The SNR W51~C or G49.2--0.7 is located near a star forming region and houses a pulsar candidate CXO~J192318.5+140335 with compact X-ray emission observed by \cha \citep[core $\sim 1^\prime$,][]{chaw51c2005}. Observations performed by \xmm reveal a second PWN candidate for the SNR which has similar extent to the other PWN \citep{xmmw51c2014}. The SNR has a radio diameter $\sim 1.0 \,\degree$ and is the basis for modeling \fermi extended $\gamma$-ray emission in the region coincident with this system. It was shown that the SNR is likely the $\gamma$-ray emitter with compelling evidence for hadronic CR acceleration and an extension roughly consistent to the radio SNR size \citep{fermiw51c2016}. 4FGL~J1923.2+1408e represents the SNR emission as an elliptical disk with radii 0.375 and 0.26\,$\degree$, see Figure~\ref{fig:w51c_g49.2} \citep{4fgl-dr2}. The extended TeV source first discovered by HESS \citep{hess2018} and subsequently detected by MAGIC \citep{magic2012}, is identified as an interaction between the SNR and the surrounding molecular clouds (MCs), similar to the GeV emission \citep{fermiw51c2016}.

We question the TeV classification due to its much smaller extension $0.12 \,\degree$ from the SNR size observed in GeV and radio bands, as well as its compelling overlap with both of the PWN candidates G49.2--0.7 (``PWNc 1'') and G49.2--0.3 (``PWNc 2''). Furthermore, as shown in both panels of Figure~\ref{fig:w51c_g49.2}, there is significant TS $\sim 25$ residual emission coincident with both PWN candidates not accounted for by 4FGL~J1923.2+1408e or the background components. We find that 4FGL~J1923.2+1408e is required to model extended emission in the region, but that 4FGL~J1922.7+1428c is the likely counterpart to one or both PWN candidates, and is therefore removed from the global source model. In its place are two additional point sources each fixed at the PWN X-ray locations. The addition of these two sources significantly improves the fit. There is a substantial improvement when modeling residual $\gamma$-ray emission coincident with G49.2--0.3 (``PWNc 2''), which results in a point source detection TS $= 654$ and spectral index $\Gamma = 2.28 \pm 0.09$. ``PWNc 1'' or G49.2--0.7 also noticeably improves the fit resulting in a detection TS $= 70$ and spectral index $\Gamma = 2.60 \pm 0.05$. We therefore classify the two new point sources (replacing 4FGL~J1922.7+1428c) as tentative PWN detections. A deeper analysis considering multi-wavelength information is needed to determine the capacity for either PWN candidate to $\gamma$-ray emit. 

\singlespacing
\doublespacing

\smallskip
{\bf G63.7+1.1:} is the central PWN to SNR~G63.7+1.1, though the central pulsar remains unknown. X-ray observations revealed a point source embedded within a diffuse non-thermal X-ray nebula, which coincides with the bright radio core of the SNR \citep{matheson2016}. The PWN has a possible \fermi association 4FGL~J1947.7+2744. Re-analysis of the region confirms that there is point-like $\gamma$-ray emission in the vicinity of the PWN/SNR system. We consider the source detection tentative for the PWN. This is motivated by the SNR being located among dense material and may be interacting with local clouds \citep{matheson2016}. Considering the spectral properties of the $\gamma$-ray source such as its best-fit photon index $\Gamma = 2.34 \pm 0.10 $, and the age of the system \citep[$\tau \gtrsim 8\,$kyr,][]{matheson2016}, it seems equally likely for the PWN or the SNR to be the \fermi counterpart. 

\singlespacing
\begin{figure*}[htbp]
\begin{minipage}[b]{.5\linewidth}
\centering
\includegraphics[width=1.0\linewidth]{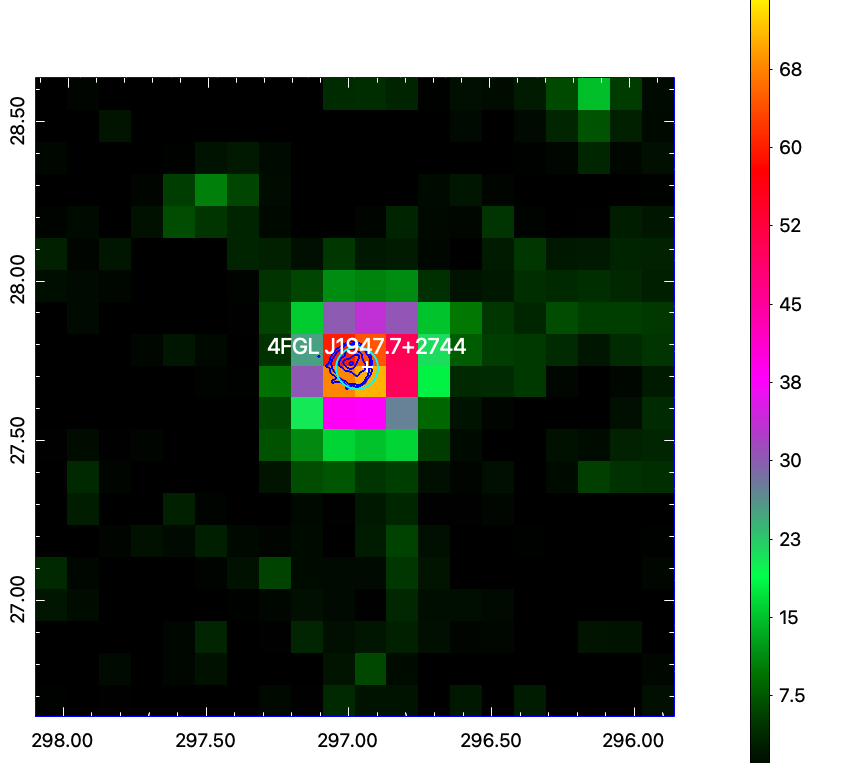}
\end{minipage}
\begin{minipage}[b]{.5\linewidth}
\centering
\includegraphics[width=1.0\linewidth]{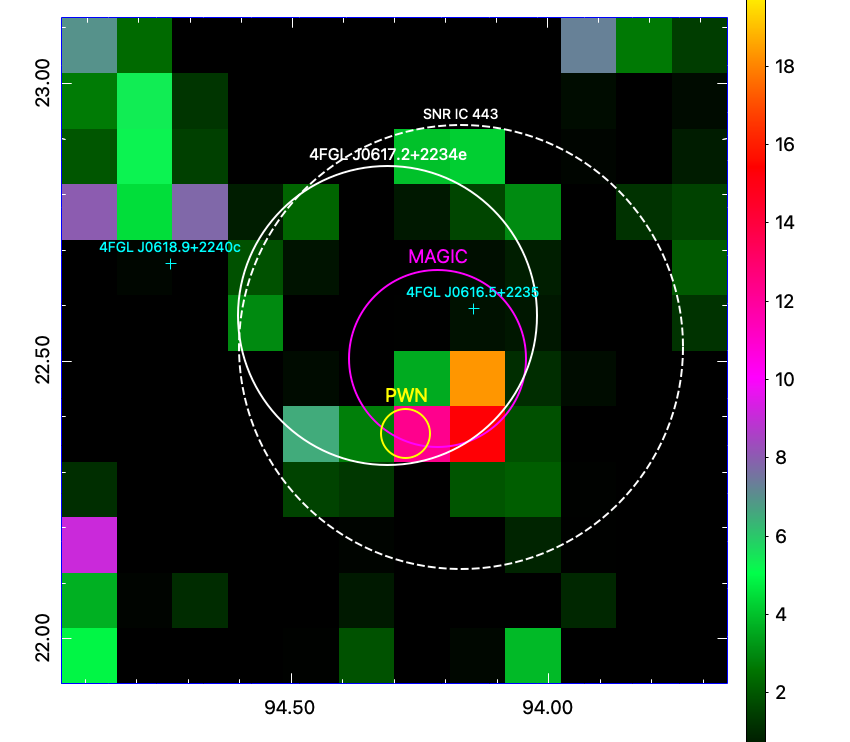}
\end{minipage}
\caption{{\it Left:} A $2\,\degree \times 2\,\degree$ 300\,MeV--2\,TeV TS map of \texttt{PSF3} events for PWN~G63.7+1.1. There is one associated \fermi source 4FGL~J1947.7+2744. Radio contours of the SNR shell and the central PWN are indicated in blue. The 95\% positional uncertainty for the point source is indicated in cyan. The maximum TS at the PWN/SNR position is $\sim$ 70. {\it Right:} A $1\,\degree \times 1\,\degree$ 10\,GeV--2\,TeV TS map of \texttt{ALL} events for PWN~G189.1+3.0, the possible PWN belonging to G189.1+3.0 (IC~443) or the overlapping SNR candidate G189.6+3.3 (not shown). The SNR IC~443 shell is both an extended GeV and TeV emitter 4FGL~J0617.2+2234e and MAGIC~J0616+225, which are displayed in solid white and magenta circles, respectively. Unrelated 4FGL sources are labeled in cyan. The X-ray position and size for the PWN is marked in yellow. The maximum TS at the PWN position is $\sim$ 16 for $E > 10$\,GeV. The size of the SNR shell observed in radio is displayed as the white dashed circle.}\label{fig:63.7_ic443}
\end{figure*} 
\doublespacing

\smallskip
{\bf G189.1+3.0 (IC~443):} The $\gamma$-ray bright SNR IC~443 is a well-known hadronic accelerator with its pion decay signature among the only ones to be observed by the \fermi  \citep{ic443w442013}. 4FGL~J0617.2+2234e corresponds to the spatially extended SNR observed by the \fermi and its TeV counterpart MAGIC~J0616+225 is also extended though only about half the size, see Figure~\ref{fig:63.7_ic443}, right panel. There is a PWN candidate G189.1+3.0 observed in X-ray that overlaps with two SNRs including IC~443 and the other being unconfirmed SNR candidate G189.6+3.3 \citep{ic4432004,ic4432018}. As shown in Figure~\ref{fig:63.7_ic443}, right panel, there is at least one source 4FGL~J0616.5+2235 in the region of the SNR that may be associated in some way to the SNR emission or perhaps the nearby PWN candidate. We test this point source by replacing it with a point source at the PWN position and find that the source is best-fit at its original 4FGL location ($\sim 0.2\,\degree$ from G189.1+3.0). Moreover, it is apparent viewing the count maps that there is a $\gamma$-ray bright point source at this location, so 4FGL~J0616.5+2235 remains in the source model. We fix all sources within $2\,\degree$ of the PWN due to the crowded region and add an additional point source to the PWN position, which yields TS $= 26$ and spectral index $\Gamma = 2.02 \pm 0.09$ for energies 300\,MeV--2\,TeV.  Extension results suggest a strong preference for an extended template, resulting in best-fit extensions $r = 1.08\,\degree$ and $\sigma = 1.24\,\degree$, respectively. 
Confirming the source detection, extent, and class is made difficult due to the crowded region, and we therefore report the point-like detection as a tentative one, noting that a deeper analysis using \fermi observations are required to understand any residual $\gamma$-ray emission coincident with the PWN candidate G189.1+3.0. 

\begin{figure*}[htb]
\begin{minipage}[b]{.5\linewidth}
\centering
\includegraphics[width=1.0\linewidth]{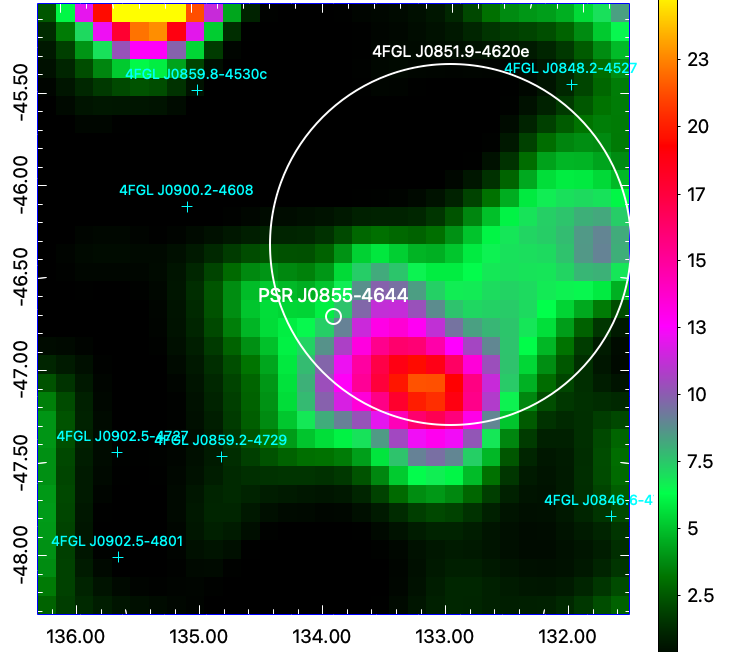}
\end{minipage}
\begin{minipage}[b]{.5\linewidth}
\centering
\includegraphics[width=1.0\linewidth]{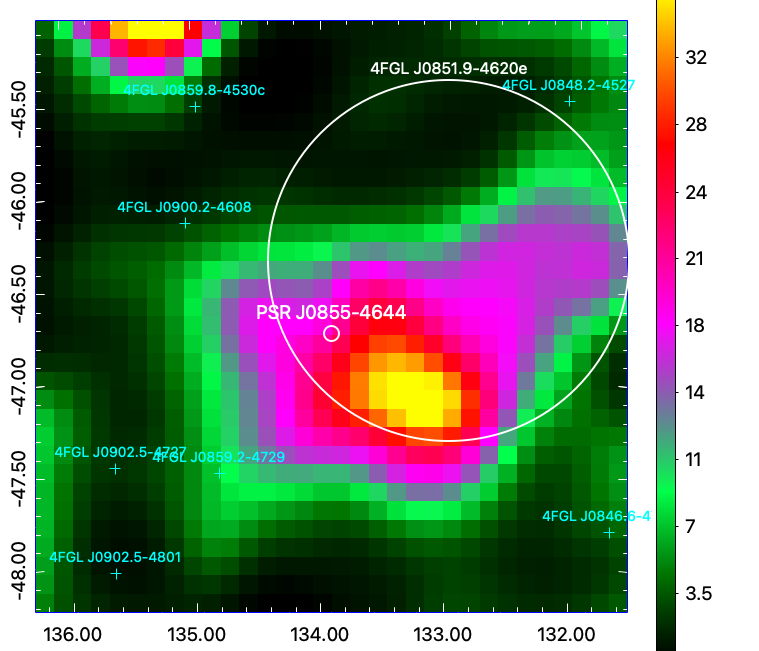}
\end{minipage}
\caption{{\it Left:} A $3\,\degree \times 3\,\degree$ 300\,MeV--1\,GeV TS map of \texttt{PSF3} events for PWN G266.9--1.10. Residual $\gamma$-ray emission is significantly detected at the PWN position that is not modeled by 4FGL~J0851.9--4620e or the backgrounds. The maximum TS for $E< 1$\,GeV occurs $\sim 0.5\,\degree$ from the PWN/PSR at TS $\sim 25$. {\it Right:} The same as the left panel but after adding a point source to the PWN/PSR position. The maximum TS for $E < 1$\,GeV occurs $\sim 0.5\,\degree$ from the PWN/PSR at TS $\sim 35$.}\label{fig:psrj0855-4644}
\end{figure*}

\smallskip
{\bf PSR~J0855--4644/G266.9--1.10:} The PWN G266.9--1.10 is powered by pulsar PSR~J0855--4644 and together they lie on the SW edge of the Vela Jr. SNR \citep{acero2013b}. Despite the overlap between locations for the PWN and Vela Jr., prior reports disfavor the Vela Jr. as the host SNR \citep{acero2013b,allen2015}. The SNR shell of Vela Jr. is a GeV emitter, 4FGL~J0851.9--4620e, and is nearly $1\,\degree$ in radius. The TeV counterpart RX~J0852.0--4622 is similarly extended to the SNR shell and GeV emission, $r = 1.0\,\degree$. There is some evidence for a PWN contribution in TS maps that include the SNR source in the background (see Figure~\ref{fig:psrj0855-4644}), but the residual emission is exclusively detected below $E < 1\,$GeV. This is not unusual for \fermi pulsars, but may also be a source of contamination such as residual SNR emission or Galactic diffuse emission structure that is not being accounted for properly by the background. Additionally, the intensely bright $\gamma$-ray Vela pulsar and its large extended host SNR are closeby, which may explain the inconclusive extension test results that imply the source is spatially extended $1\,\degree$ away from the PWN position. 

The spectral index for the tentative source is $\Gamma = 3.14 \pm 0.16$, which is indicative of a cut-off in the low-energy $\gamma$-rays, similar to what is seen in \fermi emission for pulsars \citep{2pc}. If the distance to PSR~J0855--4644 can be confirmed, it would be the second most energetic pulsar within $d \sim 1\,$kpc \citep{acero2013b}. The extended TeV counterpart RX~J0852.0--4622 to the SNR was investigated for a PWN or pulsar contribution, but disentangling TeV emission components is not feasible \citep{arribas2012}.  We consider PWN~G266.9--1.10 a tentative detection based on the $\sim 4\,\sigma$ source detection in the 300\,MeV--2\,TeV energy range, but a deeper analysis exploring both the GeV and TeV extended emission may provide better insight to a PWN contribution.

\singlespacing
\doublespacing

\begin{figure*}[htbp]
\begin{minipage}[b]{.5\linewidth}
\centering
\includegraphics[width=1.0\linewidth]{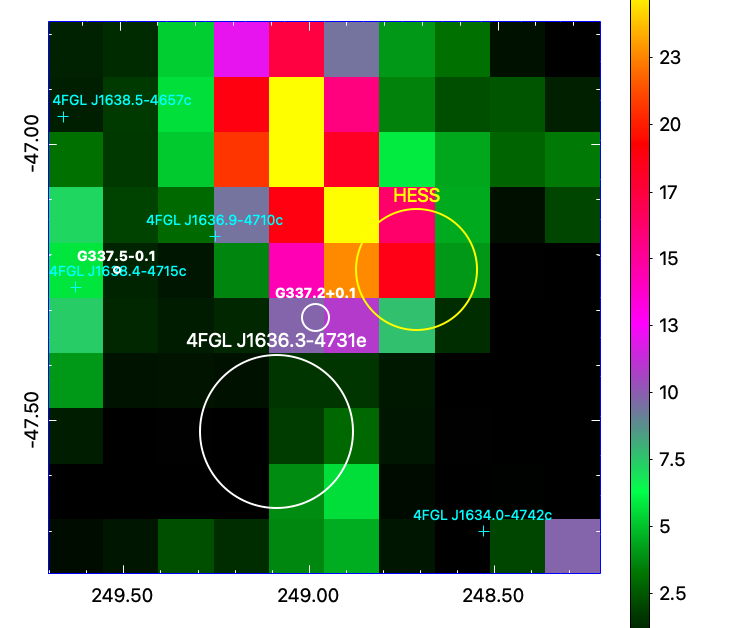}
\end{minipage}
\begin{minipage}[b]{.5\linewidth}
\centering
\includegraphics[width=1.0\linewidth]{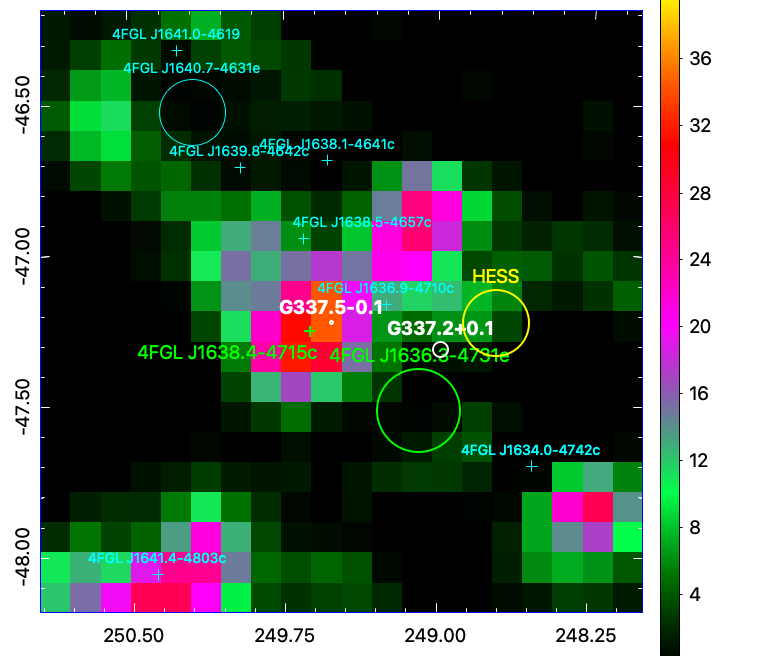}
\end{minipage}
\caption{{\it Left:} A $1\,\degree \times 1\,\degree$ 1--10\,GeV TS map of \texttt{PSF3} events for PWN G337.2+0.1. The PWN position and approximate size is centered and indicated with a white circle. 4FGL~J1636.3--4731e is the unrelated SNR~G337.1--0.1. SNR G337.5--0.1 is marked as the white circle to the far East, coincident to 4FGL~J1638.4--4715c (see right panel). The unidentified TeV source HESS~J1634--478 corresponds to the yellow circle. The maximum TS at the PWN position is TS $\sim 26$ for 1--10\,GeV. {\it Right:} A $2\,\degree \times 2\,\degree$ 1--10\,GeV TS map of \texttt{PSF3} events for G337.5--0.1, centered and indicated in white. 4FGL sources tested for association with the PWN are labeled in green, which 4FGL~J1638.4--4715c is found the likely counterpart. The unidentified TeV source HESS~J1634--478 corresponds to the yellow circle. The maximum TS at the PWN position is TS $\sim 34$ for 1--10\,GeV.
In both panels unrelated 4FGL sources are labeled in cyan. }\label{fig:g337.2_g337.5}
\end{figure*}

\smallskip
{\bf G337.2+0.1:} Another crowded region in the \fermi sky, the plerionic SNR~G337.2+0.1 may be contributing to the observed $\gamma$-ray emission. In X-ray the SNR appears Crab-like with non-thermal emission forming a central peak, presumably the pulsar and PWN, accompanied by an uncertain shell \citep{combi2006}. The diffuse X-ray emission attributed to a nebula is $\sim 1.5^\prime$ in size. An unidentified extended TeV source HESS~J1634--472 is in the vicinity and possibly associated \citep{hess2018}, see Figure~\ref{fig:g337.2_g337.5}, left panel. 

A GeV extended source 4FGL~J1636.3--4731e is also closeby, but likely unrelated as the source coincides better to  SNR~G337.0--0.1, both in size and location \citep{fges2017}. We test the possibility of association to 4FGL~J1636.3--4731e and 4FGL~J1636.9–4710c by replacing either with a point source at the location of G337.2+0.1, testing for extension of the point source, and then comparing it to that of 4FGL~J1636.3--4731e and 4FGL~J1636.9–4710c. We confirm neither of these sources improves the fit when re-located to G337.2+0.1, and instead find an additional point source to the 4FGL source model does. Adding a point source to the G337.2+0.1 position improves the global fit by $\text{TS} = 2 \log\big({\frac{L_1}{L_0}}\big) \sim 36$. The point source is detected at TS $ = 24.0$ for 2 DOF. The best-fit photon index is $\Gamma = 2.16 \pm 0.13$. There is some evidence for extension when fitting with a 2D Gaussian template, resulting in TS$_{ext}$ = 20.5 and $\sigma = 0.48\,\degree$. Given the crowded region, we consider the simplest template, the point source, as a tentative PWN detection. As can be seen in Figure~\ref{fig:g337.2_g337.5}, left panel, there is some diffuse residual $\gamma$-ray emission North of the PWN that is unlikely to be related, but is potentially being included in extension fitting tests.  

\smallskip
{\bf G337.5--0.1:} The \cha X-ray Survey unveiled a probable bow-shock PWN powered by central pulsar candidate CXOU~J163802.6--471358 \citep{jakobsen2014}. The X-ray nebula extends on the sub-arcminute scale ($\sim 40^{\prime\prime}$). The position and approximate size of the nebula is displayed in both panels of Figure~\ref{fig:g337.2_g337.5}. There is a close candidate source newly reported in the 4FGL--DR3, 4FGL~J1638.4--4715c, which we find is the likely GeV counterpart to this system and yields TS $= 55.57$ and index $\Gamma = 2.58 \pm 0.12$ when localized to the PWN, a significant improvement to the 4FGL--DR2 fit. Similar to the local but unrelated SNR~G337.2+0.1, extension tests for the new point source suggest a $\sigma = 0.50\,\degree$ 2D radial Gaussian extension can provide the best fit,  TS$_{ext} = 31$, but we cannot rule out similar contamination issues as outlined for SNR~G337.2+0.1 just above due to the complex region, see also both panels of Figure~\ref{fig:g337.2_g337.5}. We consider the $\gamma$-ray source 4FGL~J1638.4--4715c the counterpart to G337.5--0.1, but is a tentative detection due to a heavily crowded region in the Galactic plane that is among uncertain diffuse residual $\gamma$-ray emission. 

\singlespacing
\doublespacing

\subsection{Nondetections of PWNe}\label{sec:nondetect}
There are 16 PWNe and PWN candidates that have been identified at radio, X-ray, and TeV bands that are not significantly detected in the \fermi data.  All of the undetected ROIs are listed in Table~\ref{tab:nondetect} along with their measured TS and the 95\% C.L. flux upper limit for the 300\,MeV--2TeV energy range. Another source is listed in Table~\ref{tab:nondetect}: G25.1+0.02. The reported nondetection refers to the removal of a second extended PWN candidate (4FGL~J1838.0--0704e) in the region associated with TeV PWN HESS~J1837--069 (see also HESS~J1837--069 in Section~\ref{sec:new}).

\begin{figure}[htbp]
\centering
\includegraphics[width=0.85\linewidth]{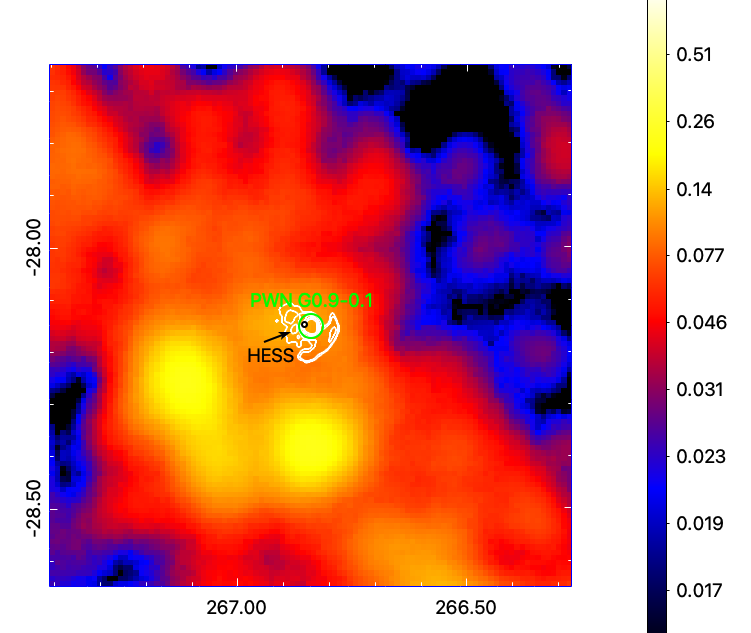}
\caption{ A $1\,\degree \times 1\,\degree$ count map for PWN~G0.9+0.1 including \texttt{PSF3} events for $E > 6\,$GeV, such that the containment angle for the image is $0.1\,\degree$ when smoothed to $\sigma = 10$ and using $0.01\,\degree$ pixel$^{-1}$. The 68\% confidence region for TeV PWN HESS~J1747--281 is marked in black. In white are the radio contours for the PWN and SNR. The green circle highlights the PWN region.}\label{fig:g0.9+0.1}
\end{figure}

PWN~G0.9+0.1 is among the undetected PWNe, even though it has a TeV counterpart HESS~J1747--281 and a particularly bright $\gamma$-ray signal in positional agreement with the radio position \citep{g0.92004}. The bright $\gamma$-ray emission is apparent in the 0.1\,$\degree$ \texttt{PSF3} event count map, see Figure~\ref{fig:g0.9+0.1}. Placing a point source to the PWN position, however, yields a TS $= 0$, indicating that the emission structure overlapping G0.9+0.1 is part of the diffuse $\gamma$-ray emission from the Galactic plane. The TS maps for G0.9+0.1 indeed show that this emission is attributed to the Galactic diffuse background model.

G23.5+0.1 is interpreted as a non-detection despite measuring $\text{TS} = 30$ at the PWN/PSR position. The maximum likelihood for the 4FGL model (e.g., no point source modeling residual emission at G32.5+0.1 location) has a noticeable statistical preference $\text{TS} = 2 \Delta\log L = 41$ over adding a point source to the 23.5+0.1 location. The PWN is located in a crowded region for \fermi $\gamma$-rays, including the $\gamma$-ray bright SNR~W41 $\sim 0.5\,\degree$ away, which prevents a reliable analysis for residual emission possibly related to G23.5+0.1. Given the 4FGL--DR2 model remains to be the best-fit for this ROI, we report G23.5+0.1 as a likely nondetection.

\begin{figure}[htbp]
\centering
\includegraphics[width=0.85\linewidth]{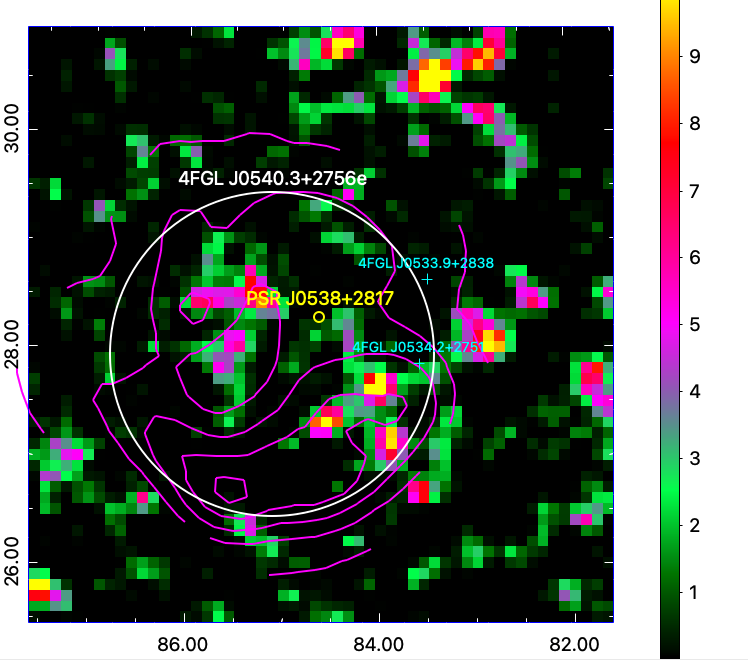}
\caption{A $5\,\degree \times 5\,\degree$ 1--10\,GeV TS map for SNR S~147. The H-$\alpha$ emission that outlines the SNR shell corresponds to the magenta contours. The H-$\alpha$ emission is used as the spatial template to characterize the GeV extended emission for 4FGL~J0540.3+2756e (solid white). PSR~J0538+2817 and its compact X-ray nebula are highlighted as the yellow circle.}\label{fig:s147}
\end{figure}

Finally, S~147 houses a pulsar and PWN that has been identified in radio and X-ray bands \citep[e.g.,][]{romani2003}. H-$\alpha$ emission outlines the SNR shell of S~147 and has radius $r \sim 2.5\,\degree$ \citep{s1472012}. Further, the SNR is a $\gamma$-ray emitter with extension that is best characterized assuming its H-$\alpha$ morphology. Several 4FGL counterparts include 4FGL~J0540.3+2756e (the SNR shell), 4FGL~J0533.9+2838, and 4FGL~J0534.2+2751. The latter two sources are plausibly associated to the SNR in some way since they coincide with pockets of residual $\gamma$-ray emission in the SNR, likely from the under-modeling of SNR emission in these areas using the H-$\alpha$ morphology as the spatial template for the extended $\gamma$-ray source, see Figure~\ref{fig:s147}. The imperfect SNR spatial template makes it difficult to search for a PWN counterpart in the \fermi data. 

Aside from the uncertainties in the ROIs just described, the regions for undetected PWNe and PWN candidates are relatively sparse, making any source detection fairly straight forward. In all cases, a point source is added to the PWN position. The TS results for each source are provided in Table~\ref{tab:nondetect}. In the Appendix (Table~\ref{tab:spectral_ul_fluxes}), we provide the 95\% C.L. upper limits on the flux for nine energy bins of each undetected ROI. 


\singlespacing
\begingroup
\begin{table*}
\centering
\begin{tabular}{cccccc}
\hline
\hline
Galactic PWN Name & 4FGL Name & R.A. &  Dec. & TS & Flux Upper Limit (MeV cm$^{-2}$ s$^{-1}$) \\
\hline
G0.9+0.10 & -- & 266.836 & --28.153  & 0.05 & $1.62 \times 10^{-7}$ \\
G23.5+0.10 & -- & 278.408 & --8.454 & 29.32$^\ddag$ & $9.30 \times 10^{-7}$ \\
G25.1+0.02 (HESS~J1837--069) & J1838.0--0704e & 279.370 & --6.960 & 2.44$^\dag$ & $3.83 \times 10^{-7}$ \\
G32.64+0.53 & -- & 282.255 & --0.024  & 11.16 & $5.13 \times 10^{-7}$ \\
G47.4--3.90 & -- & 293.182 & +10.929  & 7.85 & $1.85 \times 10^{-7}$ \\
G74.0--8.50 & -- & 312.334 & +29.018  & 0.05 & $8.09 \times 10^{-8}$ \\
G93.3+6.90c & -- & 313.058 & +55.289  & 7.43 & $1.77 \times 10^{-7}$ \\
G108.6+6.80 & -- & 336.419 & +65.602  & 3.74 & $4.48 \times 10^{-8}$ \\
G141.2+5.00 & -- & 54.290 & +61.840  & 0.00 & $4.34 \times 10^{-8}$ \\
G179.7--1.70 (S~147) & -- & 84.600 & +28.280  & 10.28 & $9.91 \times 10^{-8}$ \\
G290.0--0.93 (IGR~J11014–6103) & -- & 165.445 & --61.023  & 0.00 & $9.91 \times 10^{-8}$ \\
G310.6--1.60 & -- & 210.190 & --63.420  & 0.01 & $2.02 \times 10^{-8}$ \\
G322.5--0.10 & -- & 230.860 & --57.102  & 8.62 & $2.40 \times 10^{-7}$ \\
G341.2+0.90 & -- & 251.870 & --43.750  & 0.00 & $4.24 \times 10^{-8}$ \\
G350.2--0.80 & -- & 260.782 & --37.554  & 14.92 & $4.96 \times 10^{-7}$ \\
G358.3+0.24 & -- & 265.320 & --30.380  & 0.00 & $1.85 \times 10^{-7}$ \\
G358.6--17.2 & -- & 284.146 & --37.908  & 0.00 & $2.18 \times 10^{-8}$ \\
\hline
\hline
\end{tabular}
\caption{{\bf Sources not detected by the LAT:} Results of the maximum likelihood fits for PWNe and PWN candiates not detected by the LAT along with the ROI name (PWN Name), right ascension (R.A.) and declination in J2000 equatorial degrees, and the detection significance (TS) of a point source at the specified location. The last column provides the 95\% C.L. flux upper limit for the 300\,MeV--2\,TeV energy range. \footnotesize{$^\ddag$ This source model is interpreted as a likely nondetection (see text for details).
$^\dag$ This source is classified as a potential PWN in the 4FGL--DR2 catalog associated to TeV PWN HESS~J1837--069, but a detailed analysis of this region shows only one extended source (4FGL~J1836.5--0651e) is required to model residual emission here (see Section~\ref{sec:new} and Section~\ref{sec:nondetect} for details).}}
\label{tab:nondetect}
\end{table*}
\endgroup
\doublespacing

\singlespacing
\doublespacing

\singlespacing
\doublespacing

\section{Discussion and Conclusions}\label{sec:conclude}
We have systematically characterized the 300\,MeV--2\,TeV
emission from 11.5\,years of \fermi data in 58 ROIs containing known
PWNe and PWN candidates identified at other wavelengths, and includes six firm \fermi PWNe and 7 PWN associations. For most regions, we characterize emission within $10\,\degree$ of each PWN and then localize the candidate $\gamma$-ray PWN, starting from their positions observed in other wavelengths (Table~\ref{tab:all_rois}) and tested for extension.

From the method described above, we report 11 new PWN classifications of previously unidentified $\gamma$-ray sources, which doubles the PWN population detected by the \fermi \citep{4fgl-dr2} and are categorized based on their extension and listed in Tables~\ref{tab:extent} and \ref{tab:points} alongside relevant properties. An additional 22 previously unidentified $\gamma$-ray sources are considered PWN candidates or tentative PWN detections, mostly dependent on multiwavelength analyses that can better identify the most likely $\gamma$-ray emitter of the system. These tentative detections are outlined in the bottom panels of Tables~\ref{tab:extent} and \ref{tab:points}. We report 16 ROIs where there is no significant $\gamma$-ray emission coincident in location and are in Table~\ref{tab:nondetect}. Another undetected source is listed as G25.1+0.02, which represents 4FGL~J1838.0--0704e no longer being a required source in the source model after re-analyzing extended GeV emission corresponding to 4FGL~J1836.5--0651e for the coincident TeV PWN HESS~J1837--069.

$\sim$ 38\% of the source detections reported here are found to be extended. Only two sources are reported as extended that are not considered extended in previous \fermi catalogs: 4FGL~J1818.6--1533 and 4FGL~J1844.4--0306. 4FGL~J1818.6--1533 has TS$_{ext} = 20.60$ when modeled as a radial Gaussian with $\sigma = 0.19\,\degree$ (Table~\ref{tab:extent}). G15.4+0.10 is coincident in location with the extended emission as well as the point-like composite TeV SNR HESS~J1818--154. The TeV source is localized to within the host SNR and strongly suggests a PWN origin. The X-ray counterpart is $\sim 0.1\,\degree$ in radius. 4FGL~J1844.4--0306 has TS$_{ext} = 18.77$ when modeled as a radial Gaussian with $\sigma = 0.27\,\degree$. G29.4+0.10 is plausibly associated to 4FGL~J1844.4--0306 and its extended emission. G29.4+0.10 is an unconfirmed SNR candidate, but its radio morphology resembles an SNR ($r \sim 0.15\,\degree$). A potential TeV PWN is the unidentified HESS~J1844--030, which is also coincident to the radio position but point-like \citep{hess2018}. Furthermore, we find that the tentative detection for PWN G338.2--0.0 is point-like whereas the 4FGL--DR3 finds some extension $r \sim 0.1\,\degree$, based on the extension of its probable TeV counterpart HESS~J1640--465 \citep{mares2021}. However, our extension results for G338.2--0.0 are in line with what is reported in the 4FGL--DR3, where the 2D Gaussian yields the best-fit with $\sigma = 0.08\,\degree$ and TS$_{ext} = $ 7.65.  Finally, only one new $\gamma$-ray source classified as a PWN is reported in this catalog that was not reported in any previous \fermi catalogs which corresponds to the PWN in the LMC: B0453--685 (Figure~\ref{fig:n157b_b0453}, right panel). It is the second extragalactic $\gamma$-ray PWN to be detected at such high energies after N~157B. Four tentative $\gamma$-ray sources are first reported here that have no prior \fermi counterpart: G34.6--0.50 (W44), G189.1+3.0 (IC~443), G266.9--1.10, and G337.2+0.1.

In summary, we have verified the characterization and classification of the six \fermi PWNe that lack any detectable $\gamma$-ray pulsar, and show that all of them, with the exception of MSH 15--56, can be characterized using radial Gaussian templates that either provide comparable fits to a radial disk template or better. We note that two extended \fermi PWNe, 4FGL~J0836.5--0651e and 4FGL~J1616.2--5054e, have more than one possible PWN counterpart identified in other wavelengths (see Section~\ref{sec:new}). 
Further, we analyze seven possible PWN associations from the \fermi catalogs and find that most of them (5/7) are firm $\gamma$-ray PWNe based on their multi-wavelength properties, each discussed in Section~\ref{sec:new}. However, we argue one of the associations, 4FGL~J1838.9--0704e, a second extended source possibly modeling emission from the PWN~HESS~J1837--069, is no longer required. 4FGL~J1836.5--0651e, when modeled as a radial Gaussian extended source, can adequately model any extended residual $\gamma$-ray emission coincident to HESS~J1837--069. 

There are four extended 4FGL sources that we report as strong PWN candidates including: 4FGL~J1810.3--1925e, 4FGL~J1813.1--1737e, 4FGL~J1818.6--1533, and 4FGL~J1844.4--0306. 4FGL~J1810.3--1925e overlaps two radio PWN candidates that are plausible counterparts (Section~\ref{sec:pwnc} and Figure~\ref{fig:g11.0_g11.1_g11.2}). There are at least seven point-like 4FGL sources that we find are strong PWN candidates including: 4FGL~J1829.4--1256, 4FGL~J1828.0--1133, 4FGL~J1840.0--0411, 4FGL~J1903.8+0531, 4FGL~J1947.7+2744, 4FGL~J2016.2+3712, and 4FGL~J1640.6--4632 (DR1, now 4FGL~J1640.7--4631e in DR3). In many of the tentative cases the source classification depends on the confirmation of the lower-energy counterparts being confirmed as PWNe rather than candidates such as G11.1+0.08, G11.0--0.05, and G18.9--1.10. For the youngest systems such as G11.2--0.3 and G20.2--0.20, an additional challenge is ruling out contribution from the central pulsar and/or host SNR shell. 

The 11 new PWN classifications of previously unidentified $\gamma$-ray sources doubles the PWN population detected by the \fermi from 11 to 22 \citep{4fgl-dr2}. An additional 22 previously unidentified $\gamma$-ray sources are considered PWN candidates or tentative PWN detections. This catalog now represents the most complete representation of \fermi detected PWNe to date, particularly for systems with no $\gamma$-ray emitting pulsar. We have demonstrated that the \fermi data set is robust with a much larger PWN population present than was previously known. Nevertheless, many of the new source classifications are limited by the current capabilities of the \fermi and hence require a justification from a thorough broadband analysis. In the final chapters, we present detailed multiwavelength investigations on two of the newly classified \fermi PWNe: G327.1--1.1 (Chapter~\ref{sec:g327}) and B0453--685 (Chapter~\ref{sec:b0453}).  \clearpage

    \pdfoutput=1
\chapter{G327.1--1.1: PWN Evolution through Broadband Modeling}\label{sec:g327}
\normalsize
In this chapter, we report the detection of faint $\gamma$-ray emission coincident with the middle-aged supernova remnant SNR~G327.1--1.1. In Section \ref{sec:select} we describe the SNR~G327.1--1.1 system.  We present a multi-wavelength analysis considering the radio observations (Section~\ref{sec:radiog327}) followed by the re-analysis of archival \cha X-ray observations to spatially separate the X-ray nebula components and their X-ray spectra in Section \ref{sec:xray}. We review the \fermi data analysis in Section \ref{sec:fermi-analysis}.  Broadband modeling incorporating the spatially separated multi-wavelength data and the resulting best-fit spectral energy distribution (SED) model is described in Section \ref{sec:model}. We discuss implications of observations and modeling in Section \ref{sec:discuss} and present our final conclusions in Section \ref{sec:concludeg}.

\section{SNR~G327.1--1.1}\label{sec:select}
\begin{figure}
\centering
\includegraphics[width=0.65\linewidth]{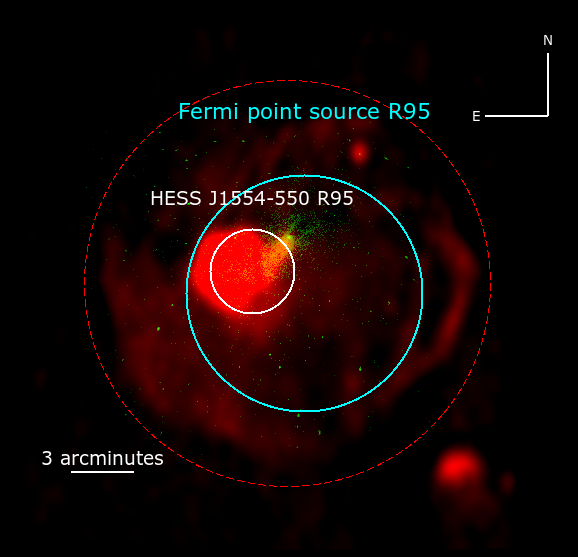}
\caption{SNR~G327.1--1.1 as observed by the Molonglo Observatory Synthesis Telescope (MOST) at 843\,MHz (red), \cha X-rays between 0.5--7\,keV (green), and the HESS TeV and \fermi best-fit positions with their 95\% uncertainty radii, $R_{95} = 0.035 \degree$ and $R_{95} = 0.099 \degree$ for the TeV and MeV--GeV positions, respectively. The faint radio SNR shell is outlined with the red-dashed circle.}\label{fig:rgb}
\end{figure}

SNR~G327.1--1.1 is a relatively old remnant with an estimated age $\tau \sim 18,000$\,yrs based on the Sedov--Taylor solution for an evolved SNR and a distance of $9$\,kpc \citep{sun1999,temim2009}. The SNR is visible as a faint radio shell at 843\,MHz accompanied by a bright, asymmetric PWN, see Fig. \ref{fig:rgb}. A similar structure is observed in the X-rays with thermal emission ($kT \sim 0.3$\,keV) present from the SNR shell and strong nonthermal emission (photon index $\Gamma_X \sim 2.1$) from the PWN \citep{bepposax2003, sun1999, temim2009, temim2015}. Radio and X-ray observations suggest the reverse shock, generated from the pressurized ISM gas swept up from the SNR forward shock, has recently interacted with the PWN asymmetrically, compressing and displacing the PWN to the East. As displayed in Figure \ref{fig:rgb}, the radio nebula is displaced from the X-ray emission while the X-ray emission has a cometary morphology, suggesting the pulsar is the point source embedded within the X-ray emission peak. \citep{temim2009} searched for X-ray pulsations within the \cha data, but the PWN outshines the pulsar, accounting for 94\% of the \cha flux for the compact region of the X-ray nebula. The pulsar is likely moving at a high velocity ($\sim 400$\,km s$^{-1}$) to the North based on its location with respect to the geometric center of the SNR. The pulsar's velocity in combination with an asymmetric reverse shock interaction from an inhomogeneous ambient ISM could explain the Southeastern displacement of the PWN \citep{temim2009, temim2015}. This scenario is consistent with the morphology, since the X-ray cometary structure is pointed Northwest and elongates to the Southeast where it meets the radio relic PWN that is displaced to the East. 


It is possible that the formation of the PWN is from the pulsar's high velocity, creating a bow shock with a comet shaped X-ray nebula.  However, the presence of the X-ray prongs ahead of the pulsar, in addition to evidence for a torus-like structure embedded inside the cometary nebula, suggests the reverse shock is responsible for the observed morphology. The impact of the reverse shock from the Northern direction could explain the observed displacement of the radio relic and the trailing compressed X-ray nebula that is freshly produced by the fast-moving pulsar \citep{temim2015}.

\citep{acero2011} reported the first detection of point-like TeV $\gamma$-ray emission in the direction of SNR~G327.1--1.1. The TeV source, HESS~J1554--550, is uniquely positioned inside the radio SNR shell and overlaps with the radio and X-ray PWN locations, with an uncertainty radius $R < 0.035$\,$\degree$ (see Figure \ref{fig:rgb}), confirming the high-energy nature of the PWN. MeV--GeV $\gamma$-ray emission from the PWN was recently detected by the \fermi \citep{xiang2021}.

\section{Radio}\label{sec:radiog327}
The SNR~G327.1--1.1 is well-known for its composite radio structure which was first revealed in the MOST radio survey \citep{whiteoak_green1996}. The SNR exhibits a faint non-thermal radio shell and a bright core of emission with a bulge extending out of the core in the northern direction sometimes referred to as the ``finger." The bulge of radio emission also outlines the X-ray cometary structure and presumed pulsar, supporting the idea that the SNR reverse shock (RS) has crushed the PWN and the pulsar has begun exiting the nebula. If this is the case, the pulsar has left its birthplace and a radio relic PWN behind while generating a freshly energized X-ray nebula in its tracks. 

More recently, \citep{ma2016} reported Australia Telescope Compact Array (ATCA) radio observations at 6 and 3 \,cm that resolves the complex structure for the PWN into two distinct components: a radio ``body'' which is the relic PWN and the radio ``head'' which is the radio ``finger'' structure that overlaps with the compact X-ray nebula \citep[see Fig.~2 and Tab.~2 of][]{ma2016}. The radio ``body'' or relic component dominates the overall radio spectrum, which is consistent with the picture that the radio relic contains the oldest particles, whereas the radio ``head'' represents a younger nebula with fainter radio emission, but brighter X-ray emission. The total radio spectrum is typical for other observed PWNe with a spectral index $\alpha = -0.3 \pm 0.1$. Additionally, the PWN is found to be highly linearly polarized with polarization fractions between 15--20\% at 6 and 3\,cm. The implied magnetic field direction is aligned along the ``finger'', with at least a partially azimuthal magnetic field structure in the relic nebula.  

\section{X-ray}\label{sec:xray}
SNR~G327.1--1.1 has been studied in detail in the X-ray band with the Einstein Observatory, ROSAT, ASCA, BeppoSAX, {\it Chandra}, and \xmm \citep[see e.g.,][]{bepposax2003, lamb1981, seward1996, sun1999, temim2009, temim2015}. In Figures~\ref{fig:rgb} and \ref{fig:xray_regions}, the PWN/SNR in X-rays as seen by \cha is displayed between energies 0.5--7\,keV. The X-ray morphology consists of the diffuse SNR thermal emission which overlaps with the fainter SNR shape in radio. A slender nonthermal component in the X-ray, which overlaps well with the radio ``finger,'' corresponds to the PWN and the presumed pulsar, which is most likely the point source embedded within the bright X-ray nebula. Two prong-like structures extend from the Northwest of the X-ray peak. It has been suggested that these structures result from an interaction between the PWN and re-heated SN ejecta, though no thermal emission has been detected \citep{temim2009,temim2015}. While the origin of the X-ray prongs remain unclear, it could alternatively pinpoint enhancements in the magnetic field \citep{temim2015}. In the next section, we extract and measure the X-ray spectra corresponding to the radio ``body'' or relic and radio ``head'' or finger components.  

\subsection{\cha X-ray Data Reduction and Analysis}
We re-analyze deep archival X-ray observations of the SNR~G327.1-1.1 performed with the Advanced CCD Imaging Spectrometer, ACIS-I, on board the Chandra X-ray Observatory for a total exposure time of 350 ks.  The standard data reduction and cleaning were performed using the Chandra Interactive Analysis of Observations \citep[CIAO v.4.12,][]{ciao2006} software package, resulting in a total clean exposure time of 337.5 ks.  An identical data reduction procedure has already been performed and reported \citep[see][for details]{temim2015}. We re-analyze the X-ray data incorporating two different source extraction regions, following the source regions as those used to measure the distinct radio ``finger'' and ``body'' PWN components from \citep{ma2016}. Figure~\ref{fig:xray_regions} shows the regions used for the spectral extraction. A spectrum for each component is extracted using the \texttt{specextract} tool in CIAO and modeled using SHERPA within CIAO \citep{sherpa2001}. We measure two X-ray source spectra corresponding to the spatially separated radio components and their radio spectra. 

\subsection{Chandra X-ray Data Analysis Results}
\begin{figure}
\centering
\hspace{-0.85cm}
\includegraphics[width=0.9\linewidth]{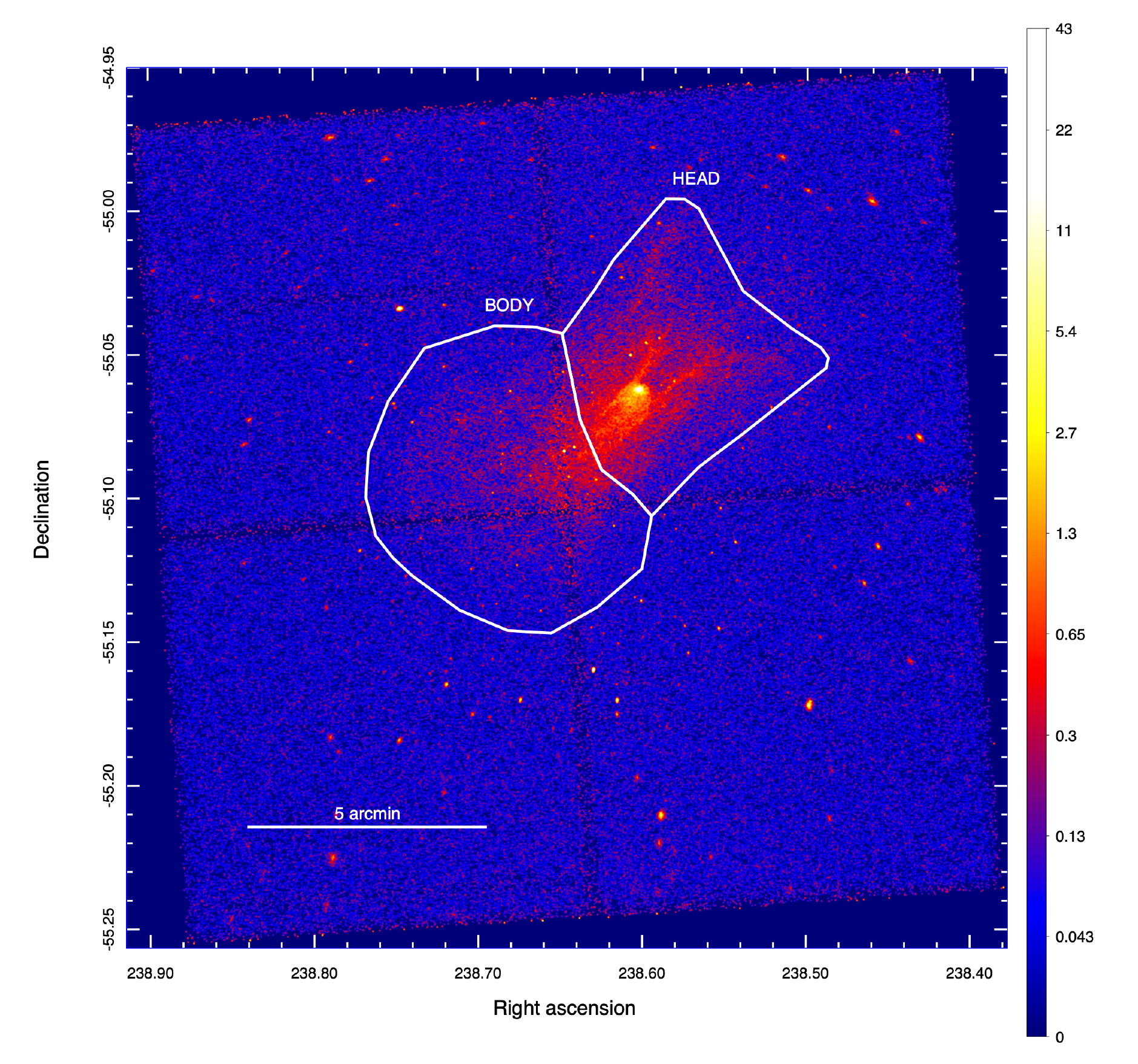}
\caption{The unsmoothed Chandra ACIS-I image of the PWN region (2.6$^\prime$ across) with the ``body'' and ``head'' regions used for the spectral extraction, which are chosen to be the same regions of the radio relic (``body'') and radio finger (``head'') spectral extraction regions used in \citep[][see their Fig.~2, bottom right panel]{ma2016}.}\label{fig:xray_regions}
\vspace{0.05cm}
\end{figure}

\begin{figure*}
\begin{minipage}{0.5\textwidth}
\hspace{-1.5cm}
\includegraphics[width=1.15\linewidth]{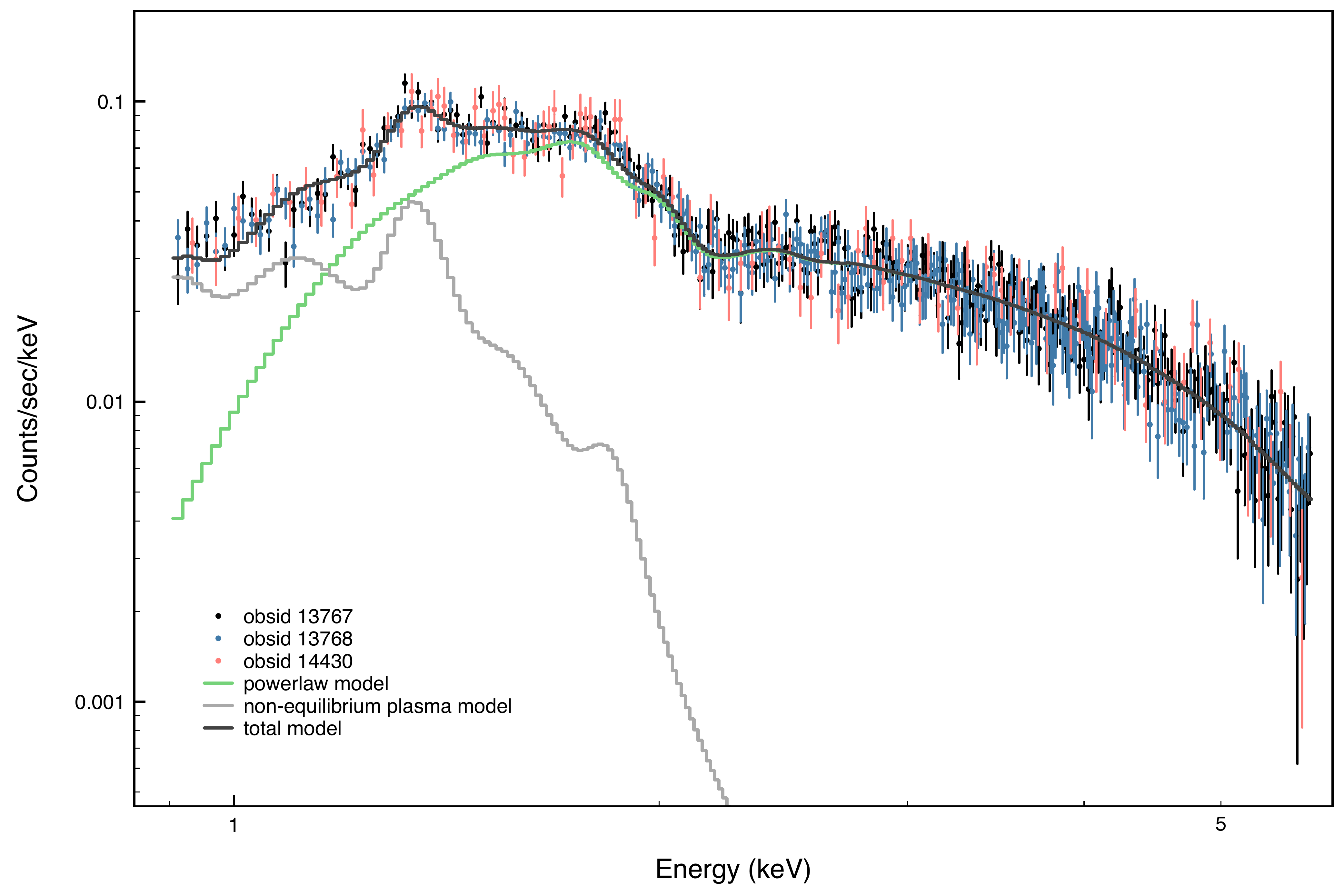}
\end{minipage}
\begin{minipage}{0.5\textwidth}
\centering
\includegraphics[width=1.15\linewidth]{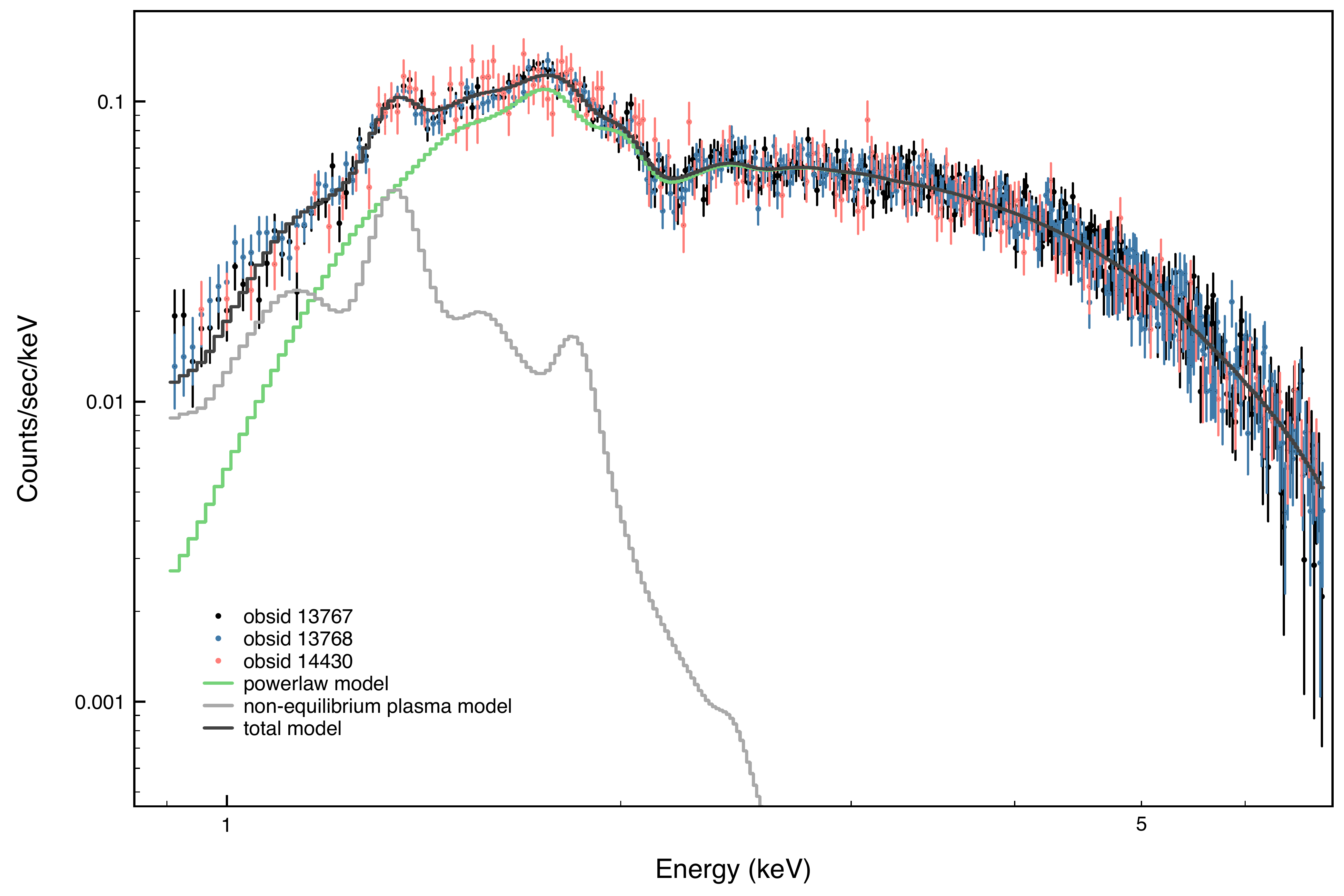}
\end{minipage}
\caption{ {\it Left:} The data and best-fit model for the ``body'' PWN component illustrated in Figure~\ref{fig:xray_regions}. {\it Right:}  The data and best-fit model for the ``head'' PWN component illustrated in Figure~\ref{fig:xray_regions}. In both panels, the nonthermal spectrum originating from the PWN is plotted in green and the thermal spectrum originating from the SNR is plotted in grey. The total X-ray spectral model is plotted in black. }\label{fig:xray_spectra}
\vspace{0.05cm}
\end{figure*}

In order to study the properties of the two distinct structures observed for the PWN, we fitted spectra from two different source regions, shown in Figure~\ref{fig:xray_regions}. Since the SNR shell covers most of the field of view, we used the ACIS blank-sky background files adapted to our observation\footnote{\url{http://cxc.harvard.edu/ciao/threads/acisbackground/}} to extract background spectra from each extraction region. We used the high-energy data from 10 to 12\,keV to compare the particle background in each ACIS-I chip in our observations to the particle background in the blank-sky data. This procedure is identical to the one performed and outlined in \citep{temim2015}. 
To fit the source spectra from each extraction region, we first subtracted the corresponding blank-sky spectrum. 
Both regions were fitted by a power-law model describing the synchrotron emission from the PWN and a non-equilibrium ionization thermal model \texttt{xsvnei} with solar abundances of \citep{wilms2000} to characterize any thermal emission from the SNR. The results for each spectral region are fully consistent with the findings of prior works \citep[e.g.,][see Table~2 for the ``diffuse'' and ``relic'' regions which roughly correspond to the regions employed here]{temim2015}. The results of the fits are listed in Table~\ref{tab:xrayparams}. The spectra and best-fit models for the regions are shown in Figure~\ref{fig:xray_spectra}.

It is apparent that the X-ray body is entering the synchrotron cut-off regime, but the X-ray spectrum observed from the ``head'' is much harder. This distinction indicates that the two particle populations are under different external conditions, such as different magnetic field strengths and structures that ultimately change the evolutionary history of each population. The average X-ray spectrum \citep{temim2015} for the entire PWN demonstrates that there are important contributions from both populations to the total X-ray spectrum. We explore the particle properties from the spatially separated radio and X-ray spectra in Section~\ref{sec:model}.

\begin{table}
\begin{minipage}[b]{0.5\linewidth}
\centering
\begin{tabular}{|c c c |}
\hline
\ Component & Parameter & Best-Fit Value \\
\hline
\ Body or relic 
 & Reduced $\chi^2$ & 1.06 \\
\hline
\ \texttt{tbabs} & $n_H (10^{22})$\,cm$^{-2}$ & 2.30$_{-0.10}^{+0.10}$ \\
\ \texttt{powlaw} (PWN) & $\Gamma$ & 2.43$_{-0.05}^{+0.04}$ \\
\ & Amplitude & 1.60$_{-0.20}^{+0.10}\times10^{-3}$\\
\hline
\ \texttt{vnei} (SNR) & $kT$(keV) & 0.22$_{-0.02}^{+0.04}$ \\
\ & $\tau$ ($10^{11}\,$cm$^{-3}$s) & 2.00$_{-1.00}^{+2.00}$ \\
\ & Normalization & 0.07$_{-0.02}^{+0.08}$ \\
\hline
\end{tabular}
\end{minipage}
\begin{minipage}[b]{0.5\linewidth}
\centering
\begin{tabular}{|c c c |}
\hline
\ Component & Parameter & Best-Fit Value \\
\hline
\ Head or finger 
 & Reduced $\chi^2$ & 1.10 \\
\hline
\ \texttt{tbabs} & $n_H (10^{22})$\,cm$^{-2}$ & 1.90$_{-0.10}^{+0.10}$ \\
\ \texttt{powlaw} (PWN) & $\Gamma$ & 2.09$_{-0.04}^{+0.04}$ \\
\ & Amplitude & 2.50$_{-0.10}^{+0.20}\times10^{-3}$\\
\hline
\ \texttt{vnei} (SNR) & $kT$(keV) & 0.27$_{-0.04}^{+0.04}$ \\
\ & $\tau$ ($10^{11}\,$cm$^{-3}$s) & 1.00$_{-0.60}^{+1.00}$ \\
\ & Normalization & 0.04$_{-0.01}^{+0.05}$ \\
\hline
\end{tabular}
\end{minipage}
\caption{Summary of the 90\% C.L. statistics and parameters for the best-fit model of each component measured in the X-ray analysis.}
\label{tab:xrayparams}
\end{table}

\section{Gamma-ray}\label{sec:fermi-analysis}
\subsection{\fermi Data Analysis}
We re-analyze the \fermi data using an optimized technique in order to maximize the source detection and to most accurately measure the MeV--GeV $\gamma$-ray spectrum. We perform a binned likelihood analysis using 11.5\,years (from 2008 August to 2020 January) of \texttt{P8R3\_SOURCE\_V3} photons with energy between 300\,MeV--2\,TeV from all events. We utilize the latest Fermitools package (v.2.0.8) and FermiPy Python 3 package \citep[v.1.0.1,][]{py3,fermipy2017} to perform data reduction and analysis. 
We organize the events by PSF type, as described in Chapter~\ref{sec:fermi_analysis}, using \texttt{evtype=4,8,16,32} to represent \texttt{PSF0, PSF1, PSF2}, and \texttt{PSF3} components. A binned likelihood analysis is performed on each event type and then combined into a global likelihood function for the ROI to represent all events.  Photons detected at zenith angles larger than 100\,$\degree$ were excluded to limit the contamination from $\gamma$-rays generated by CR interactions in the upper layers of Earth's atmosphere. The data were additionally filtered to remove time periods when the instrument was not online (e.g., when flying over the South Atlantic Anomaly). The $\gamma$-ray data are modeled with the latest comprehensive \fermi source catalog, 4FGL-DR2 \citep{4fgl-dr2}, the LAT extended source template archive for the 4FGL catalog, and the latest Galactic diffuse and isotropic diffuse templates (\texttt{gll\_iem\_v07.fits} and \texttt{iso\_P8R3\_SOURCE\_V3\_v1.txt}, respectively). 

\begin{figure*}
\begin{minipage}[b]{.5\linewidth}
\centering
\includegraphics[width=0.97\linewidth]{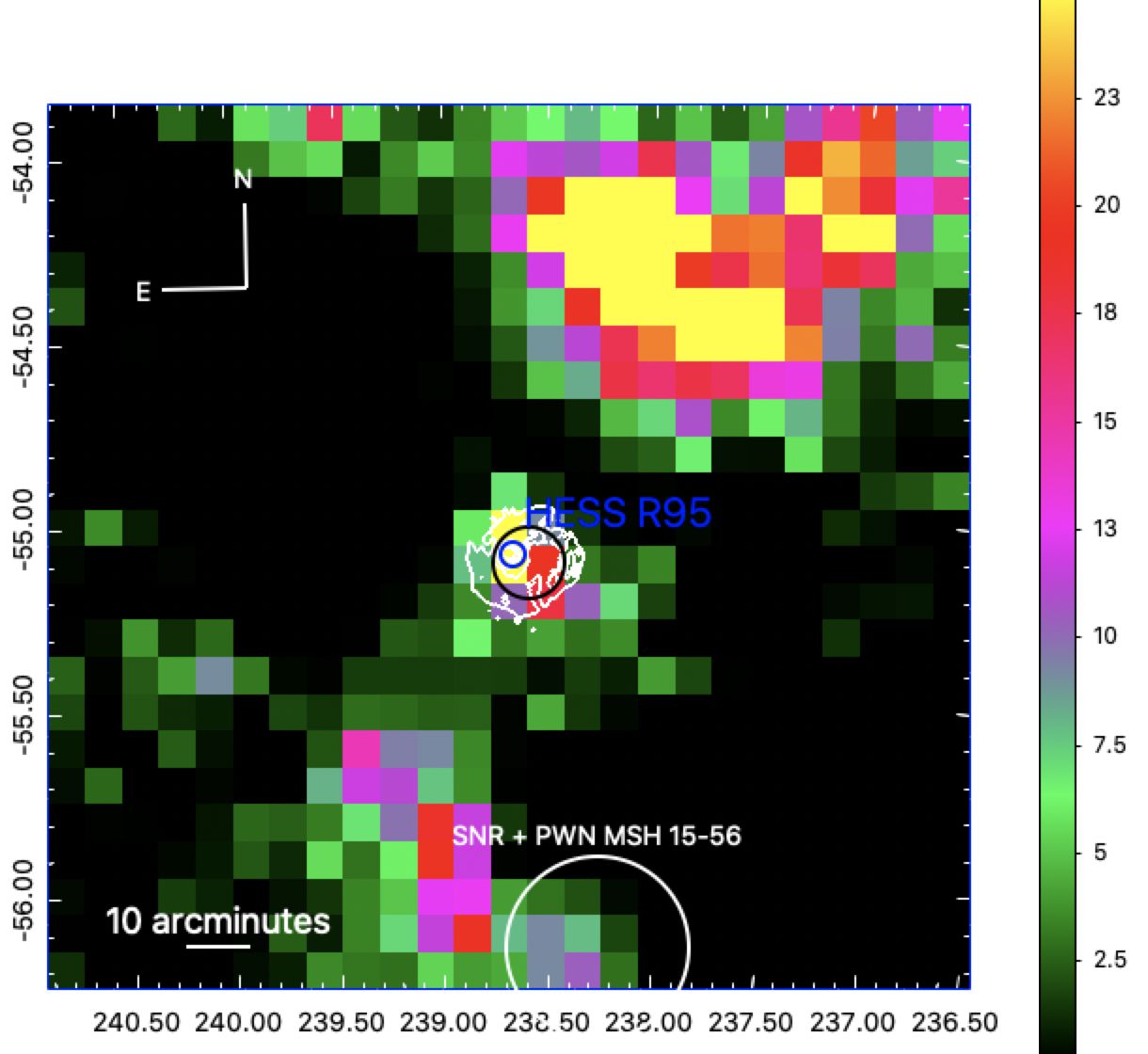}
\end{minipage}
\begin{minipage}[b]{.5\linewidth}
\hspace{-.4cm}
\centering
\includegraphics[width=1.10\linewidth]{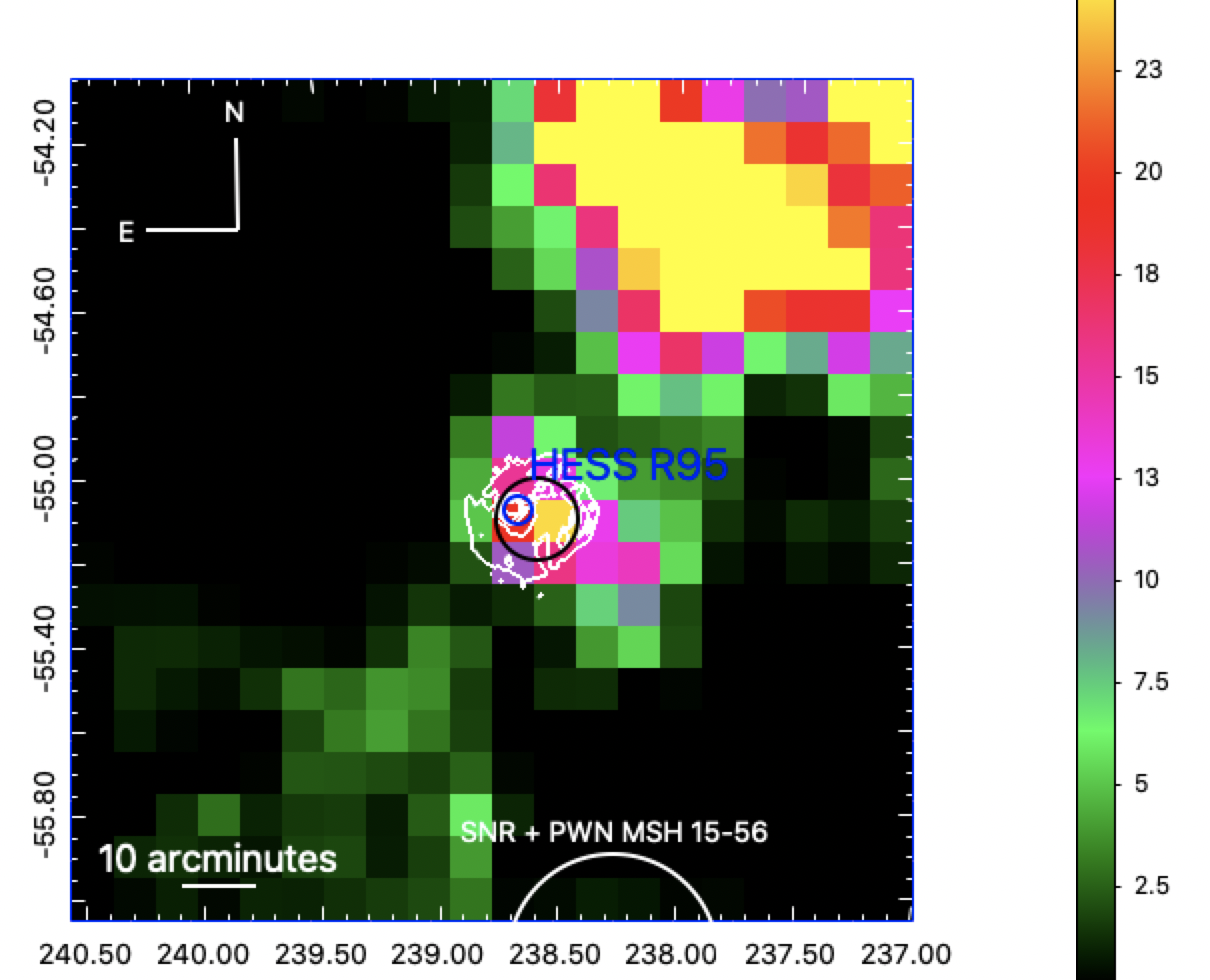}
\end{minipage}
\caption{{\it Left:} $2.5\,\degree \times 2.5\,\degree$ TS map of \texttt{ALL} events from 300\,MeV--2\,TeV. {\it Right:} $2\,\degree \times 2\,\degree$ 1--10\,GeV TS map of \texttt{ALL} events. The maximum TS at the PWN/SNR position is $\sim$26 and $\sim$24 in the 300\,MeV--2TeV and 1--10\,GeV energy ranges, respectively. The source model used only includes the background components and four bright 4FGL sources within 2\,$\degree$ of the PWN/SNR. The four bright 4FGL sources included in the source model are:  4FGL~J1553.8--5325e and 4FGL~J1550.4--5354c (associated to MSH 15-57 SNR complex), and 4FGL J1552.4-5612e and 4FGL J1552.9-5607e (the PWN and SNR of MSH 15-56, respectively). 843\,MHz (white) radio contours and HESS (blue) and \fermi (black) 95\% uncertainty regions are plotted in both panels. }\label{fig:ts_maps}
\end{figure*}

We fit the 10$\degree$ region of interest (ROI) with 4FGL point sources and extended sources that are within 15$\degree$ of the ROI center along with the diffuse Galactic and isotropic emission backgrounds. We allow the background components and sources with TS $\geq 25$ and a distance from the ROI center (chosen to be the TeV PWN position) $\leq3.0$\,$\degree$ to vary in spectral index and normalization. We computed a series of diagnostic test statistic (TS) and count maps in order to search for and understand any residual $\gamma$-ray emission in the following energy ranges: 300\,MeV--2\,TeV, 1--10\,GeV, 10--100\,GeV, and 100\,GeV--2\,TeV. The motivation for increasing energy cuts stems from the improving PSF of the \fermi instrument with increasing energies. We inspect the maps for additional sources, promptly finding a faint point-like gamma-ray source coincident with SNR~G327.1--1.1 in a relatively uncrowded region with no known 4FGL counterpart nearby\footnote{The closest 4FGL sources are more than 1$\degree$ away.}. TS maps between energies 300\,MeV--2\,TeV and 1--10\,GeV are shown in Figure \ref{fig:ts_maps}. The TS maps are generated from a source model considering only the background and four bright 4FGL sources within $2\,\degree$ of the center ROI such that remaining TS values in this map are detection measurements for $\gamma$-ray emission not attributed to background or the four nearby bright sources. As shown in Figure~\ref{fig:ts_maps}, left panel, significant residual emission coincides with the position and size of SNR~G327.1--1.1, with a maximum TS value $\sim 26$ in the 300\,MeV--2\,TeV energy range. It is apparent the residual emission is probably point-like, with little evidence for extension beyond the PSF of the \fermi.

\begingroup
\begin{table*}
\hspace{-1cm}
\scalebox{0.80}{
\begin{tabular}{cccccccc}
\hline
\hline
\ Spectral Model & $\log{L}$ & $\Gamma$ & $\alpha$ or $\Gamma_1$ & $\beta$ or $\Gamma_2$ & $N_0$ (MeV$^{-1}$ cm$^{-2}$ s$^{-1}$) & $E_b$ (GeV) or $a$ & TS \
\\
\hline
Power Law & 2143037.68 & $2.50\pm0.12$ & -- & -- & $(9.25\pm 1.79) \times10^{-13}$ &  -- & 33.76 \\
Log Parabola & 2143040.43 & -- & $2.41\pm0.20$ & $0.20\pm5.62\times10^{-5}$ & $(2.07\pm 1.28) \times10^{-13}$ & 2.00 & 38.61  \\
Power Law with Super Exp. Cut-Off & 2143039.96 & -- & $1.04\pm0.44$ & $0.67$ (fixed) & $(3.49\pm 5.62) \times10^{-12}$ & $0.01\pm 2.70 \times 10^{-4}$ & 37.56 \\
\hline
\hline
\end{tabular}}
\caption{Summary of the best-fit parameters and the associated statistics with 1-$\sigma$ statistical uncertainties for each of the \fermi point source tests.}
\label{tab:gamma_spec}
\end{table*}
\endgroup

\subsection{\fermi Data Analysis Results}\label{sec:fermi-results}
To model the $\gamma$-ray emission coincident with the PWN inside SNR~G327.1--1.1, we add a point source to the 300\,MeV--2\,TeV source model in addition to known 4FGL sources and the background emission templates for the diffuse Galactic and isotropic components. We first fix the point source with a power law spectrum,
\begin{equation}
    \frac{dN}{dE} = N_{0} \big(\frac{E}{E_0}\big)^{-\Gamma}
\end{equation}
and photon index $\Gamma=2$. We localize the point source using \texttt{GTAnalysis.localize} to find the best-fit position is at (R.A., Dec.) = (238.578$\degree$, --55.110$\degree$) (J2000), an offset of 0.071 $\degree$ from the exact PWN location, and a 95\% positional uncertainty radius $r=0.099 \degree$, which overlaps well with the radio and X-ray positions of the PWN. With the new position, we then allow the spectral index and normalization to vary. The TS for a localized point source with a power law spectrum and photon index $\Gamma=2.50\pm0.12$ is $33.76$.

\begin{figure}[!t]
\centering
\includegraphics[width=1.0\linewidth]{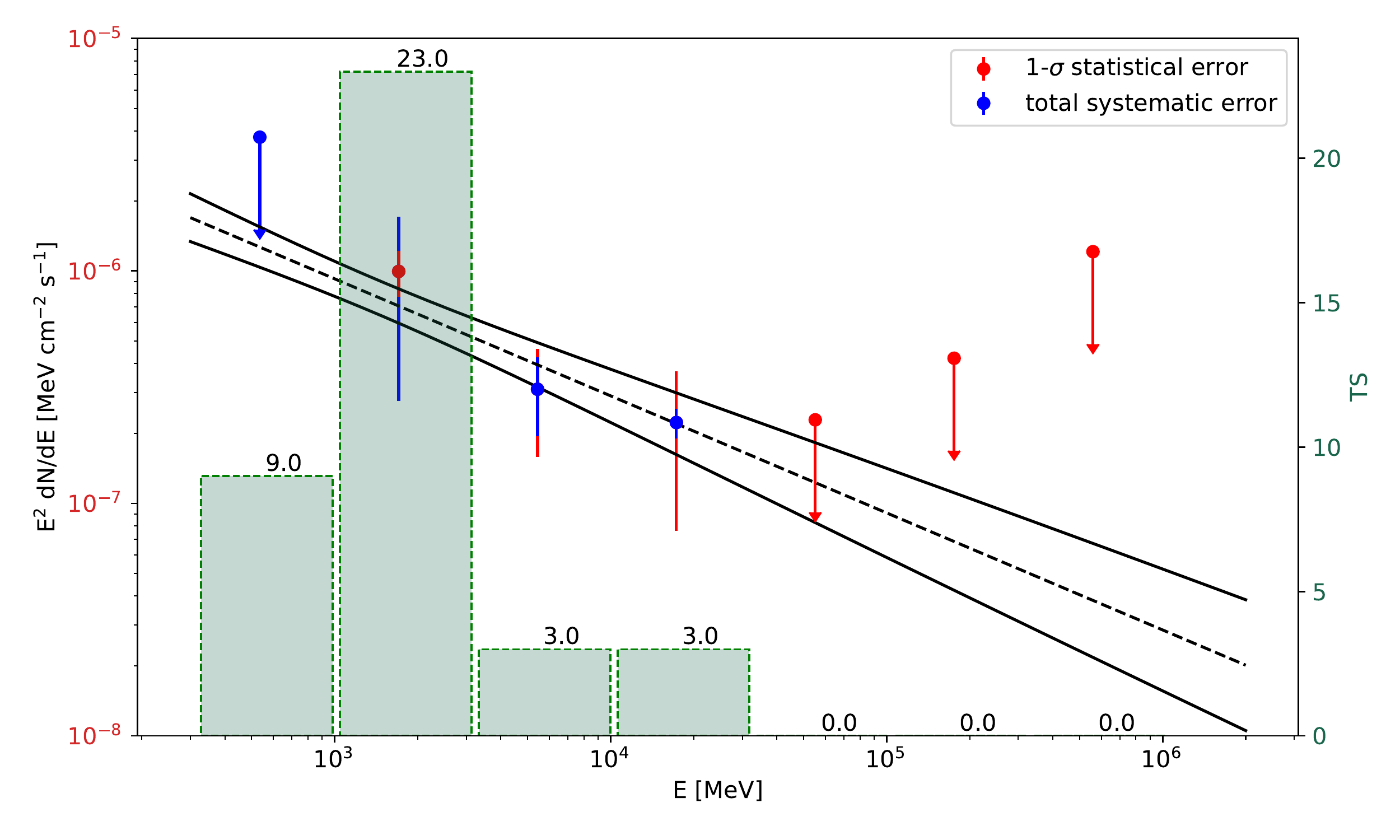}
\caption{The best-fit $\gamma$-ray SED between 300\,MeV--2\,TeV energies with 1-$\sigma$ statistical uncertainties (red) and the quadratic sum of systematic errors (blue, see Section \ref{sec:sys_err}). TS values for each spectral bin are plotted as the green histogram. The best-fit power-law spectrum is fit with a photon index $\Gamma_\gamma = 2.50\pm 0.12$.}\label{fig:gamma_sed}
\vspace{0.05cm}
\end{figure}

We investigate the spectral properties of the $\gamma$-ray emission by replacing the point source with a new point source at the same location but with a log parabola spectrum:
\begin{equation}
    \frac{dN}{dE} = N_{0} \big(\frac{E}{E_b}\big)^{-(\alpha+\beta\log{E/E_b})}
\end{equation}
We repeat the same method a third time, replacing the log parabola point source with a point source described by a power law with a super exponential cut-off (PLEC),
\begin{equation}
    \frac{dN}{dE} = N_{0} \big(\frac{E}{E_0}\big)^{-\Gamma_1} \exp{\big(-a E^{\Gamma_2}\big)}
\end{equation}
See Table \ref{tab:gamma_spec} for a summary of the best-fit parameters for each point source test. Given that all three spectral models provide similar statistical fits, we model the emission with the power-law spectrum since it is the simplest and has the fewest degrees of freedom. As shown in the $\gamma$-ray spectral energy distribution (SED) in Figure~\ref{fig:gamma_sed}, almost all of the emission is observed between 1--10\,GeV.

\begingroup
\begin{table*}
\centering
\begin{tabular}{cccccccc}
\hline
\hline
\ Spatial Template & TS & TS$_{ext}$ & R.A., Dec. (J2000) & Best-fit radius or sigma ($\degree$) & 95\% radius upper limit ($\degree$) \
\\
\hline
Point Source & 33.76 & -- & 238.58, --55.11 & -- & -- \\
Radial Disk & 35.10 & 0.77 & 238.65, --55.07 & 0.003 & 0.085 \\
Radial Gaussian & 35.10 & 0.77 & 238.65, --55.07 & 0.003 & 0.079 \\
\hline
\hline
\end{tabular}
\caption{Summary of the best-fit parameters and the associated statistics for each spatial template tested.}
\label{tab:extent1}
\end{table*}
\endgroup

We run extension tests for the best-fit point source in FermiPy utilizing \texttt{GTAnalysis.extension} and the two spatial templates supported in the FermiPy framework, the radial disk and radial Gaussian. Both of these extended templates assume a symmetric 2D shape with width parameters radius and sigma, respectively. We allow the position to vary when finding the best-fit spatial extension for both templates. The summary of the best-fit parameters for the extended templates are listed in Table \ref{tab:extent1}. The $\gamma$-ray emission shows no evidence for extension.

The origin of the emission is unlikely to be the SNR, given the distinct location inside the SNR shell in addition to a general lack of evidence for an energetic interaction between the SNR and the ISM that would produce $\gamma$-rays. The 300\,MeV--2\,TeV energy flux $F_{\gamma} = 5.12 (\pm 0.97) \times 10^{-12}$\,ergs cm$^{-2}$ s$^{-1}$ gives a $\gamma$-ray luminosity $L_\gamma = 4.94 \times 10 ^{34} \big(\frac{d}{9kpc}\big)^2$\, ergs s$^{-1}$. Taking the predicted current spin-down power $\dot{E} \sim 3.1 \times 10^{36}$\,ergs s$^{-1}$ of the pulsar from \citep{temim2015}, 
we measure the $\gamma$-ray conversion efficiency $\eta_\gamma \sim 0.0159$, which is comparable to other GeV PWNe \citep{acero2013}. However, we cannot exclude a pulsar contribution, as several $\gamma$-ray pulsars have similar values of $\eta$ \citep{2pc}. The $\gamma$-ray emission is not bright enough to perform a reliable pulsation search and we cannot rule out a pulsar contribution from the $\gamma$-ray SED alone. The results reported here are consistent with those reported in \citep{xiang2021}.

\subsection{Systematic Error}\label{sec:sys_err}
We account for systematic uncertainties introduced by the choice of the interstellar emission model (IEM) and the effective area, which mainly affect the spectrum of the faint $\gamma$-ray emission. We have followed the prescription developed in \citep{acero2016,depalma2013} and is outlined in Section~\ref{sec:sys} of Chapter~\ref{sec:fermi_analysis}. 
We find that the systematic errors are most important in the first two energy bins of the $\gamma$-ray spectrum, which effectively converts the first data point to an upper limit, but are negligible in the remainder of energy bins.   The corresponding quadratic sum of the systematic errors in the first four energy bins are plotted in Figure~\ref{fig:gamma_sed} in blue.



\section{Broadband Modeling}\label{sec:model}
A detailed broadband model was derived from a semi-analytic simulation that predicts the evolution of a PWN inside a nonradiative SNR and was reported in \citep{temim2015}. Assuming a broken power law (BPL) particle injection spectrum and given a set of input parameters, the model simultaneously predicts the magnetic field, PWN and SNR sizes, and PWN and SNR expansion velocities as a function of age. The final particle distribution is estimated from evolving the BPL injection spectrum with time taking into account the history of the evolution of the PWN and host SNR. The simulation evolves until it corresponds to the age of the SNR at which the simulated SNR radius matches that of the observed SNR size \citep[see][for details]{gelfand2009,temim2015}. The corresponding broadband spectrum from \citep{temim2015} 
suggests the remnant age is old $\tau \sim 18,000\,$yrs with distinct particle components apparent in the spectral energy distribution (SED). 

The resolved PWN morphology in radio and X-ray give a unique opportunity to study the individual properties of the particle populations through their distinct broadband spectral features, first investigated by \citep{temim2015}. In this section we attempt to model the particle populations independently in an effort to characterize each component that contributes to the observed broadband emission, incorporating the new \fermi data, and compare to the model presented in \citep{temim2015}. We use the NAIMA Python package \citep{naima}, designed to compute the non-thermal radiation from relativistic particle populations. 
To explore the possible particle distributions, we assume a power law shape with an exponential cut-off (PLEC),
\begin{equation}
f(E) = A \big(\frac{E}{E_0}\big)^{-\Gamma} 
\exp\big({-{\frac{E}{E_{break}}}\big)} \\
\end{equation}
where $A$ is in eV$^{-1}$. 
We then input similar values as those estimated in \citep{temim2015}, such as the particle index and energy break in the particle spectrum, and also consider the same three photon fields for the Inverse Compton Scattering (ICS) component. The photon fields consist of the cosmic microwave background (CMB) with energy density $\Omega_{CMB} = 0.261$\,eV cm$^{-3}$ and temperature of 2.72\,K, and two warmer photon fields from nearby starlight and heated dust, with energy densities 2.0\,$\Omega_{CMB}$ and 1.5\,$\Omega_{CMB}$ and temperatures $T = 50\,$K and $T=3.5\times10^{3}\,$K, respectively.  
We then test the combination of free PLEC parameters that can best explain the broadband spectrum. 

While a single particle distribution can generally reproduce the observed synchrotron and IC spectral shapes, it cannot reproduce the radio spectrum nor explain the MeV--GeV $\gamma$-rays adequately. It has been established that the observed morphology and estimated age of the system makes it likely that there are distinct particle components under different physical conditions as a result of the asymmetric crushing of the PWN. The sum of two leptonic populations under the same ambient photon fields thus represents the broadband spectrum of the old PWN~G327.1--1.1: one component belonging to the older, larger and displaced radio relic and the second component belonging to the compact, younger X-ray nebula.  
We additionally allow each lepton population to vary in magnetic field independently.

\begingroup
\begin{table*}
\centering
\begin{tabular}{ccccccc}
\hline
\hline
\ Particle Spectrum & $\Gamma_1$ & $\Gamma_2$ & $E_{break}$ (TeV) & $B$ ($\mu$G) \
\\
\hline
PLEC1 (low-energy pop.) & 1.61 & -- & 0.99 & 101.00 \\
PLEC2 (high-energy pop.) & -- & 2.15 & 41.69 & 10.76 \\
BPL \citep{temim2015} & 1.48 & 2.20 & 0.31 & 10.76 \\
\hline
\hline
\end{tabular}
\caption{Summary of the best-fit parameters of the broadband models in Figure~\ref{fig:broadband_sed} with those of \citep{temim2015} for comparison. }
\label{tab:broadband_params}
\end{table*}
\endgroup

\begin{figure*}
\hspace{-.4cm}
\centering
\includegraphics[width=1.0\linewidth]{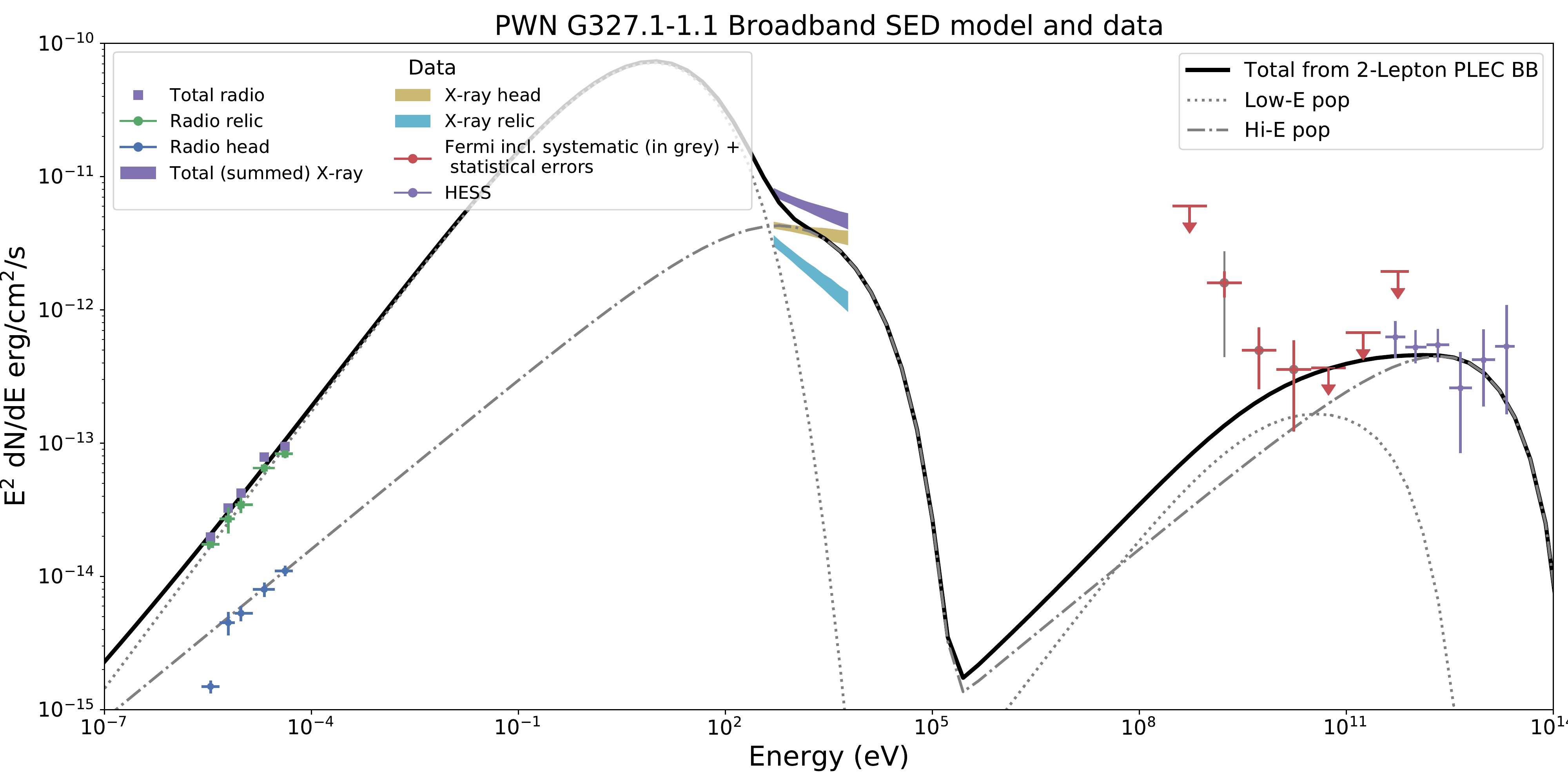}
\caption{Best-fit broadband model fits for emission from the evolved PWN in G327.1--1.1 along with radio observations of the PWN \citep[][]{ma2016}, X-ray fluxes and spectral indices (this work), new \fermi data (this work), and H.E.S.S. $\gamma$-ray emission \citep[][]{acero2011}. 
}\label{fig:broadband_sed}
\end{figure*}

Table~\ref{tab:broadband_params} lists the best-fit parameters for the two-lepton PLEC particle spectrum investigated here along with the best-fit parameters from the evolved BPL particle spectrum in \citep{temim2015} for comparison. The corresponding best-fit broadband spectrum for the two PLEC lepton populations is displayed in Figure~\ref{fig:broadband_sed}. We find the low-energy (i.e., older) population is best characterized by a particle index before the energy break $\Gamma = 1.61$. We find the high-energy (i.e., younger) population is best-fit with a particle index before the energy break $\Gamma = 2.15$. The implied magnetic field is much higher for the lower-energy population $B = 101\,\mu$G than the higher-energy population $B = 10.8\,\mu$G. 
The low-energy break is $E_{break}=0.99\,$TeV and the high-energy break is $E_{break}=41.69\,$TeV.

The low- and high-energy particle indices reported here are similar to the low- and high-energy particle indices 1.48 and 2.20 found in \citep{temim2015} which were motivated by the observed average radio and X-ray indices. Because the low-energy population clearly dominates the radio emission, the comparable low-energy index is not surprising. Similarly for the X-ray emission, as it is a combination of the two populations as shown in our Figure~\ref{fig:broadband_sed}, the similarity for the high-energy index found here and in \citep{temim2015} is expected. Moreover, the X-ray spectrum of each particle component shows two important distinctions: the oldest population is entering the synchrotron cut-off regime while the youngest population does not show strong evidence for synchrotron cut-off in the 0.5--6\,keV X-ray band. This suggests the highest-energy particles are not suffering the same synchrotron losses as the lowest-energy particles and is supported by the high magnetic field value inferred from the lower-energy population. The high-energy magnetic field component is much lower, but matches the magnetic field strength inferred from \citep{temim2015}. 

Finally, it is clear that the X-ray spectrum from the relic is underpredicted by the low-energy population in the broadband model. This could be explained by the assumption that both particle populations are characterized by a power law with an exponential cut-off distribution. The particle spectrum of an evolved PWN such as G327.1--1.1 has been shown to deviate beyond a simple power-law distribution even if the initial injection spectrum is a power-law \citep{gelfand2009}, such that if the low-energy population truly represents the relic nebula, the particle spectrum may not be best described as a power-law distribution. This limitation is not present in the \citep{temim2015} model, however, since the particle spectrum evolution is considered. The synchrotron emission peaks at a flux about an order of magnitude higher than what is predicted in the \citep{temim2015} model and is shifted to lower energy by roughly one order of magnitude. This is due to the much stronger magnetic field strength allowed combined with the low-energy break that characterizes the oldest particle population. The second most striking difference is the \citep{temim2015} model can accurately predict the MeV--GeV $\gamma$-ray emission, intersecting with all of the most constraining data points in the \fermi SED. The IC emission in the \citep{temim2015} model has a two-peak spectral shape that manifests from the evolutionary history of the particles throughout the PWN's lifetime, which accounts for the total energy output of the PWN decreasing with time once the pulsar exits the nebula. 

The divergences in the model fits, especially for the oldest particles, demonstrate the importance in considering the evolutionary history for broadband studies of PWNe. 
Despite the use of simplified radiative models to develop a new broadband model,  both models depict a physical scenario where the reverse shock has compressed the relic PWN while the exiting pulsar generates a new X-ray nebula which may have minimal to no impact from the reverse shock and subsequent compression. We discuss the physical implications of the models in the following section.

\section{Discussion}\label{sec:discuss}

We have presented a broadband model for the PWN emission in G327.1--1.1 that is characterized by the sum of two-lepton PLEC spectra and compare to the evolved particle spectrum derived in \citep{temim2015}. The best-fit model estimated from the evolved particle spectrum is based on the dynamical model for PWN evolution inside a non-radiative supernova remnant, first described in \citep{gelfand2009}. We further investigate the physical implications of the presented model by comparing the predicted observable properties from the evolutionary phases expected for PWNe from \citep{gelfand2009} with those observed from the PWN inside SNR~G327.1--1.1. 

The initial expansion stage is the first phase of PWN evolution and occurs from the PWN pressure being much greater than the pressure of the surrounding SN ejecta. 
Throughout the PWN expansion, the pulsar radiates more energy than the PWN loses from adiabatic expansion and synchrotron losses. The expansion phase introduces a rapid decline in the PWN magnetic field, which results in a steep decline in synchrotron luminosity with time. As a result, adiabatic losses will dominate over radiative losses until the reverse shock crushes the PWN, which also marks the end of the first expansion stage.  
The particle spectrum is dominated by previously injected particles at all energies for the majority of the PWN expansion. 
Similarly, the corresponding photon spectrum is dominated by previously injected particles at all energies, forming a single peak structure for both synchrotron and Inverse Compton emission. The initial expansion stage occurs within the first $\sim 1-5$\,kyrs of the PWN lifetime, but depends on the ambient ISM density and the host SNR evolution. The initial expansion phase ends when the SNR reverse shock first collides with the PWN.

The particle spectrum and consequent photon spectrum during the first contraction phase will see dramatic changes once collision with the reverse shock has begun. The synchrotron lifetime of the highest energy particles will become significantly less than the PWN age and will introduce an energy break in the particle spectrum. The most recently injected particles dominate beyond this energy break. During the PWN compression, the synchrotron luminosity will increase due to the strengthening magnetic field. Meanwhile, the IC emission depends mostly on the central pulsar output, which will peak in the GeV $\gamma$-rays if the pulsar exits the nebula during compression, leading to an increase in GeV luminosity and perhaps a modest decrease in TeV $\gamma$-ray luminosity. The corresponding photon spectrum sees the synchrotron emission peak decrease in energy due to the overall energy decreasing with time from synchrotron losses. The IC emission peak also expects to decline in the first contraction stage both in part to the central pulsar output and to the short synchrotron lifetime of the particles, such that only the highest energy particles contribute to the IC emission. %

After the PWN undergoes significant compression, the maximum particle energy and magnetic field strength will eventually begin to decrease. The corresponding photon spectrum will see another shift in the synchrotron and IC peaks such that the majority of the broadband emission occurs in the radio, soft and hard X-ray bands. If the pulsar has not yet fully exited the PWN then the injection of high-energy particles results in two distinct components visible in both the particle and photon spectra: a lower energy population composed of particles injected at earlier times and a higher energy component composed of recently injected particles. 
The corresponding photon spectrum features the double-peaked structure in both synchrotron and IC emission processes, akin to the \citep{temim2015} model for G327.1--1.1 and similar to the one presented in Section~\ref{sec:model}. At this stage, the radio emission is attributed to synchrotron radiation from the oldest particles, while the rest of the emission is dominated by recently injected particles. 
The pulsar may contribute to soft and hard X-ray luminosities, while the IC emission from recently injected particles extends to TeV energies, which increases the $\gamma$-ray luminosity from the high-energy particles injected by the central pulsar.  In the case where the pulsar does not re-enter the PWN, the pulsar still continues to inject high-energy particles into its surroundings, generating a new high-energy PWN that is separate from the relic PWN it has left behind. 

Much of the evolutionary cycle depicted here depends on the unique physical properties of the system and surroundings such as the progenitor characteristics, the ISM density, and the pulsar spin-down. The observed morphology and spectral characteristics of G327.1--1.1 suggest it to be an evolved system that has already encountered or is currently encountering the passage of the reverse shock through the SNR interior, crushing the PWN in its first contraction. The size of the SNR ($\sim 22$\,pc) at the estimated distance $\sim 9$\,kpc and the displaced PWN radio and X-ray counterparts support the scenario of an older system that has experienced an interaction with the SNR reverse shock. It is likely the reverse shock has displaced the PWN to the East and, combined with the velocity of the presumed pulsar \citep{temim2009,temim2015}, has enabled the pulsar to exit the PWN leaving behind a radio relic PWN while generating a new high-energy nebula in its wake. 
The X-ray spectrum further validates the scenario depicted here, where significant synchrotron losses are apparent, especially for the oldest particles in the relic, and likely being the direct result of the reverse shock interaction. The observed morphology also implies that the highest-energy emission, which is observed close to the putative pulsar, is near the maximum particle energy, and indicates a steepening in the particle spectrum \citep{temim2015}. 


\section{G327.1--1.1: Summary}\label{sec:concludeg}
We have reported the detection of faint $\gamma$-ray emission discovered coincident in location with the evolved SNR~G327.1--1.1. The position and extent is consistent with an origin from the high-energy PWN which is also known to emit TeV $\gamma$-rays as HESS~J1554--550. The faint $\gamma$-ray emission reported here may have a pulsar contribution, particularly for energies $E < 10$\,GeV, but cannot be ruled out by the $\gamma$-ray data alone. We perform a multi-wavelength investigation to explore the broadband models supported by observations, utilizing a one-dimensional radiative modeling package \texttt{NAIMA}. We compare the results to the in-depth semi-analytic simulation that predicts the evolved particle injection spectrum based on hydrodynamically derived quantities such as supernova explosion energy and ejecta mass and reported in \citep{temim2015}. We investigate the broadband emission structure based on our understanding of the PWN evolutionary sequence outlined in \citep{gelfand2009}. We find that the broadband model presented here and the one reported in \citep{temim2015}, although very different techniques, still provide consistent and plausible physical implications. Comparison of the two broadband models explored here also suggest that particle injection history is required for accurate $\gamma$-ray emission models of an evolved system like G327.1--1.1. The Cherenkov Telescope Array (CTA)\footnote{\url{https://www.cta-observatory.org/}} will be $\sim 100$ times more sensitive than HESS at 1\,TeV with an angular resolution that could resolve the PWN TeV structure, which could indicate how the different particle populations contribute to the lower-energy $\gamma$-ray emission.  

 \clearpage

    \pdfoutput=1
\chapter{B0453--685: PWN Evolution through Broadband Modeling}\label{sec:b0453}
\normalsize
In this chapter, we focus on the \fermi detection of faint $\gamma$-ray emission coincident to the middle-aged SNR~B0453--685, located in the Large Magellanic Cloud (LMC). In Section \ref{sec:select1} we describe the SNR~B0453--685 system.  We present a multi-wavelength analysis considering the radio observations (Section~\ref{sec:radio}) followed by an X-ray analysis using archival \cha observations in Section \ref{sec:xray1}, and the \fermi data analysis in Section \ref{sec:gamma}. We present simple broadband models investigating the $\gamma$-ray origin in Section~\ref{sec:naima_model1}. We further simulate a broadband spectral model using a semi-analytic model for PWN evolution, which incorporates known properties of the system and report the resulting best-fit spectral energy distribution (SED) in Section \ref{sec:yosi_model}. We discuss implications of observations and modeling and we outline our final conclusions in Sections \ref{sec:discuss1} and \ref{sec:conclude1}.

\begin{figure*}
\begin{minipage}[b]{1.0\textwidth}
\hspace{-1.35cm}
\includegraphics[width=0.52\linewidth]{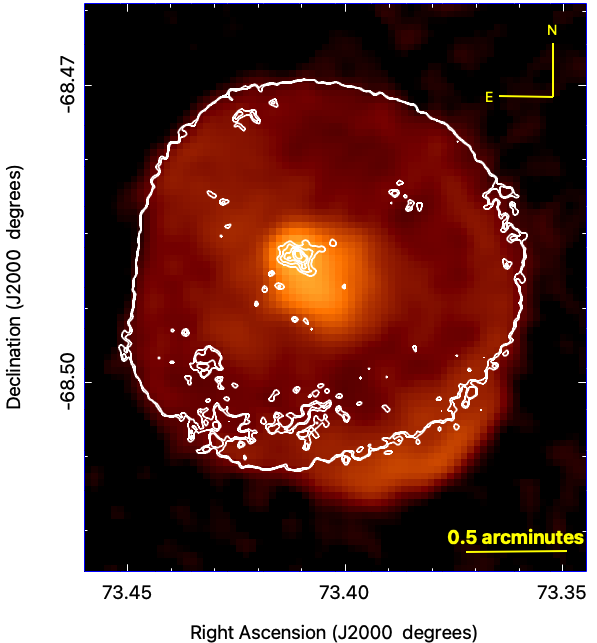}
\hspace{0.45cm}
\includegraphics[width=0.57\linewidth]{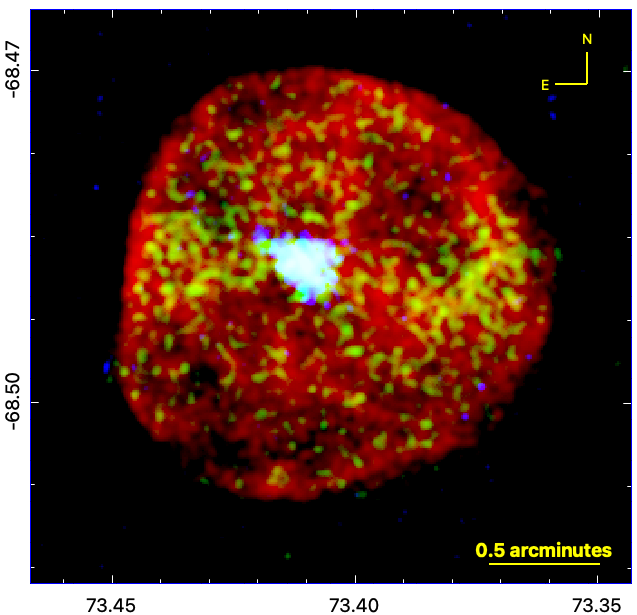}
\end{minipage}
\begin{minipage}[b]{1.0\textwidth}
\centering
\includegraphics[width=0.65\linewidth]{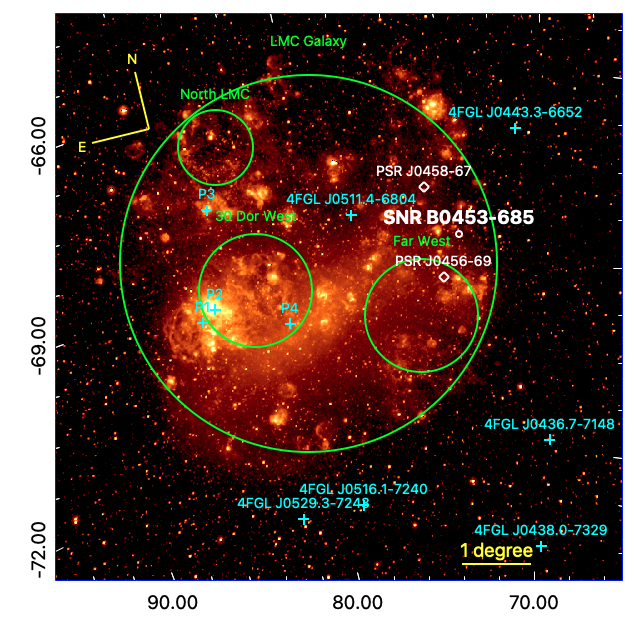}
\end{minipage}
\caption{{\it Top Left:} The 1.4\,GHz radio emission observed from SNR~B0453--685 \citep{gaensler2003}. The white contours correspond to the central PWN and the outer SNR shell as observed in X-ray (next panel). {\it Top Right:} Tri-color flux map. Red is soft X-ray emission between 0.5--1.2\,keV, green is medium flux between 1.2--2\,keV, and blue is hard flux from 2--7\,keV. Soft and medium X-ray emission outlines and fills the entire SNR while the hard X-ray emission is heavily concentrated towards the center of the SNR where the PWN is located. {\it Bottom:} The LMC in the H--$\alpha$ band from the Southern H-Alpha Sky Survey Atlas \citep[SHASSA,][]{halpha2001}. The P1--P4 labels identify the four brightest 4FGL point sources in the LMC, following the naming convention used in \citep{lmc2016}. P1 is the most energetic pulsar ever detected, PSR~J0540-6919, which lies $< 0.5\,\degree$ from P2. P2 is the possible \fermi PWN N~157B. P3 is a high mass binary (HMB) system and P4 is the SNR N~132D located near the 30 Doradus region. The four extended templates used to describe the diffuse $\gamma$-ray emission from the LMC \citep[components E1--E4 in][]{lmc2016} are indicated as the green circles. The location of SNR~B0453--685 is marked in white with radius $r=0.05\degree$. The two closest known radio pulsars near SNR~B0453--685 are labeled as white diamonds. Both are located too far from the SNR to be a reasonable central pulsar candidate. \footnotesize{We used the ATNF radio pulsar catalog \url{https://www.atnf.csiro.au/research/pulsar/psrcat/} \citep{manchester2005}.}}\label{fig:radio_xray_halpha}
\end{figure*}

\section{SNR~B0453--685}\label{sec:select1}
\normalsize
SNR~B0453--685 is located in the LMC at a distance $d\approx50$\,kpc \citep{clementini2003}. The LMC has an angular size of nearly 6 degrees in the sky where SNR~B0453--685 (angular size $<$ 0.05\,$\degree$) is positioned on the western wall of H--$\alpha$ emission as shown in Figure~\ref{fig:radio_xray_halpha}, bottom panel. SNR~B0453-685 was identified as a middle-aged ($\tau \sim 13\,$kyr) composite SNR hosting a bright, polarized central core in \citep{gaensler2003} at 1.4 and 2.4\,GHz frequencies and in X-rays between 0.3--8.0\,keV, see the top left and right panels of Figure~\ref{fig:radio_xray_halpha}. A thin, faint SNR shell is visible in both radio and X-ray (0.3--2.0\,keV) as softer, diffuse X-ray emission filling the SNR. Within the radio SNR shell is a much brighter, large, and polarized, central core: the PWN. The PWN also dominates the hard X-ray emission (2.0--8.0\,keV, Figure~\ref{fig:radio_xray_halpha}). While the new radio and X-ray observations reported by \citep{gaensler2003} confirmed the composite morphology of the SNR, no pulsations from a central pulsar have been detected. 

\citep{manchester2006} performed a deep radio pulsar search in both of the Magellanic Clouds using the Parkes 64--m radio telescope and reported 14 total pulsars, 11 of which were located within the LMC, but none were associated to SNR~B0453--685. It is reported in later investigations \citep[e.g.,][]{lopez2011,mcentaffer2012} using the same \cha X-ray observations as those in \citep{gaensler2003} that an X-ray point source is detected inside the central PWN core using the \texttt{wavdetect} tool in the \cha data reduction software package, CIAO. This remains the most promising evidence for the central pulsar. 

Displayed in Figure \ref{fig:radio_xray_halpha}, bottom panel, are the few known sources within the LMC that emit $\gamma$-rays in the \fermi band, labeled P1--P4 following the convention from \citep{lmc2016}. Only one LMC PWN, N~157B (P2), is identified as a GeV \citep{4fgl-dr2} and TeV \citep{n157b2012} $\gamma$-ray source and is located on the opposite (Eastern) wall of the LMC with respect to SNR~B0453--685. N~157B is located in a very crowded area, accompanied by two bright $\gamma$-ray sources nearby, SNR~N132~D and PSR~J0540--6919. SNR~B0453--685, however, is conveniently located in a much less crowded region of the LMC, making its faint point-like $\gamma$-ray emission detectable even against the diffuse LMC background, diffuse Galactic background, and the isotropic background emissions.


\section{Radio}\label{sec:radio}
\normalsize
Australia Telescope Compact Array (ATCA) observations at 1.4 and 2.4\,GHz were performed on SNR~B0453--685, revealing the composite nature of the SNR and confirmation for the presence of a PWN \citep{gaensler2003}. The PWN is visible as a bright central core that is surrounded by the SNR shell roughly 2$^{\prime}$ in diameter. \citep{gaensler2003} measures the flux density of the radio core to be 46$ \pm 2$\,mJy at both 1.4 and 2.4\,GHz. The PWN radio spectrum is flat, $\alpha = -0.10\pm0.05$ \citep{gaensler2003}. No central point source such as a pulsar is seen, but authors place an upper limit on a point source of 3\,mJy at 1.4\,GHz and 0.4\,mJy at 2.4GHz at the location of the emission peak and suggest the PWN to be powered by a Vela-like pulsar that has a spin period of $P \approx 100$\,ms, a surface magnetic field $B \approx 3\times10^{12}$\,Gauss, and a spin-down luminosity $\dot{E} \approx 10^{37}$\, ergs s$^{-1}$. 

\citep{haberl2012} observed SNR~B0453--685 with ATCA at 8640 and 4800\,MHz, providing radio flux density measurements of both the SNR and PWN. The authors measure a flat radio spectrum for the PWN, $\alpha_{pwn} =- 0.04 \pm 0.04$, along with significant polarization from the PWN core at 1.4\,GHz, 2.4GHz, 4800\,MHz, and 8640\,MHz frequencies. The outer SNR shell, excluding the PWN contribution, has a radio spectrum characterized as $\alpha_{shell} = -0.43 \pm 0.01$, which is a typical value for radio SNR shells.

\begin{figure}
\centering
\includegraphics[width=0.9\linewidth]{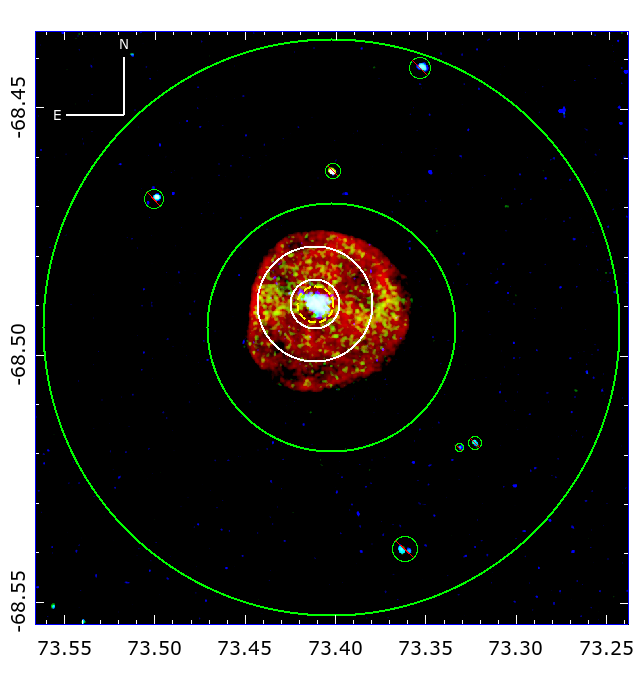}
\caption{Tri-color X-ray flux map. Red = 0.5--1.2\,keV, green is 1.2--2\,keV and blue is 2--7\,keV. The source and background regions used for spectral analysis are indicated. The yellow dashed circle corresponds to the PWN region, the white annulus corresponds to the SNR region, and the large green annulus excluding six bright X-ray point sources corresponds to the background region.}\label{fig:chandra}
\end{figure} 

\section{X-ray}\label{sec:xray1}
\subsection{Chandra X-ray Data Reduction and Analysis}\label{sec:chamethod1}
\normalsize
SNR~B0453--685 has been analyzed in X-rays in great detail \citep{gaensler2003,haberl2012, lopez2009, mcentaffer2012} considering data from \xmm and \cha X-ray telescopes. Thermal X-ray emission dominates the soft X-rays and is largely attributed to the SNR while the hard X-ray emission is concentrated towards the center of the remnant where the PWN is located (see Figure \ref{fig:chandra}). In order to understand the $\gamma$-ray origin, we must combine the new \fermi data with available multi-wavelength data for the region. Therefore, we re-analyzed archival Chandra X-ray observations (ObsID: 1990) taken with the Advanced CCD Imaging Spectrometer (ACIS) on board the Chandra X-ray Observatory. The observation exposure is 40\,ks and was completed on December 18, 2001. The entire SNR is imaged on one back-illuminated chip (called ``S3'', see Figure \ref{fig:chandra}, left panel). Back illuminated chips are more sensitive to soft X-ray emission which is ample within the SNR. Data reprocessing was conducted using the standard processing procedures in the Chandra Interactive Analysis of Observations \citep[CIAO v.4.12,][]{ciao2006} software package. The cleaned spectra are then extracted and background-subtracted using one large annulus-shaped region surrounding the remnant (see Figure \ref{fig:chandra}, right panel). We model both SNR and PWN emission components using data extracted from the regions indicated in Figure \ref{fig:chandra}, right panel, and perform a spectral analysis. A spectrum for each component is extracted using the \texttt{specextract} tool in CIAO and modeled using SHERPA within CIAO \citep{sherpa2001}.

\subsection{Chandra X-ray Data Analysis Results}\label{sec:cha-results1}
\begin{figure*}
\begin{minipage}[b]{.5\textwidth}
\hspace{-1.15cm}
\includegraphics[width=1.0\linewidth]{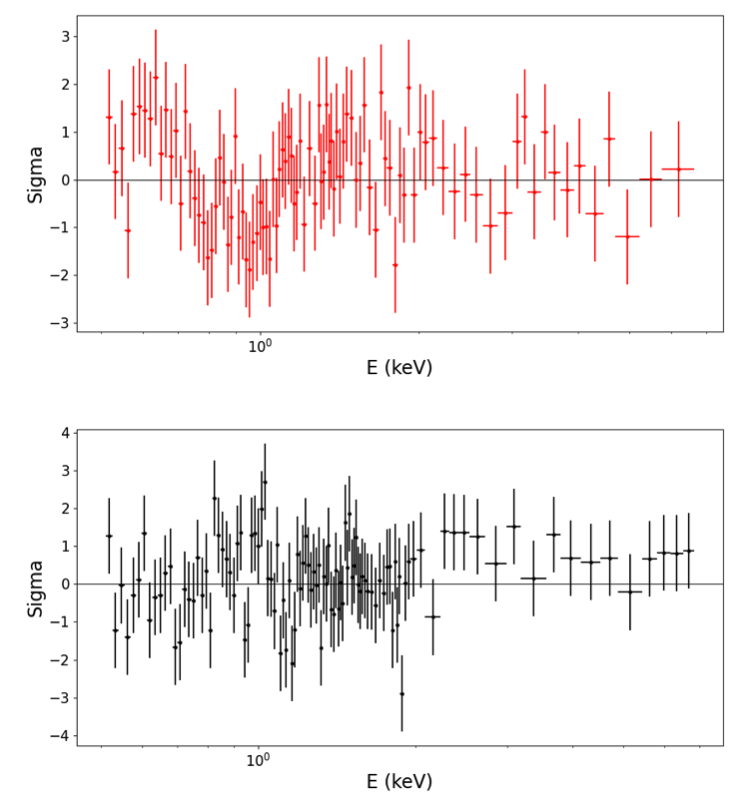}
\end{minipage}
\begin{minipage}[b]{.5\textwidth}
\hspace{-1cm}
\includegraphics[width=1.3\linewidth]{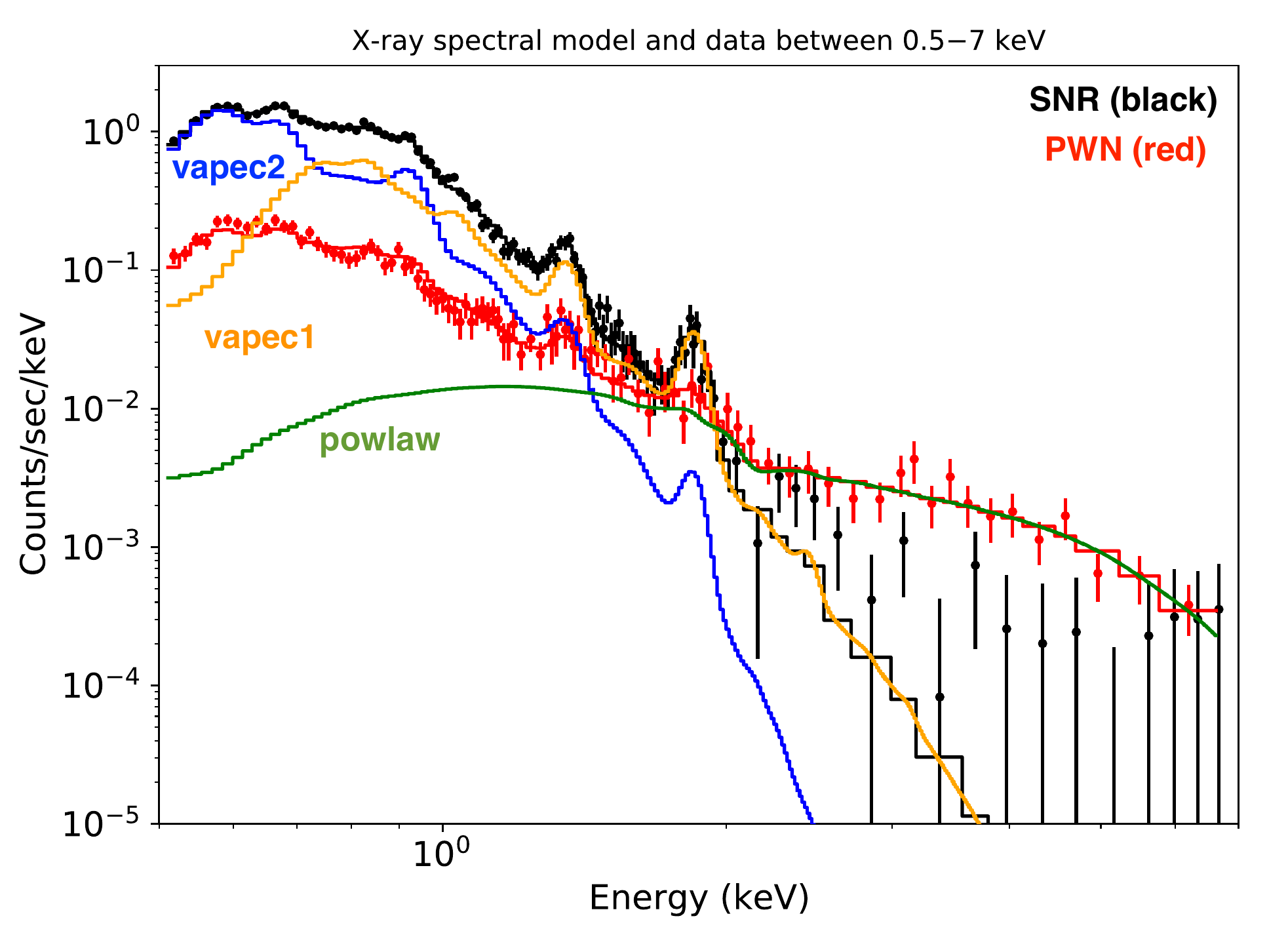}
\end{minipage}
\caption{{\it Left:} The residuals of the difference in the best-fit model and data for the PWN spectral fit (top) and the SNR spectral fit (bottom) in units of $\sigma$. {\it Right:} 0.5--7\,keV X-ray data and best-fit models for the two source models. The green solid line represents the non-thermal component from the PWN and the solid orange and blue lines represent the first and second thermal components of the SNR spectrum, respectively.}\label{fig:xray}
\end{figure*}

Data between 0.5--7\,keV are used to model observed emission and is binned to at least 20 counts per bin. We fit the two source regions for the SNR and PWN components simultaneously and the best-fit model is displayed in Figure \ref{fig:xray}, right panel. A two-component collisionally ionized plasma model (\texttt{xsvapec}) is found to best describe the emission from the SNR and one nonthermal \texttt{powlaw1d} model is preferred for the PWN component \citep[similar to prior works, e.g.][]{haberl2012, mcentaffer2012}. We account for interstellar absorption along the line of sight by including the \texttt{tbabs} hydrogen column density parameter which uses the abundances estimated from \citep{wilms2000}. The best-fit parameters are listed in Table \ref{tab:xrayparams1} and are calculated for 90\% confidence intervals using the \texttt{conf} tool in Sherpa. 

The initial values of elemental abundances are set to those estimated for the LMC in \citep{lmc1992} and are allowed to vary one by one in each fit iteration. We keep the element free if it significantly improves the fit, otherwise the value remains frozen at the following abundances relative to solar: He 0.89, C 0.26, N 0.16, O 0.32, Ne 0.42, Mg 0.74, Si 1.7, S 0.27, Ar 0.49, Ca 0.33, Fe 0.50, and Ni 0.62. Aluminum is not well constrained \citep[see Section 4.3 in][]{lmc1992} so we freeze its value to 0.33.  

\begin{table*}[t!]
\centering
\begin{tabular}{|c c c|}
\hline
\ Data points & d.o.f.$^a$ & Reduced $\chi^2$ \\
\hline 
204 & 192 & 1.03 \\
\hline
\hline
\ Component & Model & \\
\hline
\ SNR & {\texttt{tbabs$\times$(vapec$_1$+vapec$_2$)}} &  \\
\ PWN & {\texttt{tbabs$\times$(c$_1 \times$(vapec$_1$+vapec$_2$) + powlaw)}} &  \\
\hline
\end{tabular}
\caption{Summary of the statistics and best-fit model for the SNR and PWN components in the X-ray analysis. The thermal component of the PWN spectrum is linked to the SNR model with the free coefficient $c_1$. \footnotesize{$^a$ Degrees of freedom}}
\label{tab:xraymodel}
\end{table*}

The PWN spectrum consists of two thermal components from the SNR emission and a nonthermal component described best as a power law. Because SNR emission contaminates the PWN emission, we link the thermal parameters of the two models using the \texttt{const1d} parameter in Sherpa. We leave the amplitude, $C_0$, free to vary in the fit.

\begin{table*}
\centering
\begin{tabular}{|c c c |}
\hline
\ SNR & & \\
\ Component & Parameter & Best-Fit Value \\
\hline
\ {\texttt{tbabs}}$^a$ & N$_{H}$(10$^{22}$ cm$^{-2}$) & 0.37$_{-0.09}^{+0.11}$ \\
\hline
\ \texttt{vapec$_1$} & $kT$(keV) & 0.34$_{-0.05}^{+0.02}$ \\
\ & Normalization & 3.67$_{-0.97}^{+2.55}\times10^{-3}$ \\
\ \texttt{vapec$_2$} & $kT$(keV) & 0.16$_{-0.01}^{+0.01}$ \\
\ & O &  0.35$_{-0.11}^{+0.26}$ \\
\ & Ne & 0.39$_{-0.13}^{+0.32}$ \\
\ & Mg & 0.56$_{-0.33}^{+0.50}$ \\
\ & Fe & $<$ 0.70 \\
\ & Normalization & 0.05$_{-0.03}^{+0.06}$ \\
\hline
\hline
\ PWN & & \\
\ Component & Parameter & Best-Fit Value \\
\hline
\ $c_1$ & $C_0$ & 13.70$_{-0.61}^{+0.60}$ \\
\ \texttt{powlaw} & $\Gamma$ & 1.74$_{-0.20}^{+0.20}$ \\
\ & Amplitude & 5.28$_{-1.01}^{+1.18}\times10^{-5}$\\
\hline
\end{tabular}
\caption{Summary of the 90\% C.L. statistics and parameters for the best-fit model for each component in the X-ray analysis. Metal abundances are reported in solar units. \footnotesize{$^a$ Absorption cross section set to \citep{vern1996}.}}
\label{tab:xrayparams1}
\end{table*}

The hydrogen column density is $N_H = 3.7^{+1.1}_{-0.9} \times 10^{21}$\,cm$^{-2}$, the PWN power law index is $\Gamma_X = 1.74^{+0.20}_{-0.20}$, and the unabsorbed X-ray flux of the PWN component between 0.5--7\,keV is $f_x = 2.68\times10^{-13}$\,erg cm$^{-2}$ s$^{-1}$. The $N_H$ value is consistent to what is measured in the direction of the LMC\footnote{Using the \texttt{nh} tool from the HEASARC FTOOLS package \url{http://heasarc.gsfc.nasa.gov/ftools}.}, $N_H = 2.2\times10^{21}$\,cm$^{-2}$ \citep{ftools1995}. The best-fit model is consistent to other X-ray analyses \citep{haberl2012, mcentaffer2012}, with the largest differences being the elemental abundances, which can be explained by the use of \citep{wilms2000} and the \citep{vern1996} cross sections in this work, in addition to slight differences in choice of model components for the thermal emission and detector capabilities. In particular, \citep{haberl2012} analyzed \xmm observations of the entire SNR, but the PWN is not resolved and thus only one global spectrum was used to characterize any SNR and PWN emission. The SNR is much brighter than the PWN in the X-rays so the nonthermal component from the PWN in the \xmm X-ray spectrum is not well constrained. \citep{mcentaffer2012} used \citep{anders} abundances and \citep{bal1992} cross-sections, and instead of two thermal equilibrium models, \texttt{vapec}, the best-fit model assumes a two-component structure from a \texttt{vapec+vnei} combination, where the \texttt{vnei} models the second thermal component without equilibrium conditions. 

The best-fit temperatures for the two-component thermal model used to describe SNR emission is $kT = 0.34^{+0.02}_{-0.05}$\,keV and $kT = 0.16^{+0.01}_{-0.01}$\,keV, similar to what is reported in \citep{mcentaffer2012}. 
The PWN spectrum is non-thermal and best-fit as a power law and photon index, $\Gamma_X = 1.74^{+0.20}_{-0.20}$. The PWN's spectral index is slightly harder than what is reported in \citep{mcentaffer2012}, where an index $\Gamma_X \sim 2$ across the PWN region is measured, but is still in agreement within the 90\% C.L. uncertainties. 
No synchrotron component is attributed to the SNR, but we estimate the 0.5--7\,keV 90\% C.L. upper limit of the flux for a nonthermal component to the SNR spectrum to be $F_X < 5.49 \times 10^{-13}$\,erg cm$^{-2}$ s$^{-1}$.


\section{Gamma-ray}\label{sec:gamma}
\subsection{\fermi Data Analysis}\label{sec:fermethod1}
We perform a binned likelihood analysis using 11.5\,years (from 2008 August to 2020 January) of \texttt{P8R3\_SOURCE\_V3} photons with energy between 300\,MeV-–2\,TeV from all events. We utilize the latest Fermitools package (v.2.0.8) and FermiPy Python~3 package \citep[v.1.0.1,][]{fermipy2017} to perform data reduction and analysis. 
We organize the events by PSF type, as described in Chapter~\ref{sec:fermi_analysis}, using \texttt{evtype=4,8,16,32} to represent \texttt{PSF0, PSF1, PSF2}, and \texttt{PSF3} components. A binned likelihood analysis is performed on each event type and then combined into a global likelihood function for the ROI to represent all events\footnote{See FermiPy documentation for details: \url{https://fermipy.readthedocs.io/en/0.6.8/config.html}}. Photons detected at zenith angles larger than 100\,$\degree$ were excluded to limit the contamination from $\gamma$-rays generated by cosmic ray (CR) interactions in the upper layers of Earth's atmosphere. The data were additionally filtered to remove time periods where the instrument was not online (i.e. when flying over the South Atlantic Anomaly). The $\gamma$-ray data are modeled with the latest comprehensive \fermi source catalog, 4FGL \citep[data release 2 (DR2),][]{4fgl-dr2}, the LAT extended source template archive for the 4FGL catalog, and the latest Galactic diffuse and isotropic diffuse templates (\texttt{gll\_iem\_v07.fits} and \texttt{iso\_P8R3\_SOURCE\_V3\_v1.txt}, respectively).

\begin{figure}
\centering
\includegraphics[width=1.0\textwidth]{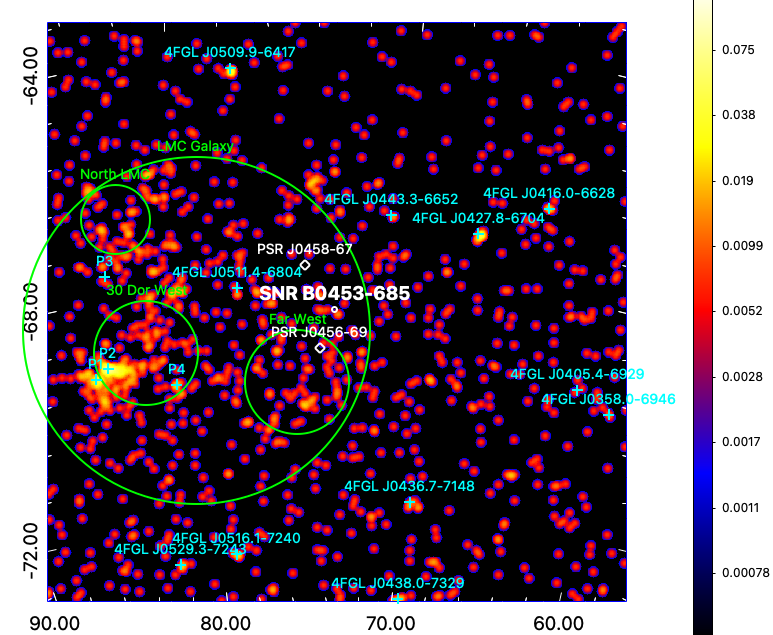}
\caption{{Adaptively smoothed ($\sigma = 10$) $10\,\degree\times10\,\degree$ count map for events with energies $>$ 60\,GeV classified as \texttt{PSF1}, $>20$\,GeV for \texttt{PSF2}, and $>6$\,GeV for \texttt{PSF3} type. The pixel size is 0.01 deg pixel$^{-1}$ such that the angular resolution for the image is $\sim0.1$\,$\degree$. 4FGL sources are indicated in cyan. The 4FGL point sources located in the LMC are labeled P1--P4 as in Figure \ref{fig:radio_xray_halpha}, right panel. The four extended templates used to describe the diffuse $\gamma$-ray emission from the LMC \citep[components E1--E4 in][]{lmc2016} are indicated with the green circles. The two closest known radio pulsars are plotted as white diamonds. The location of SNR~B0453--685 is marked in white with radius $r=0.05\degree$.}}\label{fig:6gev_cmap}
\end{figure} 

We fit the 10$\degree$ region of interest (ROI) using a pixel bin size $0.05\,\degree$ with 4FGL point sources and extended sources that are within 15$\degree$ of the ROI center along with the diffuse Galactic and isotropic emission backgrounds. Because B0453--685 is located in the LMC, we need to properly account for all background components: the isotropic component, the galactic diffuse component, and any diffuse emission from the LMC. We employ in the 4FGL source model four additional extended source components to reconstruct the {\it{emissivity model}} developed in \citep{lmc2016} to represent the diffuse LMC emission. The four additional sources are 4FGL~J0500.9--6945e (LMC Far West), 4FGL~J0519.9--6845e (LMC Galaxy), 4FGL~J0530.0-6900e (30 Dor West), and 4FGL~J0531.8--6639e (LMC North). These four extended templates along with the isotropic and Galactic diffuse templates define the total background for the ROI.

With the source model described above, we allow the background components and sources with TS $\geq 25$ and a distance from the ROI center (chosen to be the PWN position) $\leq3.0$\,$\degree$ to vary in spectral index and normalization. We computed a series of diagnostic test statistic (TS) and count maps in order to search for and understand any residual $\gamma$-ray emission. We generated the count and TS maps by the following energy ranges: 300\,MeV--2\,TeV, 1--10\,GeV, 10--100\,GeV, and 100\,GeV--2\,TeV. The motivation for increasing energy cuts stems from the improving PSF of the \fermi instrument with increasing energies We inspected the TS maps for additional sources, finding a faint point-like $\gamma$-ray source coincident in location with B0453--685 and no known 4FGL counterpart\footnote{The closest 4FGL source is the probable unclassified blazar 4FGL J0511.4--6804 $\sim2$\,$\degree$ away.}.

\begin{figure*}
\begin{minipage}[b]{.5\textwidth}
\centering
\includegraphics[width=1.0\linewidth]{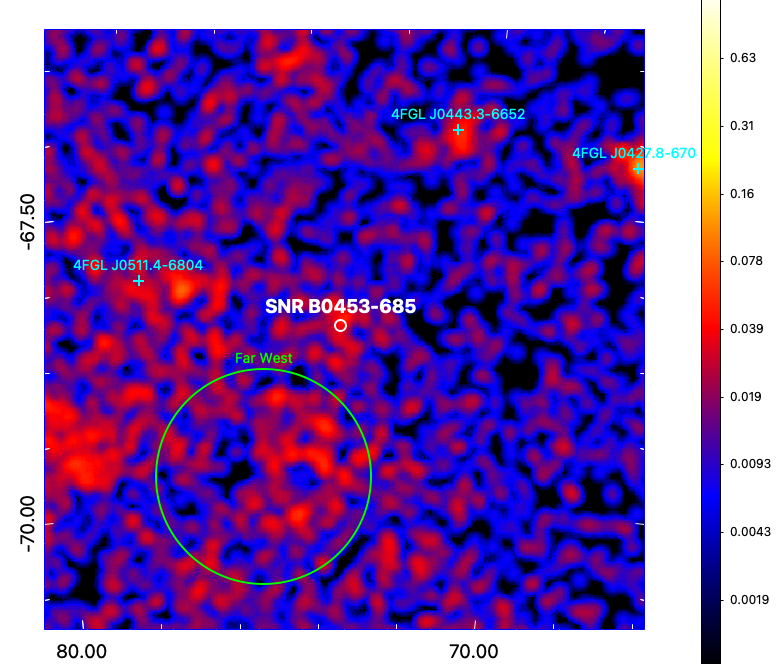}
\end{minipage}
\begin{minipage}[b]{.5\textwidth}
\hspace{-.4cm}
\centering
\includegraphics[width=1.0\linewidth]{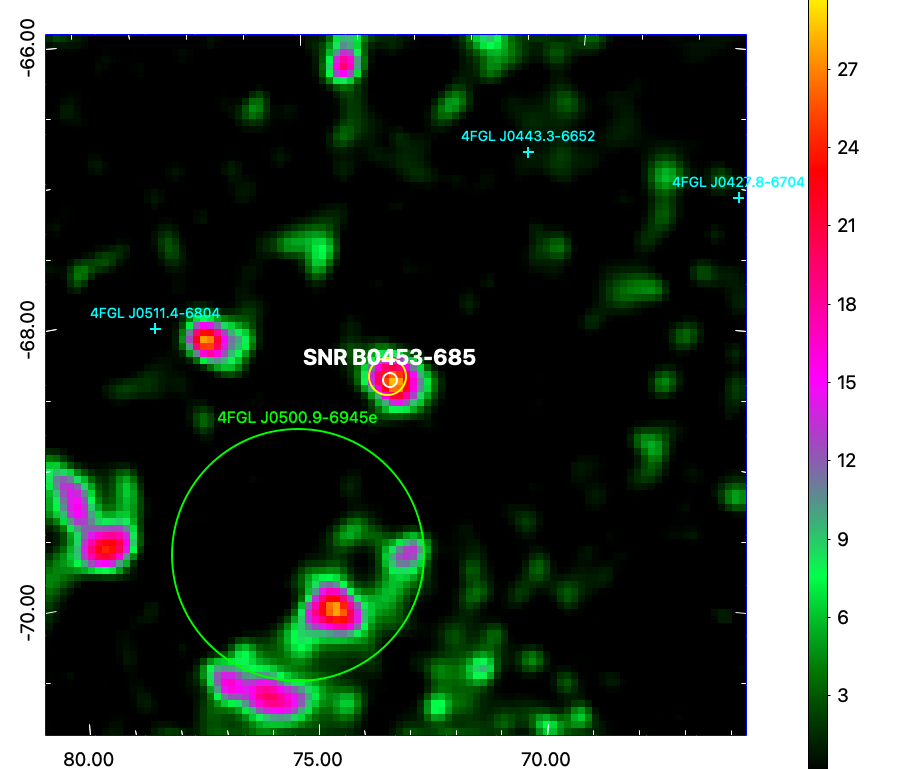}
\end{minipage}
\caption{{\it Left:} 5\,$\degree\times 5$\,$\degree$ count map of \texttt{PSF3} events between 1--10\,GeV, displaying only the 4FGL sources in the field of view. {\it Right:} 5\,$\degree\times 5$\,$\degree$ TS map between 1--10\,GeV that only considers the background components and seven of the brightest 4FGL sources in the source model. The maximum TS value at the SNR position is $\sim$ 28. The 95\% positional uncertainty for the best-fit $\gamma$-ray point source is in yellow. The 7 4FGL sources included in the source model for the TS map in the right panel are 4FGL~J0540.3--6920 (P1), 4FGL~J0537.8--6909 (P2), 4FGL~J0535.2--6736 (P3), 4FGL~J0524.8--6938 (P4), 4FGL~J0511.4--6804, 4FGL~J0443.3--6652, and 4FGL~J0427.8--6704. Only 4FGL~J0511.4--6804, 4FGL~J0443.3--6652, and 4FGL~J0427.8--6704 are indicated. P1--P4 are outside the field of view but would fall to the Southeast of the ROI. }\label{fig:gamma_maps}
\end{figure*}

A 10\,$\degree$ $\times$ 10\,$\degree$ count map for events with energies $>$60\,GeV classified as \texttt{PSF1}, $>20$\,GeV for \texttt{PSF2}, and $>6$\,GeV for \texttt{PSF3} event type is displayed in Figure \ref{fig:6gev_cmap} to demonstrate the total source model used in the analysis (except the isotropic and Galactic diffuse templates): the four diffuse LMC emission components are displayed in green, the LMC point sources are labeled accordingly (P1--P4), and unrelated 4FGL sources are displayed as cyan crosses with their 4FGL identifiers. A count and TS map between energies 1--10\,GeV are shown in Figure \ref{fig:gamma_maps} where the TS map, right panel, utilizes a source model where only the seven most significant 4FGL sources (including backgrounds) are considered such that remaining TS values are values of detection significance not attributed to background or the seven 4FGL sources. Faint $\gamma$-ray emission is visible and coincident with the SNR~B0453--685.

\begingroup
\begin{table*}
\hspace{-1cm}
\scalebox{0.85}{
\begin{tabular}{cccccccc}
\hline
\hline
\ Spectral Model & $\log{L}$ & $\Gamma$ & $\alpha$ or $\Gamma_1$ & $\beta$ or $\Gamma_2$ & $N_0$ (MeV$^{-1}$ cm$^{-2}$ s$^{-1}$) & $E_b$ or $a$ & TS \
\\
\hline
Power law & --505675.13 & $2.27\pm0.18$ & -- & -- & $ 1.56 (\pm 0.44) \times10^{-13}$ & -- & 22.92 \\
Log Parabola & --505673.39 & -- & $2.34 \pm0.20$ & $0.2 \pm 3.0 \times 10^{-4}$ & $ 8.87 (\pm 7.12) \times10^{-15}$ & 4000 & 26.24 \\
Power Law with Exponential Cut-Off & --505672.84 & -- &  $0.86\pm 0.99$ & $0.67$ (fixed) & $3.17 (\pm 2.10) \times10^{-13}$ & $0.0085 \pm 0.0065$ &  27.05\\
\hline
\hline
\end{tabular}}
\caption{Summary of the best-fit parameters and the associated statistics for all point source models tested. The units for $E_b$ are MeV. The units for the exponential factor $a$ are MeV$^{-\Gamma_2}$.}
\label{tab:gamma}
\end{table*}
\endgroup

\subsection{\fermi Data Analysis Results}\label{sec:fermi-results1}
To model the $\gamma$-ray emission coincident with B0453--685 we add a point source to the PWN location (R.A., Dec.) J2000 = (73.408\,$\degree$, --68.489\,$\degree$) to the 300\,MeV--2\,TeV source model in addition to known 4FGL sources in the ROI and the background emission templates for the diffuse LMC, Galactic, and isotropic components. With a fixed location, we set the spectrum to a power law characterized by a photon index $\Gamma=2$,
\begin{equation}
    \frac{dN}{dE} = N_{0} \big(\frac{E}{E_0}\big)^{-\Gamma}
\end{equation}\label{eq:12}
We then allow the spectral index and normalization to vary. The TS value for a point source with a power law spectrum and photon index, $\Gamma=2.27\pm0.18$, is $22.92$. We investigate the spectral properties of the $\gamma$-ray emission by replacing the point source with a new point source at the same location but with a log parabola spectrum following the definition.
\begin{equation}
    \frac{dN}{dE} = N_{0} \big(\frac{E}{E_b}\big)^{-(\alpha+\beta\log{E/E_b})}
\end{equation}\label{eq:22}
We initially set $\alpha=2.0$, $\beta=0.15$, and $E_b = 4.0$\,GeV but allowed $\alpha$, $\beta$, and $N_0$ to vary in the fit. The TS value of a point source at the PWN/SNR position with a log parabola spectrum is $26.24$ and has $\alpha=2.34\pm0.20$ and $\beta = 0.2 \pm 3.0 \times 10^{-4}$. We test the spectral parameters once more using a spectrum typically observed with MeV--GeV pulsars, a power law with a super exponential cut-off (PLEC)\footnote{This follows the \texttt{PLSuperExpCutoff2} form used for the 4FGL--DR2. Details can be found here: \url{https://fermi.gsfc.nasa.gov/ssc/data/analysis/scitools/source\_models.html\#PLSuperExpCutoff2}}:
\begin{equation}
    \frac{dN}{dE} = N_{0} \big(\frac{E}{E_0}\big)^{-\Gamma_1} \exp{\big(-a E^{\Gamma_2}\big)}
\end{equation}
where $E_0$ is the scale (set to 1000\,MeV), $\Gamma_1$ is the first index, $\Gamma_2$ is the exponential index, and $a$ the exponential factor. The TS value of a point source at the position of B0453--685 with a PLEC spectrum is $27.05$ and has $\Gamma_1=0.86\pm0.99$, $\Gamma_2$ is fixed to $0.67$, and exponential factor $a = 0.0085 \pm 0.0065$. The exponential factor corresponds to an energy cut-off $E_c = 1.3\,$GeV. See Table \ref{tab:gamma} for a summary of the spectral parameters for each point source test. 

\begin{figure}
\centering
\includegraphics[width=0.9\textwidth]{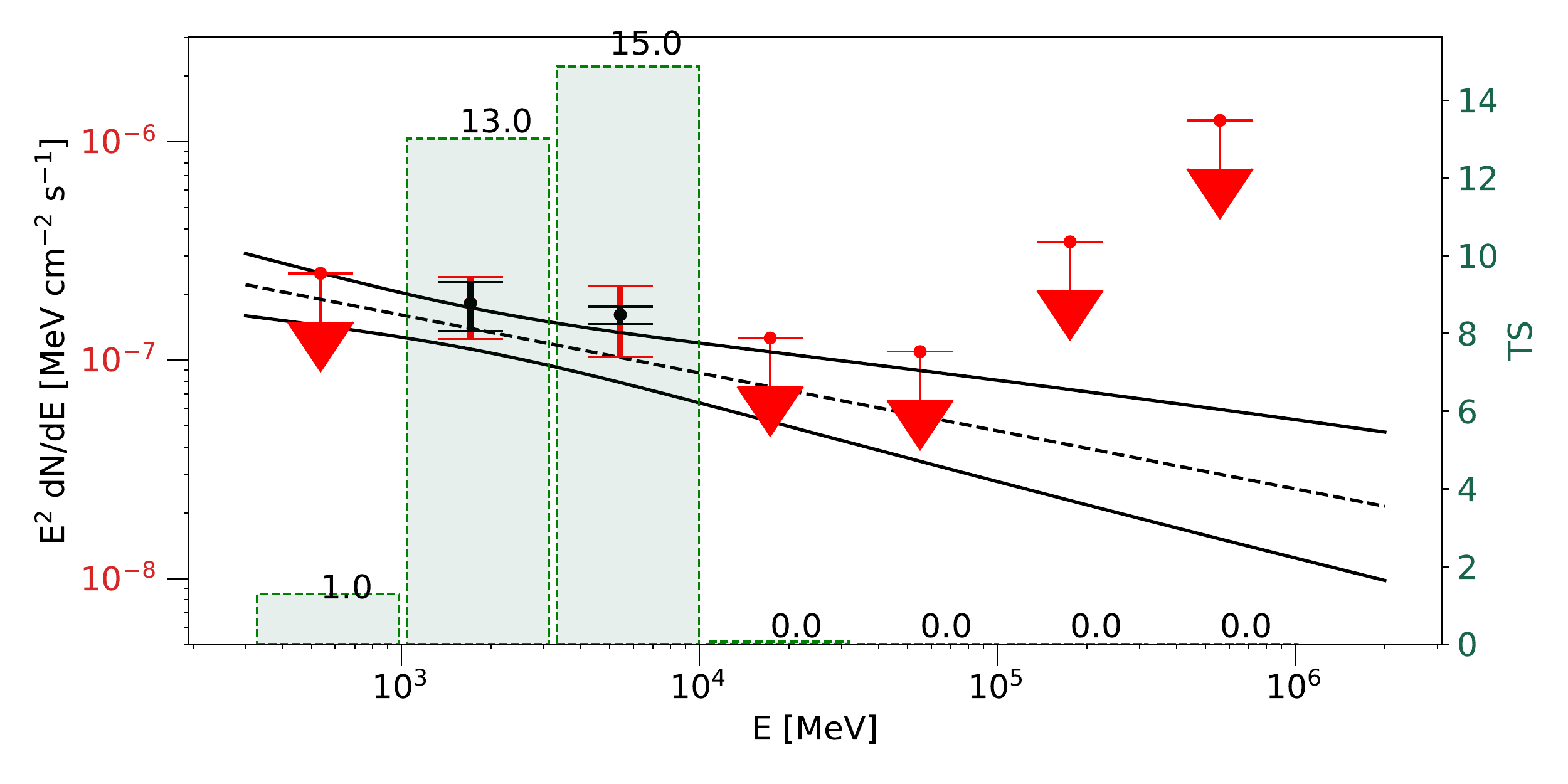}
\caption{The best-fit $\gamma$-ray SED for B0453--685 with 1-$\sigma$ statistical uncertainties in red for $TS > 1$ and 95\% confidence level (C.L.) upper limits otherwise. 
The systematic error from the choice of diffuse LMC model is plotted in black. TS values for each spectral bin are plotted as the green histogram. The data are best characterized as a power-law with $\Gamma = 2.27 \pm 0.18$.}\label{fig:gamma_sed1}
\end{figure}

\fermi pulsars are often characterized as either a power-law or a PLEC spectrum and typically cut off at energies $<10$\,GeV \citep[e.g.,][]{2pc}. While we cannot firmly rule out that the observed $\gamma$-ray emission is from the still-undetected pulsar based on the best-fit spectral parameters, it seems unlikely given the majority of the emission is measured in 1--10\,GeV. 
Between the three tested spectral models, the log parabola and PLEC are only marginally preferred (e.g., $TS_{\texttt{LogParabola}} = 2 \Delta(\ln{\mathcal{L}})=3.5$) and carry another degree of freedom with respect to the power law spectral model. We therefore conclude that the best characterization for the $\gamma$-ray emission coincident with SNR~B0453--685 is a power-law spectrum. The corresponding $\gamma$-ray SED is displayed in Figure~\ref{fig:gamma_sed1}. 

We localize the point source with \texttt{GTAnalysis.localize} to find the best-fit position and uncertainty. The localized position for the new $\gamma$-ray source is offset 0.0095\,$\degree$ from the exact position of B0453--685 and has R.A., Dec. = 73.388, --68.495 (J2000). The corresponding 95\% positional uncertainty radius is $r=0.12 \degree$.
We run extension tests for the best-fit point source in FermiPy utilizing \texttt{GTAnalysis.extension} and the two spatial templates supported in the FermiPy framework, the radial disk and radial Gaussian templates. Both of these extended templates assume a symmetric 2D shape with width parameters radius and sigma, respectively. We fix the position but keep spectral parameters free to vary when finding the best-fit spatial extension for both templates. The summary of the best-fit parameters for the extended templates are listed in Table \ref{tab:extension}. The faint extragalactic $\gamma$-ray emission shows no evidence for extension. 

\subsection{Systematic Error from Choice of IEM and IRF}
We account for systematic uncertainties introduced by the choice of the interstellar emission model (IEM) and the effective area, which mainly affect the spectrum of the faint $\gamma$-ray emission. We have followed the prescription developed in \citep{acero2016,depalma2013} and is outlined in Section~\ref{sec:sys} of Chapter~\ref{sec:fermi_analysis}. We find that the systematic errors are negligible for B0453--685 which is not surprising given the location of the Large Magellanic Cloud with respect to the bright diffuse $\gamma$-ray emission along the Galactic plane. 

\subsection{Systematic Error from Choice of Diffuse LMC}
We must also account for the systematic error that is introduced by having an additional diffuse background component. This third component is attributed to the cosmic ray (CR) population of the Large Magellanic Cloud interacting with the LMC ISM and there are limitations to the accuracy of the background templates used to model this emission, similar to the Galactic diffuse background. We can probe these limitations by employing a straightforward method described in \citep{lmc2016} to measure systematics from the diffuse LMC. This requires replacing the four extended sources that represent the diffuse LMC in this analysis \citep[the {\it emissivity model},][]{lmc2016} with four different extended sources to represent an alternative template for the diffuse LMC \citep[the {\it analytic model},][]{lmc2016}. The $\gamma$-ray point source coincident with SNR~B0453--685 is then refit with the alternative diffuse LMC template to obtain a new spectral flux that we then compare with the results of the {\it emissivity model} following equation (5) in \citep{acero2016}. The systematic error from the choice of the diffuse LMC template is largest in the two lowest-energy bins, but 
negligible in higher-energy bins. The corresponding systematic error is plotted in Figure~\ref{fig:gamma_sed1} in black.

\begingroup
\begin{table*}
\centering
\begin{tabular}{ccccccc}
\hline
\hline
\ Spatial Template & TS & TS$_{ext}$ & Best-fit radius or sigma ($\degree$) & 95\% radius upper limit ($\degree$) \
\\
\hline
Point Source & 22.92 & -- & -- & -- \\
Radial Disk & 23.02 & 0.10  & 0.05 & 0.17 \\
Radial Gaussian & 23.02 & 0.11 & 0.05 & 0.16 \\
\hline
\hline
\end{tabular}
\caption{Summary of the best-fit parameters and the associated statistics for each spatial template used in our analysis.}
\label{tab:extension}
\end{table*}
\endgroup

\section{Broadband modeling}
\subsection{Investigating Origin of Gamma-ray Emission} \label{sec:naima_model1}
\begin{figure*}[htbp]
\begin{minipage}[b]{1.0\textwidth}
\centering
\includegraphics[width=1.0\linewidth]{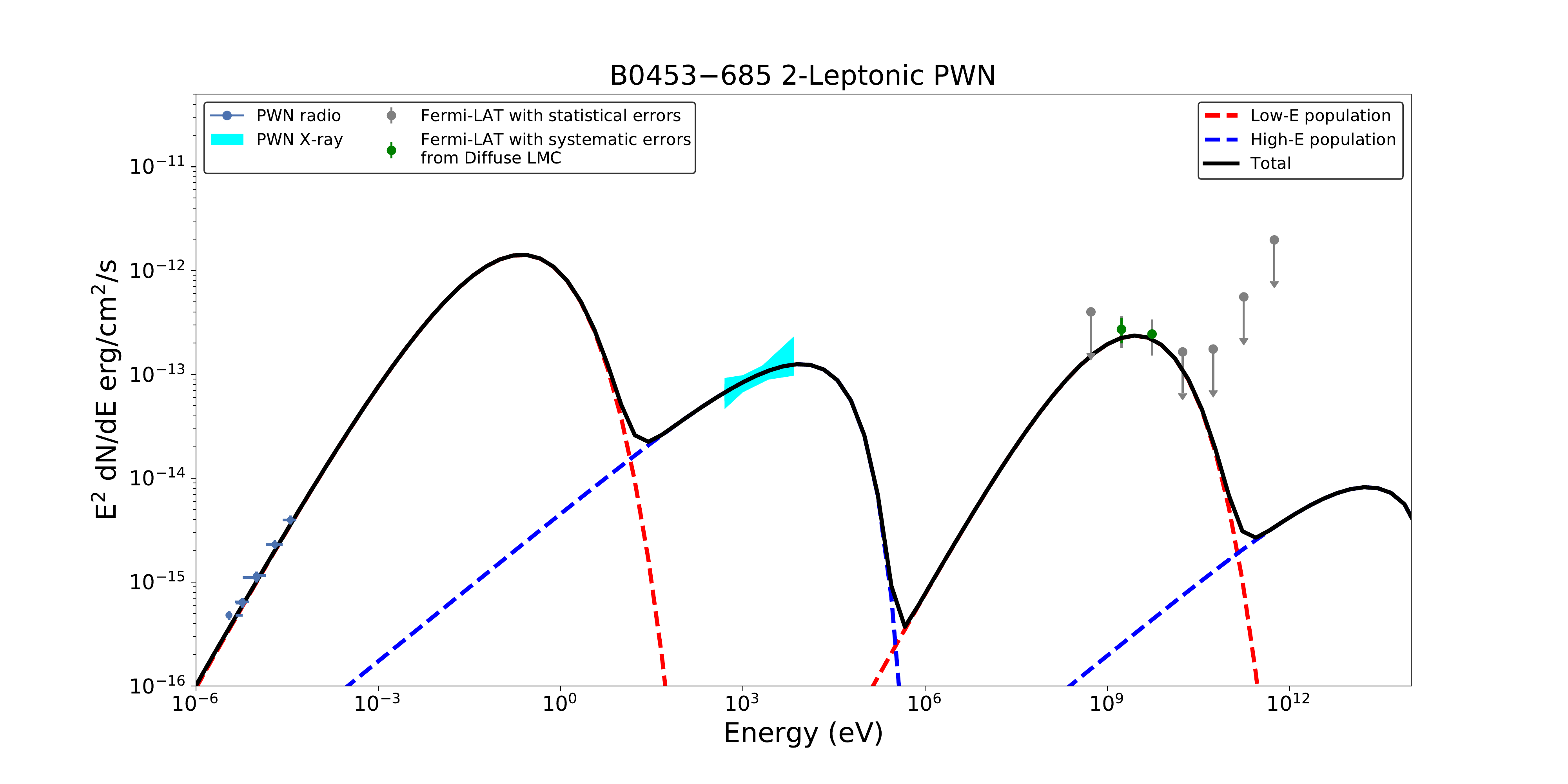}
\vspace{-0.35cm}
\end{minipage}
\begin{minipage}[b]{1.0\textwidth}
\hspace{-2cm}
\includegraphics[width=1.2\linewidth]{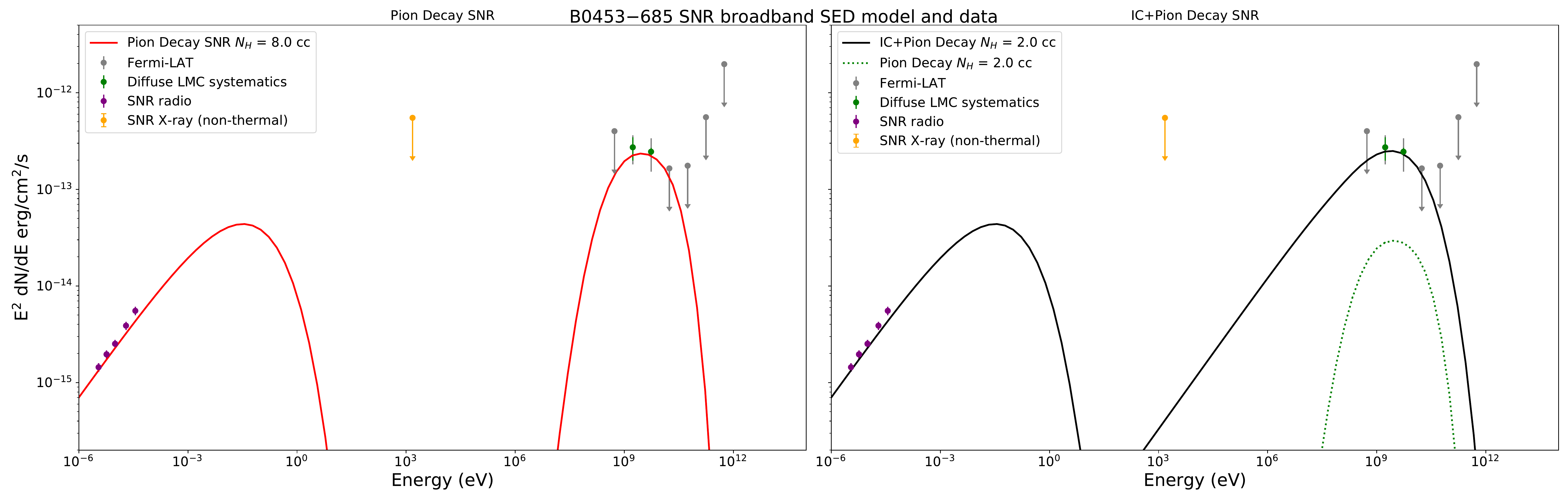}
\end{minipage}
\vspace{-0.5cm}
\caption{The best-fit broadband models for the three scenarios investigated to understand $\gamma$-ray origin. {\it Top:} Two leptonic populations are required to explain the broadband PWN emission. {\it Bottom Left:}  a single leptonic population describing SNR synchrotron emission combined with a single hadronic population describing the $\gamma$-ray emission via pion decay from the SNR. {\it Bottom Right:} The case where the leptonic population dominates over the hadronic population in the SNR. This third case represents a non-detection of pion decay (plotted as green dotted line), providing a lower limit on the ambient post-shock density for the SNR. Radio data of PWN (blue) and SNR (purple) are from \citep{haberl2012}, X-ray data of PWN (cyan) and SNR (yellow) are described in detail in Section \ref{sec:cha-results1}, and $\gamma$-ray data (grey/green) in Section \ref{sec:fermi-results1}.}\label{fig:broadbandpwnmodel}
\end{figure*} 

In order to investigate the origin of the observed $\gamma$-ray emission, we use the NAIMA Python package \citep{naima}, which computes the radiation from a single non-thermal relativistic particle population and performs a Markov Chain Monte Carlo (MCMC) sampling of the likelihood distributions \citep[using the \texttt{emcee} package,][]{emcee}. 
For the particle distribution, we assume a power law shape with an exponential cut-off (PLEC),
\begin{equation}
f(E) = A \big(\frac{E}{E_0}\big)^{-\Gamma} \exp\big({-\big({\frac{E}{E_{break}}}\big)}\big) \\
\end{equation}
where $A$ is in eV$^{-1}$. We then test a combination of free parameters (namely the normalization $A$, index $\Gamma$, energy break $E_{break}$, and magnetic field $B$) that can best explain the broadband spectra for the SNR and PWN independently. 

\subsubsection{PWN as $\gamma$-ray origin}
For the PWN, we find it infeasible to model the combined radio, X-ray, and $\gamma$-ray data using a single particle distribution, so we instead incorporate two leptonic particle populations under the same conditions (nebular magnetic field and ambient photon fields) and sum the total to represent a two-leptonic broadband model. A two-leptonic broadband model can describe well the PWN radio, X-ray, and $\gamma$-ray data, where the lower-energy particles dominate the radio and $\gamma$-ray emission while the higher energy particles are losing more energy in synchrotron radiation than they are in IC radiation, and therefore dominate in X-ray. We allow Population~1, the lower-energy population, to constrain the magnetic field strength, given it is likely the oldest particles dominate the synchrotron emission \citep{gelfand2009}. It is possible each population is interacting with magnetic field regions of varying strength, but for simplicity, we fix the magnetic field value to the best-fit found from the lower-energy population's broadband model when searching for a model fit for the higher-energy population, $B \sim 8 \mu$G. We consider only one photon field in all Inverse Compton Scattering calculations in this section, the Cosmic Microwave Background (CMB). The best-fit parameters for the low-energy population are $\Gamma = 0.879$ and $E_{break} \sim 355$\,GeV. The best-fit parameters for the high-energy population are $\Gamma = 2.05$ and $E_{break} \sim 224$\,TeV. The best-fit two-leptonic broadband model for the PWN is displayed in the top panel of Figure~\ref{fig:broadbandpwnmodel} and the corresponding best-fit parameters for both particle populations are listed in Table \ref{tab:naima}. 

The two-leptonic broadband model for the PWN has an estimated total particle energy $W_e = 2.86 \times 10 ^{49}$\,erg. The lower-energy population is responsible for $W_e = 2.84 \times 10 ^{49}$\,erg and the higher-energy population with the remainder, $W_e = 1.43 \times 10 ^{47}$\,erg.


\begin{table*}[t!]
\centering
\begin{tabular}{| c c c c |}
\hline
\ & Two-Leptonic PWN & Pion Decay & IC+Pion Decay SNR \\
\ & Population~1 \, Population~2 & & (Leptons Only)  \\
\hline 
Maximum Log Likelihood & --2.07\, --8.71 & --0.02 & --1.67 \\
\hline
\hline
\ Maximum Likelihood values & & &\\
\hline
\ $W_e$ or $W_p^{a}$ & $2.84 \times 10^{49}$ \, $1.43 \times 10^{47}$ & $2.08 \times 10^{50}$ & $2.71 \times 10^{50}$ \\
\  Index & 0.88 $\pm\, 0.13$ \, 2.05 $\pm\, 0.62$ & 1.31 $\pm\, 0.54$ & 1.95 $\pm\, 0.05$ \\
\  $\log_{10}{E_{c}}^{b}$ & --0.45 $\pm\, 0.11$ \, 2.35 $\pm\, 0.26$ & --1.18 $\pm\, 1.70$ & --0.17 $\pm\, 0.15$\\
\  $B^{c}$ & 8.18 $\pm\, 4.25$ \, \, 8.18 \text(fixed) & 1.47  \text(fixed) & 1.47$\pm\, 0.29$ \\
\hline
\end{tabular}
\caption{Summary of the statistics and best-fit models for the PWN and SNR broadband models displayed in Figure~\ref{fig:broadbandpwnmodel}.\footnotesize{$^a$ The total particle energy $W_e$ or $W_p$ in unit ergs, $^b$ Logarithm base 10 of the cutoff energy in TeV, $^c$ magnetic field in units $\mu$Gauss}}
\label{tab:naima}
\end{table*}

\subsubsection{SNR as $\gamma$-ray origin}
There are two possible scenarios for the SNR to be responsible for the $\gamma$-ray emission. The first is a single lepton population that is accelerated at the SNR shock front, generating both synchrotron emission at lower energies and IC emission at higher energies in $\gamma$-ray. The second scenario is a single lepton population emitting synchrotron radiation at lower energies and a single hadron population emitting $\gamma$-rays through pion decay. Under the assumption of the leptonic-dominant scenario and $k_{ep} = 0.01$ \citep{castro_2013}, we additionally estimate a lower limit to the target proton density. We describe both of these models and their implications below. 

To model the lower energy SNR emission together with the newly discovered \fermi emission using a single lepton population, we require a particle index $\Gamma = 1.95$, an energy break at 671\,GeV, and an inferred magnetic $B = 1.47\,\mu$G, comparable to the coherent component of the LMC magnetic field $B \sim 1\,\mu$G \citep{gaensler2005}. For the hadronic scenario, we model the broadband SNR emission assuming a single lepton and single hadron population, fixing the magnetic field value for the synchrotron component just described, and characterizing the $\gamma$-ray emission via pion decay through proton-proton collisions at the SNR shock front. We find a particle index $\Gamma = 1.31$, an energy break at 66\,GeV, and an inferred target proton density $n_h = 8.0\,$cm$^{-3}$ can reasonably reproduce the observed $\gamma$-ray spectrum. The post-shock proton density $n_h = 8.0\,$cm$^{-3}$ corresponds to four times the pre-shock LMC ISM density $n_0 \sim 2.0\,$cm$^{-3}$ \citep{kim2003}. We can estimate a lower limit to the target proton density, assuming the electron-to-proton ratio is $k_{ep} = 0.01$ \citep{castro_2013} and combining the best-fit leptonic and hadronic models of the SNR. We assume the same hadronic and leptonic populations as before, but scaling the normalization of the hadronic population by $\frac{1}{k_{ep}}$ and testing values of $n_h$ until the IC emission dominates the $\gamma$-ray spectrum and matches what is observed by the \fermi. We find the lower limit to the target density is $n_h = 2.0$\,cm$^{-3}$, the same as the mean value for pre-shock ISM in the LMC \citep{kim2003}. 
The best-fit broadband models for the SNR are displayed in the lower panels of Figure~\ref{fig:broadbandpwnmodel} and the corresponding parameters are listed in Table \ref{tab:naima}. 


The best-fit leptonic model for the SNR yields a total electron energy $W_e = 2.02 \times 10^{50}$\,erg. This implies, assuming $k_{ep} = 0.01$, the total proton energy from undetected pion decay emission is $W_p = W_e \times 100 = 2.02 \times 10^{52}$\,erg, requiring 20 times the canonical expectation $E = 10^{51}$\, ergs be in total SNR CR energy alone. 
Similarly, the best-fit pion decay model requires a total proton energy $W_p = 2.08 \times 10 ^{50}$\,erg, corresponding to an energy density $n_h \times W_p = 8.0\,$cm$^{-3} \times W_p = 1.67 \times 10^{51}$\,erg cm$^{-3}$, and requires twice the amount of explosion energy to be carried away in SNR CRs than is typically expected. 

Furthermore, the inferred magnetic field in both models is lower than one would expect at the SNR shock front, where shock compression can amplify the magnetic field strength 4--5 times the initial value   \citep[see e.g.,][and references therein]{castro_2013}. This contradicts the best-fit target proton density in the SNR pion decay model as it corresponds to the expected shock-compression ratio of 4 for the mean LMC ISM value \citep[$n \sim 2$\,cm$^{-3}$,][]{kim2003, mao2012}. For the SNR to accelerate hadronic CRs that generate the observed $\gamma$-ray emission via pion decay, the SNR must be interacting with dense ambient material \citep[e.g., W44 and IC443,][]{ic443w442013,chen2014,slane2015}. The radio and X-ray observations of the SNR show a fainter, limb-brightened shell compared to the bright, compact central PWN, providing little evidence of the SNR forward shock colliding with ambient media. 
Moreover, magnetic field amplification is a likely requirement for efficient particle acceleration \citep[e.g.,][]{vink2003} which is at odds with the thermal nature of the SNR in X-ray and with both of the best-fit broadband models for the SNR. If there were evidence for the SNR shock interacting with dense material, which is not the case for B0453--685, no synchrotron emission would only be expected if there were only a small fraction of leptonic CRs being generated, making a combined synchrotron and pion-decay broadband model unlikely. 

In summary, whether leptonic or hadronic, if particles are being accelerated at the SNR shock front, one would expect nonthermal X-ray emission from the relativistic electrons as the magnetic field at the shock is amplified. Given the low magnetic field inferred in both SNR broadband models, we find it unlikely for the SNR to be responsible for the observed $\gamma$-ray emission. The energetics inferred by the SNR models combined with the lack of observational evidence favor the two-leptonic PWN broadband model as the most likely explanation for the $\gamma$-ray emission reported here. We explore the most accurate representation to the PWN broadband data while also exploring the likelihood of a pulsar contribution in the following section.

\subsection{PWN Evolution through Semi-Analytic Modeling}\label{sec:yosi_model}
\normalsize

\begin{figure}
\centering
\includegraphics[width=0.9\textwidth]{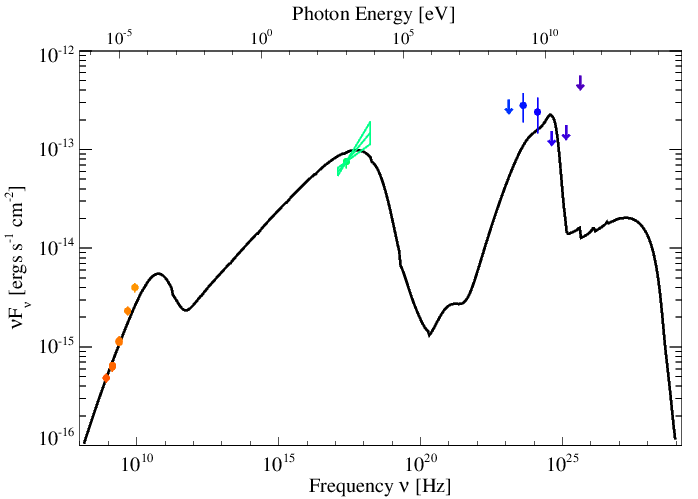}
\caption{The best-fit SED assuming all \fermi emission is non-magnetospheric in origin (i.e., PWN only) obtained through the evolutionary model method described in Section~\ref{sec:yosi_model}. The coloured points represent the values of observed data that the model used as comparison points for fitting and are the same values as those in the top panel of Figure~\ref{fig:broadbandpwnmodel}. }\label{fig:gelfand_sed}
\end{figure} 

\small{
\begin{table*}
\hspace{-2cm}
\begin{tabular}{|c c c c c|}
\hline
\/Shorthand & Parameter & PWN Best-Fit & PWN+PSR Best-Fit & Units \\
\hline
\hline
\ \texttt{loglh} & Log Likelihood of Spectral Energy Distribution & -19.8857 & -17.5866 & - \\
\hline
\ \texttt{esn} & Initial Kinetic Energy of Supernova Ejecta & 5.24 & 5.21 & $10^{50}$ ergs \\
\hline
\ \texttt{mej} & Mass of Supernova Ejecta & 2.24 & 2.42 & Solar Masses \\
\hline
\ \texttt{nism} & Number Density of Surrounding ISM & 0.97 & 1.00 & cm$^{-3}$ \\
\hline
\ \texttt{brakind} & Pulsar Braking Index & 2.89 & 2.83 & - \\
\hline
\ \texttt{tau} & Pulsar Spin-down Timescale & 172 & 166 & years \\
\hline
\ \texttt{age} & Age of System & 13900 & 14300 & years \\
\hline
\ \texttt{e0} & Initial Spin-down Luminosity of Pulsar & 6.95 & 6.79 & $10^{39}$ ergs s$^{-1}$ \\
\hline
\ \texttt{velpsr} & Space Velocity of Pulsar & 0 & 0 & cm$^{-1}$ \\
\hline
\ \texttt{etag} & Fraction of Spin-down Luminosity lost as Radiation & 0 & 0.246 & - \\
\hline
\ \texttt{etab} & Magnetization of the Pulsar Wind & 0.0006 & 0.0007 & - \\
\hline
\ \texttt{emin} & Minimum Particle Energy in Pulsar Wind & 1.77 & 2.26 & GeV \\
\hline
\ \texttt{emax} & Maximum Particle Energy in Pulsar Wind & 0.90 & 0.73 & PeV \\
\hline
\ \texttt{ebreak} & Break Energy in Pulsar Wind & 76 & 72 & GeV \\
\hline
\ \texttt{p1} & Injection Index below the Break & 1.47 & 1.34 & - \\
\  & (${dN}/{dE} \sim E^{-p1}$) & & & \\
\hline
\ \texttt{p2} & Injection Index below the Break & 2.36 & 2.36 & - \\
\  & (${dN}/{dE} \sim E^{-p2}$) & & & \\
\hline
\ \texttt{ictemp} & Temperature of each Background Photon Field & 1.02 & 1.13 & $10^{6}$ K \\
\hline
\ \texttt{icnorm} & Log Normalization of each Background Photon Field & -17.9209 & -17.998 & - \\
\hline
\ \texttt{kpsr} & Log Normalization of Direct $\gamma$-ray Emission from the Pulsar & 0 & -9.9874 & - \\
\hline
\ \texttt{gpsr} & Photon Index of the $\gamma$-rays Produced Directly by the Pulsar & 0 & 2.00 & - \\
\hline
\ \texttt{ecut} & Cutoff Energy from the Power Law of Pulsar Contribution & 0 & 3.21 & GeV \\
\hline
\end{tabular}
\caption{Summary of the input parameters for the evolutionary system and their best fit values considering PWN-only and PWN+PSR contributions to the \fermi emission.}
\label{tab:input_output_gelfand}
\end{table*}}

\begin{figure*}
\begin{minipage}[b]{.5\textwidth}
\hspace{-1cm}
\includegraphics[width=1.1\textwidth]{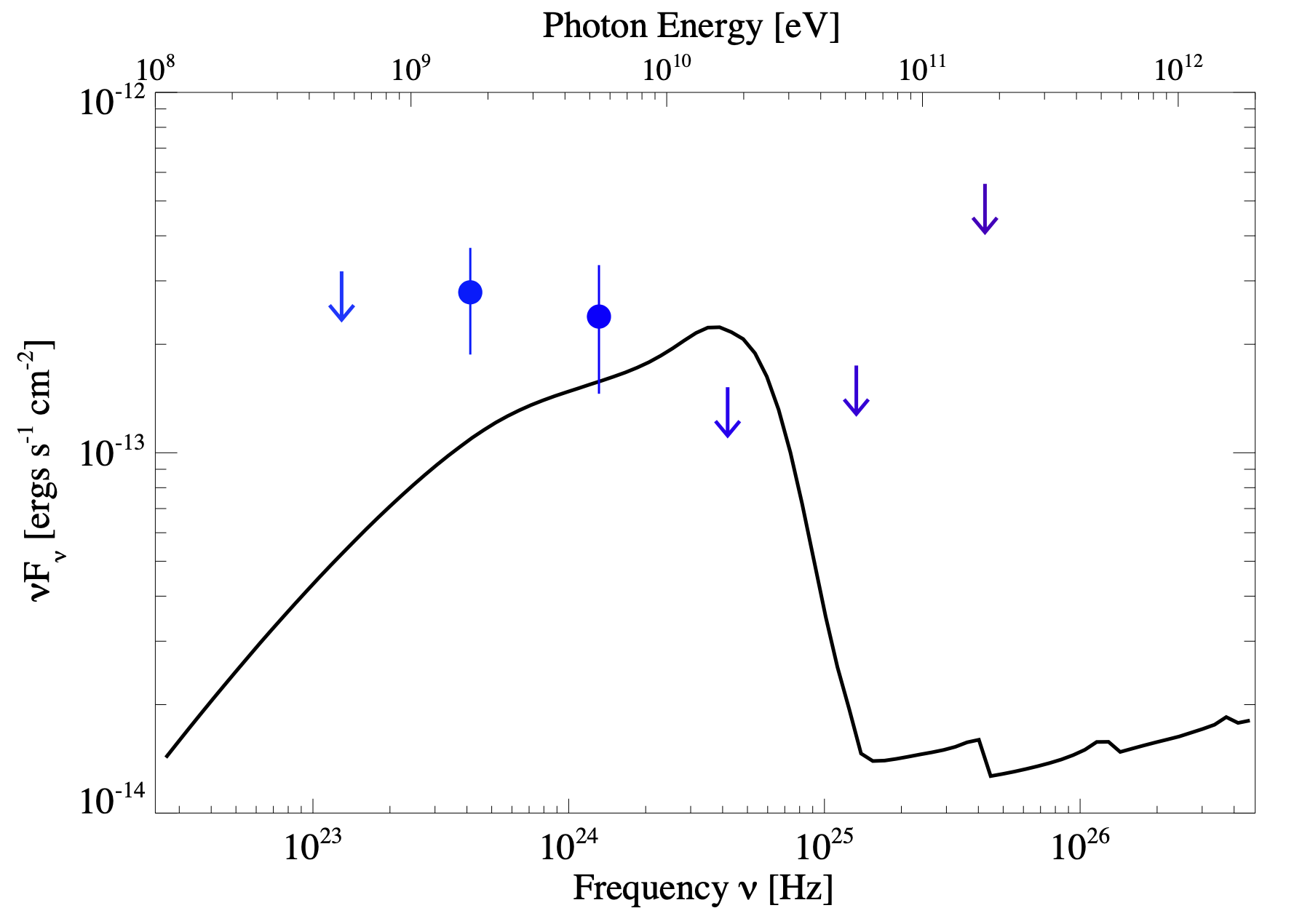}
\end{minipage}
\begin{minipage}[b]{.5\textwidth}
\centering
\includegraphics[width=1.1\textwidth]{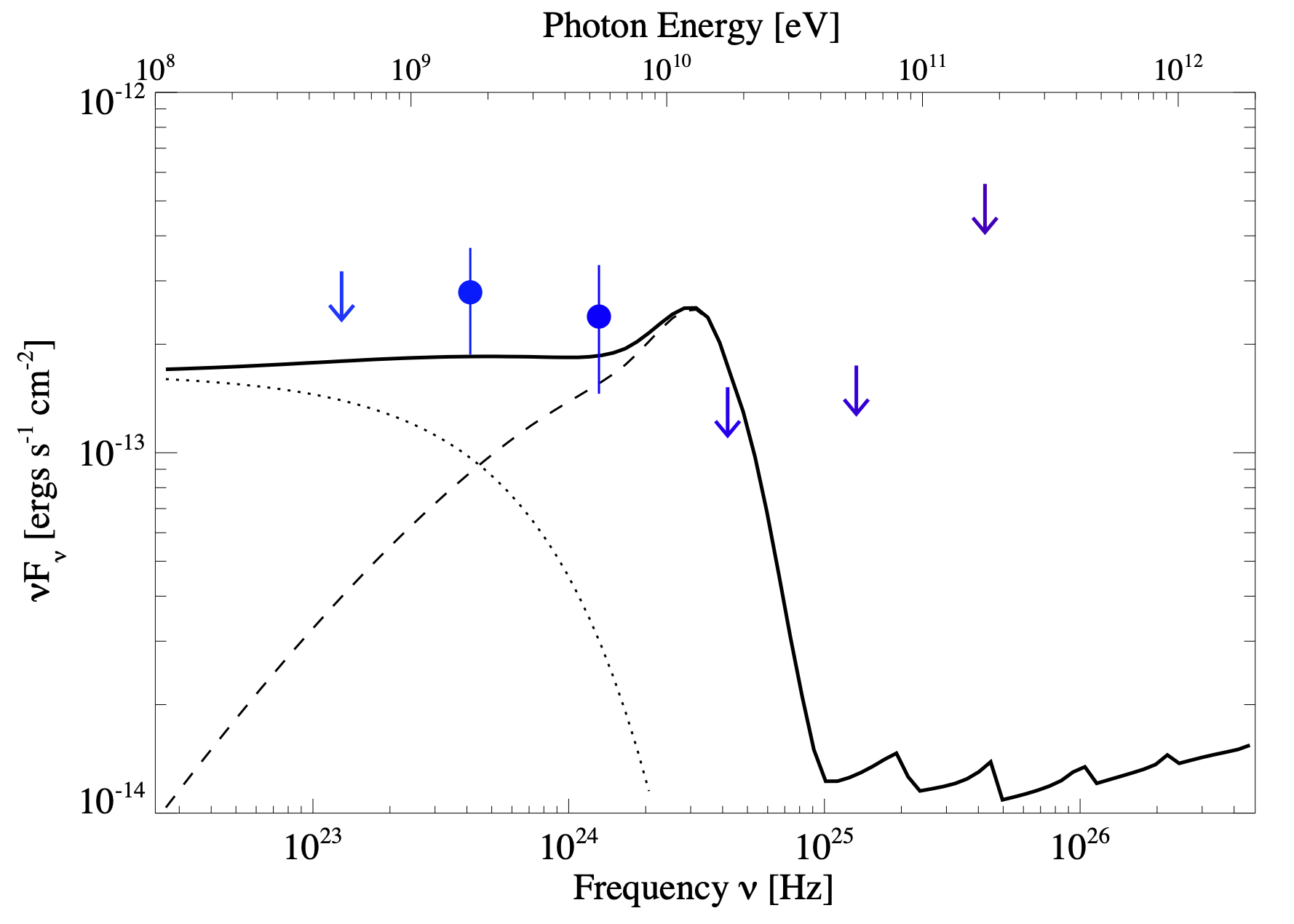}
\end{minipage}
\caption{{\it Left:} The $\gamma$-ray spectral evolutionary model assuming all \fermi emission is non-magnetospheric in origin (i.e., PWN only). {\it Right:} The $\gamma$-ray spectral evolutionary model assuming magnetospheric contribution to the \fermi emission. The dotted line indicates the pulsar contribution and the dashed line indicates the PWN contribution. The coloured points represent the values of observed data that the model used as comparison points for fitting and are the same values as those in the top panel of Figure~\ref{fig:broadbandpwnmodel}. }\label{fig:compare_gelfand_seds}
\end{figure*} 

We have established in the previous section that modeling the non-thermal broadband SED suggests that it most likely originates from two populations of leptons with different energy spectra, similar to what is expected for evolved PWNe once they have collided with the SNR reverse shock \citep[see e.g.,][]{gelfand2009, temim2015}.  To determine if the depicted scenario can explain the intrinsic properties of this system, we model the observed properties of the PWN, assuming it is responsible for the detected \fermi $\gamma$-ray emission, as it evolves inside the composite SNR B0453$-$685.

We use the dynamical and radiative properties of a PWN predicted by an evolutionary model, similar to what is described by \citep{gelfand2009}, to identify the combination of neutron star, pulsar wind, supernova explosion, and ISM properties that can best reproduce what is observed. The predicted dynamical and radiative properties of the PWN that correspond to the best representation of the broadband data are listed in Table \ref{tab:input_output_gelfand}. 
The analysis performed here is similar to what has previously been reported for MSH 15--56 \citep{temim2013}, G21.5--0.9 \citep{hattori2020}, Kes~75 \citep{gotthelf2021,straal2022}, HESS~J1640--465 \citep{mares2021}, and G54.1+0.3 \citep{gelfand2015}. For the characteristic timescale $t_{ch}$ of a pulsar \citep[see][]{pacini1973,slane2006}, the age $t_{age}$ is defined as
\begin{equation}
    t_{age} = \frac{2t_{ch}}{p-1} - \tau_{sd}
\end{equation}
and the initial spin-down luminosity $\dot{E_0}$ is defined as
\begin{equation}
    \dot{E}(t) = \dot{E_0}\big(1 + \frac{t}{\tau_{sd}}\big)^{-\frac{p+1}{p-1}}
\end{equation}
and are chosen for a braking index $p$ and spin-down timescale $\tau_{sd}$ to reproduce the pulsar's likely characteristic age and spin-down luminosity $\dot{E}$. A fraction $\eta_\gamma$ of this luminosity is responsible for $\gamma$-ray emission from the neutron star's magnetosphere, the rest $(1-\eta_\gamma)$ is injected into the PWN in the form of a magnetized, highly relativistic outflow, i.e., the pulsar wind. The pulsar wind enters the PWN at the termination shock, where the rate of magnetic energy $\dot{E}_B$ and particle energy $\dot{E}_P$ injected into the PWN is assumed to be:
\begin{eqnarray}
\label{eqn:edotb}
\dot{E}_B(t) & \equiv & (1-\eta_\gamma)\eta_{\rm B}\dot{E}(t) \\
\dot{E}_P(t) & \equiv & (1-\eta_\gamma)\eta_{\rm P}\dot{E}(t)
\end{eqnarray}
where $\eta_B$ is the magnetization of the wind and defined to be the fraction of the pulsar's spin-down luminosity injected into the PWN as magnetic fields and $\eta_P$ is the fraction of spin-down luminosity injected into the PWN as particles. We assume the PWN Inverse Compton emission results from leptons scattering off of the CMB. A second ambient photon field is also required, defined with the temperature $T_{IC}$ and normalization $K_{IC}$, such that the energy density of the photon field $u_{IC}$ is
\begin{equation}
    u_{IC} = K_{IC}a_{BB}T^4_{IC}
\end{equation}
where $a_{BB} = 7.5657 \times 10^{-15}$\,erg cm$^{-3}$ K$^{-4}$. Additionally, we assume the particle injection spectrum at the termination shock is well-described by a broken power law distribution:
\begin{equation}
\frac{d\dot{N}_{e^\pm}(E)}{dE} =
\begin{cases}
 \dot{N}_{break} \big(\frac{E}{E_{break}}\big)^{-p_1} & E_{min} < E < E_{break} \\
 \dot{N}_{break} \big(\frac{E}{E_{break}}\big)^{-p_2} & E_{break} < E < E_{max} \\
\end{cases}
\end{equation}
where $\dot{N}_{e^\pm}$ is the rate that electrons and positrons are injected into the PWN, and $\dot{N}_{break}$ is calculated using
\begin{equation}
    (1-\eta_B)\dot{E} = \int_{E_{min}}^{E_{max}} E \frac{d\dot{N}(E)}{dE} dE
\end{equation}
We show the spectral energy distribution for PWN~B0453--685 that can reasonably reproduce the observed spectrum in Figure~\ref{fig:gelfand_sed}. 

To investigate the potential for a pulsar contribution to the \fermi data, we model the broadband spectrum again by adding a second emission component from the pulsar. In this case, we assume any \fermi pulsar flux can be described by a power-law with an exponential cut off:
\begin{equation}
    \frac{dN_{\gamma}}{dE} = N_0 E^{-\Gamma} \rm{exp}\left(-\frac{E}{E_{\rm cut}}\right),
\end{equation}
We find that the pulsar can readily explain the lower-energy \fermi emission with a cut-off energy $E_c = 3.21\,$GeV and spectral index $\Gamma = 2.0$. Figure~\ref{fig:compare_gelfand_seds} displays both $\gamma$-ray SEDs for the two considered cases where the \fermi emission is PWN-only (left panel) and where there is both a PWN and pulsar contribution (right panel). If there is a pulsar contribution to the \fermi emission, it is likely to dominate for $E \lesssim 3\,$GeV whereas the PWN may only begin to dominate beyond this energy.  We discuss the physical implications of the presented broadband models in the next section.

\section{Discussion}\label{sec:discuss1}
We can determine the physical properties of the host SNR and ambient medium inferred from the most accurate broadband models presented in Sections~\ref{sec:naima_model1} and \ref{sec:yosi_model} and compare to the theoretical values expected for a middle-aged SNR in the Sedov-Taylor phase. The angular diameter of SNR~B0453--685 is 0.036\,$\degree$ which corresponds to a shock radius $R_s = 15.71$\,pc at a distance $d = 50$\,kpc. 

The pre-shock proton density $n_0$ has been estimated to be $\sim 0.4$\,cm$^{-3}$ from the SNR X-ray emission measured along the rim region \citep{gaensler2003}. The post-shock proton density $n_H$ could be at least four times as high as $n_0$, thus for a compression ratio $\frac{n_h}{n_0} = 4$, $n_h \sim 1.6\,$cm$^{-3}$. Assuming $\frac{n_e}{n_h} = 1.2$ and taking $n_H \sim 1.6\,$cm$^{-3}$, we estimate the post-shock electron density to be $n_e \sim 1.9\,$cm$^{-3}$. This result is consistent to prior works finding a range of values $n_e/f \sim 1.5 - 8.0\,$cm$^{-3}$ \citep{gaensler2003,haberl2012,mcentaffer2012}. The normalization and post-shock proton density characterizing pion decay emission are inversely proportional. If we assume the new value for $n_H$, we can sample the degeneracy of the normalization and $n_h$ parameters of the best-fit hadronic model, $\frac{n_{h, \text{naima}}}{n_{h, \text{theory}}} = 5$, which would scale the total energy in protons by a factor $\frac{1}{5}$. From this degeneracy, the total proton energy in the SNR hadronic model could be $E_p = 4.2 \times 10^{49}\,$erg or 4\% of $E_{SN} = 10^{51}$\,erg, which is closer to the canonical expectation of 10\% than the hadronic model presented in Section~\ref{sec:naima_model1}. However, X-ray observations of the SNR shell indicate an explosion energy as low as $E_{SN} \approx 10^{50}$\,erg \citep{gaensler2003,haberl2012}.

We can evaluate the SNR age assuming it is in the Sedov-Taylor phase \citep{sedov1959,taylor1950}:
\begin{equation}
    t = \bigg(\frac{R_{sh}}{2.3\text{\,pc}} \big(\frac{E}{10^{51}\text{\,ergs}}\big)^{\frac{1}{5}}\big(\frac{\rho_{0}}{10^{-24}\text{\,g cm$^{-3}$}}\big)^{-\frac{1}{5}}\bigg)^{5/2} 100\, \text{yr}
\end{equation}
The SNR age estimates range between 13\,kyr \citep{gaensler2003} using $E = 5 \times 10^{50}$\,erg and $\rho_0 = 0.4 \times 10^{-24}$\,g cm$^{-3}$ and 15.2\,kyr using $E = 7.6 \times 10^{50}$\,erg and $\rho_0 = 0.3 \times 10^{-24}$\,g cm$^{-3}$ \citep{haberl2012}. \citep{mcentaffer2012} find the largest range $\tau \sim 17-23$\,kyr using equilibrium shock velocity estimates $\sim 280-380\,$km s$^{-1}$. We adopt the SNR age $\tau \sim 13\,$kyr \citep{gaensler2003}, which corresponds to a shock velocity $v_s = 478\,$km s$^{-1}$ from $v_s = \frac{2R_s}{5t}$. The age predicted from the evolutionary method in Section~\ref{sec:yosi_model}, $t \sim 14.3$\,kyr, is in good agreement with prior work. The ambient proton density predicted in Section~\ref{sec:yosi_model}, $n_0 = 1.0$\, cm$^{-3}$, is somewhat higher than the values estimated in prior work \citep{gaensler2003,haberl2012}. In any case, the $n_0$ estimates are much lower than the average LMC ISM value $n_0 \sim 2$\, cm$^{-3}$ \citep{kim2003}, and indicate that the ambient medium surrounding SNR~B0453--685 may be less dense than the average LMC ISM. This is supported by Figure~\ref{fig:radio_xray_halpha}, right panel, where a possible density gradient decreasing from east to west is apparent. The lower ambient particle density estimate is also consistent with the observed faint SNR shell in radio and X-ray.

The explosion energy predicted by the evolutionary model, $E = 5.2 \times 10^{50}$\,erg, is very similar to the one inferred by X-ray observations, $E \sim 5-7.6 \times 10^{50}$\,erg \citep{gaensler2003,haberl2012}. Additionally, the magnetic field and total particle energy in the PWN from the evolutionary model are predicted to be $5.9\,\mu$G and $W_e = 5.4 \times 10^{48}\,$erg respectively, which is roughly consistent to the values implied by NAIMA modeling in Section~\ref{sec:naima_model1}, $8.18\,\mu$G and $W_e = 2.9 \times 10^{49}\,$erg. Lastly, one can estimate the $\gamma$-ray efficiency $\eta = \frac{L_\gamma}{\dot{E}}$ from the predicted current spin-down power of the central pulsar in the evolutionary model, $\dot{E} \sim 8.1 \times 10^{35}$\, erg s$^{-1}$. For a 300\,MeV--2\,TeV $\gamma$-ray source at $d = 50\,$kpc, the $\gamma$-ray luminosity is $L_\gamma = 2.6 \times 10^{35}$\, erg s$^{-1}$ which corresponds to $\eta = 0.32$. This efficiency value is not uncommon for $\gamma$-ray pulsars \citep[e.g.,][]{2pc}, though it is a more reasonable value to expect from evolved PWNe.

From our semi-analytic evolutionary models, we find the best representation to the data occurs with the supernova energy values $\sim 5 \times 10^{50}$\,erg, $\sim 2.3$ solar masses for SN ejecta, and $\sim$1.0 cm$^{-3}$ for the density of the ISM (see Table~\ref{tab:input_output_gelfand}). These values can then be used in combination with other models to survey the possible physical characteristics of the progenitor for SNR~B0453--685. For example, a correlation reported in \citep{ertl2020} has found that the only supernovae that have an explosion energy $\sim 5 \times 10^{50}$\,erg are those whose progenitors have a final helium core mass $< 3.5\,M_\odot$. Given an ejecta mass $\sim 2.3\,M_\odot$ from our evolutionary model, we calculate a neutron star mass $M_{NS} = 3.5\,M_\odot - 2.3\,M_\odot = 1.2\,M_\odot$, which is reasonable \citep[see e.g.,][]{kaper2006}. 

A core collapse supernova progenitor cannot have an initial mass smaller than $8\,M_\odot$. We can use the known inverse correlation between the age and mass of a main-sequence star,
\begin{eqnarray}
\label{eqn:age_mass}
    \frac{t_{MS}}{t_{Sun}} \sim \big(\frac{M}{M_{Sun}}\big)^{-2.5}
\end{eqnarray}
to get a maximum possible lifetime $\tau \sim 20\,$million\,years for any supernova progenitor. A map from \citep{harris2009} of the Large Magellanic Cloud with age and metallicity data distributions is used to acquire the age and metallicity distributions for the LMC regions closest to the location for B0453--685. 
By compiling their data, we can see that there was possibly a burst of star formation in those regions around the maximum possible lifetime estimate, as it contains many stars that are from 10$^{6.8}$ ($\sim 6.3$\,million) to 10$^{7.4}$ ($\sim 25$\,million) years old. From this, the progenitor must have had a main sequence lifetime comparable to the maximum possible lifetime for us to observe the supernova remnant today. We can use Eq.~\ref{eqn:age_mass} to estimate the mass of the precursor star of B0453--685 to be between 11 and 19\,$M_\odot$. However, as said above, our model predicts a pre-explosion helium core of 3.5 solar masses, which does not reach the 11--19\,$M_\odot$ dictated by the above analysis. The similarity between the final core mass $M_{NS} = 1.2\,M_\odot$ and the pre-explosion mass $M_{NS} = 3.5\,M_\odot$ suggested by our modeling implies that the progenitor lost its envelope before exploding.

If the models presented are correct, then there are two plausible ways to explain the loss of $\sim 7.5-15.5M_\odot$ of material before exploding: an isolated star could have lost mass by way of stellar wind, while a star that is part of a binary system could have transferred some of its mass to the other star. To account for stellar wind quantitatively, we looked at the model presented in \citep{sukh2016} where it is shown that normal ejecta mass for a 10--15$M_\odot$ star is 8--10$M_\odot$, respectively. However, stellar wind can only account for up to 3$M_\odot$ in mass loss for stars more massive than 15$M_\odot$. Additionally, it is known that low metallicity stars experience less mass loss \citep{heger2003}, and the young stars in the LMC region of B0453--685 all have metallicity $\sim$ 0.008\,$Z_\odot$. In summary, it seems plausible that the progenitor for B0453--685 was a part of a binary star system. 

\section{B0453--685: Summary}\label{sec:conclude1}
We have reported the discovery of faint, point-like $\gamma$-ray emission by the \fermi that is coincident with the composite SNR B0453--685, located within the Large Magellanic Cloud. We provide a detailed multiwavelength analysis that is combined with two different broadband modeling techniques to explore the most likely origin of the observed $\gamma$-ray emission. We compare the physical implications and energetics from the best-fit broadband models to the theoretically expected values for such a system and find that the most plausible origin is the pulsar wind nebula within the middle-aged SNR~B0453--685 and possibly a substantial pulsar contribution to the low-energy $\gamma$-ray emission below $E < 5$\,GeV. Theoretical expectation based on observational constraints and the inferred values from the best-fit models are consistent, despite assumptions about the SNR kinematics and environment such as a spherically symmetric expansion into a homogeneous ISM density. The MeV--GeV detection is too faint to attempt a pulsation search, but the $\gamma$-ray SED cannot rule out a pulsar component. We attempt to model the $\gamma$-ray emission assuming both PWN and pulsar contributions and the results indicate that any pulsar $\gamma$-ray signal is likely to be prominent below $E \leq 5\,$GeV, if present. 
Further work should explore the $\gamma$-ray data particularly for energies $E < 10\,$GeV to investigate the potential for a pulsar contribution as well as the possibilities for PWN and/or pulsar emission in the MeV band pass for a future MeV space mission such as COSI\footnote{\url{https://cosi.ssl.berkeley.edu/}} and AMEGO\footnote{\url{https://asd.gsfc.nasa.gov/amego/index.html}}. The IC emission spectra reported here may be even better constrained when combined with TeV data from ground-based Cherenkov telescopes such as H.E.S.S. or the upcoming Cherenkov Telescope Array\footnote{\url{https://www.cta-observatory.org/}}.  \clearpage

    \pdfoutput=1
\chapter{Discussion and Conclusions}\label{sec:final}
11.5\,years of observational data from the \fermi are analyzed and the results constitute the \fermi PWN catalog. 11 new $\gamma$-ray source classifications as PWNe, which doubles the current \fermi PWN population known, and 22 as PWN candidates are reported in Chapter~\ref{sec:results}. The \fermi PWN catalog together with other available multiwavelength data and semi-analytic modeling is now being used to determine the evolutionary stage, the physical particle properties of individual systems, and to establish a basic evolutionary trend for the broader PWN population.  

Detailed semi-analytic simulations that predict what we observe based on the intrinsic properties of the system (e.g. pulsar spin down and characteristic age) have been developed \citep[see][]{blondin2001, gelfand2009, swaluw2004}. Emerging trends are already apparent when applying these models to a sample of PWNe varying in age such as the young PWN systems Kes~75 \citep{gotthelf2021,straal2022} and G54.1+0.3 \citep{gelfand2015}, compared to older PWN systems such as MSH 15--56 \citep{temim2013}, G21.5--0.9 \citep{hattori2020}, and HESS~J1640--465 \citep{mares2021}. The majority of high-energy (GeV--TeV) PWNe are typically observed to be in later stages of evolution whereas it is more likely that younger PWNe are luminous in lower-energies such as the MeV band \citep[e.g.,][]{gelfand2019}. The model developed in \citep{gelfand2009} has been applied to PWNe from this project  in Chapters~\ref{sec:g327} and \ref{sec:b0453} for G327.1--1.1 and B0453--685, respectively. G327.1--1.1 and B0453--685 represent at least two other $\gamma$-ray sources that are associated to evolved PWNe where their broadband information already hint towards significant synchrotron losses, likely resulting from the SNR reverse shock interactions with the nebulae. 

Combining the \fermi PWN catalog results with available TeV data from Imaging Air Cherenkov Telescopes (IACTs) such as HESS\footnote{\url{https://www.mpi-hd.mpg.de/hfm/HESS/pages/dl3-dr1/}}, VERITAS\footnote{\url{https://veritas.sao.arizona.edu/}}, MAGIC\footnote{\url{http://vobs.magic.pic.es/fits/}}, and HAWC\footnote{\url{https://www.hawc-observatory.org/data/}} can accurately characterize the high-energy emission from PWNe, which is crucial for identifying how the relativistic particles are accelerated \citep{karg2013,renaud2009}. For the evolved systems like G327.1--1.1, where TeV emission is also detected from the PWN, we show that considering the evolution of the particles is essential for the most accurate characterization for the IC emission spectrum, especially for the oldest particles in the nebula. Further, the \fermi detections of PWNe G327.1--1.1 and B0453--685 are used to explore the parameter space that can best explain their broadband spectral features, comparing simple radiative models to semi-analytic evolutionary models. The initial results illustrate the significant radiative losses during late-phase PWN evolution and is consistent with the current understanding.  Hence, studying the PWN emission structure over the entire electromagnetic spectrum, as is done for G327.1--1.1 in Chapter~\ref{sec:g327} and B0453--685 in Chapter~\ref{sec:b0453}, offers the best way to understand the particle injection spectrum, its long term evolution, and the nature of the particles.


 \clearpage

    %
    %
    \begin{appendices}
        \begin{subappendices}
            
    \pdfoutput=1
\singlespacing
\setlength\LTcapwidth{1.0\textwidth}
\setlength{\tabcolsep}{0.03cm}
\centering
\begin{landscape}
\scriptsize{
\begin{longtable}{|c|c|c|c|c|c|c|c|}
\hline
PWN Name & $E = 535\,$MeV & $E = 1,705\,$MeV & $E = 5,432\,$MeV & $E = 17,303\,$MeV & $E = 55,116\,$MeV & $E = 175,560\,$MeV & $E = 559,204\,$MeV \\
            \hline
            \hline
			G8.4+0.15 & $8.82 \pm 1.03 \pm 7.02$ & $7.38 \pm 1.02 \pm 5.65$ & $8.33 \pm 1.27 \pm 3.9$ & $9.35 \pm 1.7 \pm 1.76$ & $12.2 \pm 2.78 \pm 1.36$ & $7.1 \pm 3.83 \pm 1.23$ & $12.8 \pm 8.15 \pm 1.06$ \\	
			\multirow{2}{*}{G11.0$-$0.05} & \multirow{2}{*}{$7.53 \pm 1.01 \pm 10.5$} & \multirow{2}{*}{$7.47 \pm 1 \pm 8.11$} & \multirow{2}{*}{$2.72 \pm 1.17 \pm 4.61$} & \multirow{2}{*}{$1.33 \pm 1.53 \pm 2.03$} & \multirow{2}{*}{$0.114 \pm 1.99 \pm 1.78$} & \multirow{2}{*}{$4.63$} & \multirow{2}{*}{$2.74 \pm 7.49 \pm 0.723$} \\
            G11.1+0.08 & & & & & & &  \\
			G11.2$-$0.35 & $1.51$ & $0.822 \pm 0.539 \pm 0.732$ & $1.58 \pm 0.47 \pm 0.352$ & $1.23 \pm 0.508 \pm 0.113$ & $1.56 \pm 0.857 \pm 0.0484$ & $1.05 \pm 1.31 \pm 0.0536$ & $5.78$ \\
			G12.8$-$0.02 & $25.2 \pm 1.14 \pm 4.91$ & $20 \pm 1.21 \pm 1.74$ & $10.5 \pm 1.45 \pm 0.611$ & $4.57 \pm 1.83 \pm 0.403$ & $4.73 \pm 2.58 \pm 0.935$ & $3.05 \pm 4.42 \pm 1.42$ & $7.91$ \\
			G15.4+0.10 & $14.9 \pm 0.905 \pm 6.52$ & $7.95 \pm 0.758 \pm 3.31$ & $2.49 \pm 0.721 \pm 1.32$ & $0.0576 \pm 0.829 \pm 0.61$ & $1.27$ & $3.57$ & $1.19 \pm 6.56 \pm 0.265$ \\
			G16.7+0.08 & $7.49 \pm 0.851 \pm 4.25$ & $3.56 \pm 0.594 \pm 1.67$ & $1.58 \pm 0.479 \pm 0.297$ & $0.354 \pm 0.458 \pm 0.044$ & $0.835$ & $1.83$ & $5.86$ \\
			G18.0$-$0.69 & $9.2 \pm 1.57 \pm 2.2$ & $13.2 \pm 1.86 \pm 5.41$ & $13.3 \pm 2.4 \pm 2.79$ & $18.3 \pm 3.28 \pm 0.888$ & $21.4 \pm 5 \pm 3.57$ & $32.3 \pm 8.76 \pm 5.45$ & $14.7$ \\
			G18.9$-$1.10 & $2.47 \pm 0.712 \pm 0.813$ & $3.89 \pm 0.511 \pm 0.243$ & $1.38 \pm 0.41 \pm 0.14$ & $0.0694 \pm 0.378 \pm 0.0441$ & $0.111 \pm 0.637 \pm 0.0396$ & $1.79$ & $5.84$ \\
			G20.2$-$0.20 & $6.51 \pm 0.861 \pm 1.22$ & $5.72 \pm 0.649 \pm 1.13$ & $1.08 \pm 0.451 \pm 0.262$ & $0.693 \pm 0.475 \pm 0.0838$ & $0.627$ & $2.21 \pm 1.87 \pm 0.15$ & $5.84$ \\
			G24.7+0.60 & $5.89 \pm 0.878 \pm 1.87$ & $7.34 \pm 0.738 \pm 1.18$ & $7.78 \pm 0.868 \pm 0.452$ & $4.61 \pm 1.11 \pm 0.285$ & $4.07 \pm 1.71 \pm 0.438$ & $6.7 \pm 3.6 \pm 0.58$ & $0.327 \pm 6.47 \pm 0.457$ \\
			\multirow{2}{*}{G25.2$-$0.19} & \multirow{2}{*}{$30 \pm 1.29 \pm 1.89$} & \multirow{2}{*}{$30.1 \pm 1.45 \pm 1.16$} & \multirow{2}{*}{$26.3 \pm 1.96 \pm 0.944$} & \multirow{2}{*}{$31.5 \pm 2.94 \pm 1.18$} & \multirow{2}{*}{$33.7 \pm 4.74 \pm 2.68$} & \multirow{2}{*}{$41.1 \pm 8.67 \pm 3.05$} & \multirow{2}{*}{$26 \pm 13.1 \pm 2.24$} \\
            G25.1+0.02 & & & & & & &  \\
			G26.6$-$0.10 & $12.8 \pm 1.06 \pm 1.81$ & $11.5 \pm 1.02 \pm 0.646$ & $8.12 \pm 1.21 \pm 0.352$ & $5.21 \pm 1.64 \pm 0.263$ & $12 \pm 2.98 \pm 1.65$ & $12.6 \pm 5.16 \pm 2.44$ & $6.45$ \\
			G27.8+0.60 & $3.31 \pm 0.725 \pm 1.57$ & $3.28 \pm 0.499 \pm 0.726$ & $2.6 \pm 0.467 \pm 0.165$ & $0.765 \pm 0.414 \pm 0.0342$ & $0.547$ & $1.85$ & $5.87$ \\
			G29.4+0.10 & $13.6 \pm 0.996 \pm 7.24$ & $8.1 \pm 0.849 \pm 4.45$ & $3.12 \pm 0.914 \pm 1.84$ & $1.68 \pm 1.13 \pm 0.693$ & $1.79 \pm 1.85 \pm 0.741$ & $0.308 \pm 2.58 \pm 0.889$ & $5.88$ \\
			G29.7$-$0.30 & $4.73 \pm 0.867 \pm 3.63$ & $2.77 \pm 0.579 \pm 2.06$ & $0.0183 \pm 0.361 \pm 0.0183$ & $0.732 \pm 0.474 \pm 0.0891$ & $1.07 \pm 0.803 \pm 0.0571$ & $1.80$ & $5.88$ \\
			G34.6$-$0.50 & $3.66 \pm 1.11 \pm 1.21$ & $2.55 \pm 0.969 \pm 1.18$ & $1.06 \pm 0.73 \pm 0.201$ & $0.910$ & $1.74$ & $1.81$ & $5.87$ \\
			G36.0+0.10 & $11.1 \pm 1.1 \pm 2.53$ & $10.1 \pm 1.08 \pm 0.763$ & $7 \pm 1.28 \pm 1.2$ & $7.55 \pm 1.79 \pm 0.915$ & $8.1 \pm 2.55 \pm 0.703$ & $6.36 \pm 3.91 \pm 1.55$ & $6.27 \pm 6.38 \pm 1.26$ \\
			G39.2$-$0.32 & $5.59 \pm 0.8 \pm 3.54$ & $2.84 \pm 0.547 \pm 1.5$ & $2.17 \pm 0.473 \pm 0.3$ & $0.523 \pm 0.408 \pm 0.057$ & $0.0484 \pm 0.546 \pm 0.0662$ & $1.87$ & $5.83$ \\
			G49.2$-$0.70 & $1.71 \pm 0.818 \pm 0.703$ & $1.93 \pm 0.721 \pm 0.981$ & $1.89 \pm 0.72 \pm 0.311$ & $1.65 \pm 0.803 \pm 0.127$ & $1.37 \pm 1 \pm 0.122$ & $1.75$ & $5.66$ \\
			G49.2$-$0.30 & $9.72 \pm 0.843 \pm 3.62$ & $8.83 \pm 0.769 \pm 2.44$ & $3.34 \pm 0.738 \pm 0.746$ & $2.14 \pm 0.82 \pm 0.269$ & $0.551 \pm 0.857 \pm 0.12$ & $2.12 \pm 1.83 \pm 0.122$ & $5.88$ \\
			G54.1+0.30 & $0.202 \pm 0.389 \pm 0.194$ & $0.601 \pm 0.22 \pm 0.226$ & $0.391 \pm 0.161 \pm 0.138$ & $0.575 \pm 0.199 \pm 0.0836$ & $0.0704 \pm 0.237 \pm 0.00977$ & $0.872 \pm 0.626 \pm 0.0903$ & $1.65$ \\
			G63.7+1.10 & $1.12 \pm 0.351 \pm 0.821$ & $1.55 \pm 0.261 \pm 0.357$ & $0.914 \pm 0.247 \pm 0.0841$ & $0.408 \pm 0.249 \pm 0.0244$ & $0.490$ & $1.64$ & $5.34$ \\
			G65.7+1.18 & $2.36 \pm 0.395 \pm 0.427$ & $1.99 \pm 0.292 \pm 0.195$ & $0.585 \pm 0.234 \pm 0.0548$ & $0.164 \pm 0.181 \pm 0.0157$ & $0.495$ & $1.51 \pm 1.47 \pm 0.0781$ & $5.33$ \\
			G74.9+1.11 & $3.25 \pm 0.546 \pm 1.3$ & $2.02 \pm 0.426 \pm 0.511$ & $1.77 \pm 0.417 \pm 0.296$ & $1.24 \pm 0.452 \pm 0.141$ & $0.229 \pm 0.41 \pm 0.0253$ & $3.38 \pm 1.97 \pm 0.153$ & $5.16$ \\
			G189.1+3.00 & $2.46 \pm 0.41 \pm 1.04$ & $3.69 \pm 0.419 \pm 19.2$ & $1.91 \pm 0.448 \pm 0.101$ & $2.03 \pm 0.523 \pm 0.0749$ & $0.945 \pm 0.599 \pm 0.051$ & $0.828 \pm 0.725 \pm 0.0542$ & $2.57$ \\
			G266.9$-$1.10 & $0.746 \pm 0.265 \pm 0.679$ & $0.535 \pm 0.16 \pm 0.409$ & $0.0866$ & $0.0654 \pm 0.157 \pm 0.0257$ & $0.179$ & $0.423 \pm 0.599 \pm 0.0787$ & $1.95 \pm 1.9 \pm 0.384$ \\
			G279.6$-$31.70 & $1.49 \pm 0.23 \pm 0.0451$ & $1.74 \pm 0.226 \pm 0.0533$ & $1.28 \pm 0.255 \pm 0.039$ & $0.613 \pm 0.279 \pm 0.0188$ & $1.24 \pm 0.647 \pm 0.0375$ & $5.45$ & $3.51 \pm 3.57 \pm 0.281$ \\
			G279.8$-$35.80 & $0.104 \pm 0.085 \pm 0.007$ & $0.162 \pm 0.0541 \pm 0.0156$ & $0.148 \pm 0.0561 \pm 0.00468$ & $0.103$ & $0.109$ & $0.348$ & $1.23$ \\
			G304.1$-$0.24 & $1.98 \pm 0.573 \pm 2.81$ & $2.98 \pm 0.498 \pm 2.09$ & $2.56 \pm 0.557 \pm 0.763$ & $3.54 \pm 0.823 \pm 0.202$ & $5.97 \pm 1.6 \pm 0.286$ & $0.899 \pm 1.6 \pm 0.203$ & $6.98 \pm 5.1 \pm 0.565$ \\
			G315.8$-$0.23 & $4.43 \pm 0.582 \pm 1.76$ & $1.41 \pm 0.378 \pm 0.587$ & $0.589 \pm 0.29 \pm 0.12$ & $0.246$ & $0.0797 \pm 0.518 \pm 0.0195$ & $1.04 \pm 1.18 \pm 0.0757$ & $4.93$ \\
			G318.9+0.40 & $0.0183 \pm 0.372 \pm 0.172$ & $1.07 \pm 0.223 \pm 0.213$ & $0.549 \pm 0.168 \pm 0.125$ & $0.363$ & $0.0329 \pm 0.214 \pm 0.0154$ & $0.924$ & $1.26$ \\
			G326.2$-$1.70 & $0.347 \pm 0.367 \pm 0.347$ & $1.56 \pm 0.325 \pm 0.49$ & $1.73 \pm 0.403 \pm 0.195$ & $1.58 \pm 0.565 \pm 0.151$ & $1.35 \pm 0.713 \pm 0.11$ & $0.745 \pm 1.01 \pm 0.13$ & $4.41$ \\
			G327.1$-$1.10 & $1.1 \pm 0.365 \pm 2.62$ & $0.946 \pm 0.212 \pm 0.738$ & $0.301 \pm 0.147 \pm 0.115$ & $0.219 \pm 0.144 \pm 0.0322$ & $0.229$ & $0.421$ & $1.21$ \\
			G328.4+0.20 & $20.1 \pm 1 \pm 4.53$ & $20.6 \pm 1.07 \pm 3.03$ & $16.5 \pm 1.4 \pm 1.79$ & $13.8 \pm 1.93 \pm 1.17$ & $12.1 \pm 2.73 \pm 1.07$ & $2.69 \pm 3.45 \pm 1.84$ & $10.3$ \\
			\multirow{2}{*}{G332.5$-$0.30} & \multirow{2}{*}{$12 \pm 0.96 \pm 2.2$} & \multirow{2}{*}{$11.4 \pm 0.907 \pm 1.32$} & \multirow{2}{*}{$11.8 \pm 1.14 \pm 0.977$} & \multirow{2}{*}{$12.5 \pm 1.7 \pm 0.578$} & \multirow{2}{*}{$15 \pm 2.78 \pm 1.02$} & \multirow{2}{*}{$12.4 \pm 4.67 \pm 2.09$} & \multirow{2}{*}{$16.1 \pm 8.29 \pm 1.46$} \\
            G332.5$-$0.28 & & & & & & &  \\
			G336.4+0.10 & $4.25 \pm 0.884 \pm 1.78$ & $6.16 \pm 0.794 \pm 0.894$ & $5.1 \pm 0.893 \pm 0.621$ & $2.33 \pm 1.08 \pm 0.559$ & $8.05 \pm 2.17 \pm 0.551$ & $11.3 \pm 4.33 \pm 0.748$ & $3.05 \pm 3.87 \pm 0.355$ \\
			G337.2+0.1 & $0.638 \pm 0.888 \pm 0.549$ & $1.16 \pm 0.645 \pm 0.16$ & $0.458 \pm 0.501 \pm 0.087$ & $0.474 \pm 0.471 \pm 0.0507$ & $1.06 \pm 0.773 \pm 0.084$ & $1.60$ & $3.48 \pm 3.81 \pm 0.281$ \\
			G337.5$-$0.1 & $4.24 \pm 0.878 \pm 1.91$ & $2.31 \pm 0.621 \pm 0.879$ & $1.79 \pm 0.502 \pm 0.189$ & $0.414 \pm 0.443 \pm 0.0655$ & $0.425 \pm 0.664 \pm 0.0954$ & $0.637 \pm 1.46 \pm 0.119$ & $5.24$ \\
			G338.2$-$0.00 & $1.71 \pm 0.86 \pm 0.266$ & $2.79 \pm 0.643 \pm 0.667$ & $3.75 \pm 0.588 \pm 0.214$ & $2.64 \pm 0.671 \pm 0.0969$ & $7.06 \pm 1.6 \pm 0.221$ & $9.02 \pm 3.32 \pm 0.414$ & $5.68$ \\
\hline
\caption{\normalsize The spectral flux $E^2 \frac{dN}{dE}$ per bin for seven logarithmically spaced energy bins for all 39 detected sources (Tables~\ref{tab:extent} and \ref{tab:points}). All flux values are in units 10$^{-6}$ MeV cm$^{-2}$ s$^{-1}$. The first quoted error is the $1 \sigma$ statistical error and the latter is the total systematic error. Flux values that lack quoted errors are instead the 95\% upper limit flux for that bin.} \label{tab:spectral_fluxes}
\end{longtable}}
\end{landscape}
\doublespacing

\nopagebreak

\singlespacing
\setlength\LTcapwidth{1.0\textwidth}
\setlength{\tabcolsep}{0.1cm}
\footnotesize{
\begin{landscape}
\begin{longtable}{|l|c|c|c|c|c|c|c|c|c|}
\hline
PWN Name & $E = 477\,$MeV & $E = 1205\,$MeV & $E = 3044\,$MeV & $E = 7690\,$MeV & $E = 19,429\,$MeV & $E = 49,086\,$MeV & $E = 124,017\,$MeV & $E = 313,325\,$MeV & $E = 791,608\,$MeV \\
\hline
\hline
G0.9+0.10 & 2.84 $\times 10^{-7}$ &  7.12 $\times 10^{-7}$ &  1.59 $\times 10^{-6}$ &  5.79 $\times 10^{-7}$ &  2.44 $\times 10^{-7}$ &  8.08 $\times 10^{-7}$ &  1.65 $\times 10^{-6}$ &  3.32 $\times 10^{-6}$ &  4.33 $\times 10^{-6}$  \\
G23.5+0.10 & $3.89 \times 10^{-6}$ & $3.07 \times 10^{-6}$ & $1.63 \times 10^{-6}$ & $1.46 \times 10^{-6}$ & $3.93 \times 10^{-7}$ & $9.18\times 10^{-7}$ & $1.10\times 10^{-6}$ & $1.71\times 10^{-6}$ & $2.71\times 10^{-6}$ \\
G32.64+0.53 & 8.87 $\times 10^{-7}$ &  1.18 $\times 10^{-6}$ &  7.53 $\times 10^{-7}$ &  5.30 $\times 10^{-7}$ &  1.16 $\times 10^{-6}$ &  1.27 $\times 10^{-6}$ &  4.51 $\times 10^{-7}$ &  9.62 $\times 10^{-7}$ &  3.79 $\times 10^{-6}$  \\
G47.4--3.90 & 4.60 $\times 10^{-7}$ &  3.94 $\times 10^{-7}$ &  4.12 $\times 10^{-7}$ &  1.44 $\times 10^{-7}$ &  3.95 $\times 10^{-7}$ &  1.65 $\times 10^{-7}$ &  4.61 $\times 10^{-7}$ &  9.0 $\times 10^{-7}$ &  2.45 $\times 10^{-6}$ \\
G74.0--8.50 & 1.13 $\times 10^{-7}$ &  2.06 $\times 10^{-7}$ &  2.16 $\times 10^{-7}$ &  4.00 $\times 10^{-7}$ &  2.43 $\times 10^{-7}$ &  5.82 $\times 10^{-7}$ &  1.48 $\times 10^{-6}$ &  3.943 $\times 10^{-6}$ &  1.02 $\times 10^{-5}$  \\
G93.3+6.90 & 2.85 $\times 10^{-7}$ &  4.05 $\times 10^{-7}$ &  2.41 $\times 10^{-7}$ &  3.62 $\times 10^{-7}$ &  2.57 $\times 10^{-7}$ &  1.54 $\times 10^{-7}$ &  8.51 $\times 10^{-7}$ &  7.49 $\times 10^{-7}$ &  2.03 $\times 10^{-6}$  \\
G108.6+6.80 & 5.80 $\times 10^{-7}$ &  2.77 $\times 10^{-7}$ &  8.44 $\times 10^{-8}$ &  7.65 $\times 10^{-8}$ &  1.29 $\times 10^{-7}$ &  3.93 $\times 10^{-7}$ &  2.67 $\times 10^{-7}$ &  7.35 $\times 10^{-7}$ &  1.82 $\times 10^{-6}$ \\
G141.2+5.00 &  4.40 $\times 10^{-7}$ &  3.03 $\times 10^{-7}$ &  5.70 $\times 10^{-8}$ &  8.61 $\times 10^{-8}$ &  1.55 $\times 10^{-7}$ &  1.32 $\times 10^{-7}$ &  3.13 $\times 10^{-7}$ &  7.0 $\times 10^{-7}$ &  1.84 $\times 10^{-6}$ \\
G179.7--1.70 & 1.27 $\times 10^{-6}$ &  3.43 $\times 10^{-7}$ &  1.05 $\times 10^{-7}$ &  7.50 $\times 10^{-8}$ &  2.35 $\times 10^{-7}$ &  6.76 $\times 10^{-7}$ &  4.03 $\times 10^{-7}$ &  1.00 $\times 10^{-6}$ &  2.53 $\times 10^{-6}$ \\
G290.0--0.93 & 6.49 $\times 10^{-7}$ &  4.83 $\times 10^{-7}$ &  4.93 $\times 10^{-7}$ &  1.11 $\times 10^{-7}$ &  2.44 $\times 10^{-7}$ &  8.61 $\times 10^{-7}$ &  3.22 $\times 10^{-7}$ &  9.64 $\times 10^{-7}$ &  2.24 $\times 10^{-6}$ \\
G310.6--1.60 &  1.28 $\times 10^{-7}$ &  4.75 $\times 10^{-8}$ &  5.32 $\times 10^{-8}$ &  9.87 $\times 10^{-8}$ &  3.30 $\times 10^{-7}$ &  3.38 $\times 10^{-7}$ &  1.11 $\times 10^{-6}$ &  9.03 $\times 10^{-7}$ &  2.13 $\times 10^{-6}$ \\
G322.5--0.10 & 1.95 $\times 10^{-6}$ &  9.32 $\times 10^{-7}$ &  5.91 $\times 10^{-7}$ &  5.45 $\times 10^{-7}$ &  1.53 $\times 10^{-7}$ &  2.70 $\times 10^{-7}$ &  3.46 $\times 10^{-7}$ &  1.16 $\times 10^{-6}$ &  2.20 $\times 10^{-6}$ \\
G341.2+0.90 & 4.44 $\times 10^{-7}$ &  2.85 $\times 10^{-7}$ &  1.61 $\times 10^{-7}$ &  1.73 $\times 10^{-7}$ &  1.62 $\times 10^{-7}$ &  2.70 $\times 10^{-7}$ &  8.44 $\times 10^{-7}$ &  1.03 $\times 10^{-6}$ &  2.43 $\times 10^{-6}$ \\
G350.2--0.80 & 1.90 $\times 10^{-6}$ &  1.13 $\times 10^{-6}$ &  8.65 $\times 10^{-7}$ &  7.39 $\times 10^{-7}$ &  5.73 $\times 10^{-7}$ &  5.49 $\times 10^{-7}$ &  6.94 $\times 10^{-7}$ &  1.01 $\times 10^{-6}$ &  2.40 $\times 10^{-6}$ \\
G358.3+0.24 &  1.01 $\times 10^{-6}$ &  8.49 $\times 10^{-7}$ &  6.52 $\times 10^{-7}$ &  6.77 $\times 10^{-7}$ &  3.69 $\times 10^{-7}$ &  4.37 $\times 10^{-7}$ &  8.43 $\times 10^{-7}$ &  9.51 $\times 10^{-7}$ &  2.58 $\times 10^{-6}$ \\
G358.6--17.2 &  1.02 $\times 10^{-7}$ &  7.12 $\times 10^{-8}$ &  5.21 $\times 10^{-8}$ &  7.15 $\times 10^{-8}$ &  2.11 $\times 10^{-7}$ &  2.06 $\times 10^{-7}$ &  3.58 $\times 10^{-7}$ &  9.57 $\times 10^{-7}$ &  2.48 $\times 10^{-6}$ \\
\hline
\caption{\normalsize The 95\% upper limit to the spectral flux $E^2 \frac{dN}{dE}$ per bin for nine logarithmically spaced energy bins for the 16 ROIs with no source detection (Table~\ref{tab:nondetect}). All flux values are in units MeV cm$^{-2}$ s$^{-1}$. \label{tab:spectral_ul_fluxes}}
\end{longtable}
\end{landscape}}
\doublespacing \clearpage

        \end{subappendices}
    \end{appendices}

    \singlespacing                             

    %
    %
    %
    %
    %
    %
    %
    \bibliographystyle{plain}
    \addtotoc{Bibliography}{\bibliography{thesis.bib}}
\end{document}